\documentclass[aoas]{imsart}
\RequirePackage{amsthm,amsmath,amsfonts,amssymb}
\RequirePackage{natbib}
\RequirePackage[colorlinks,citecolor=blue,urlcolor=blue]{hyperref}
\RequirePackage{graphicx}
\usepackage{mathrsfs}
\usepackage{enumerate}
\usepackage{tabularx}

\usepackage[T1]{fontenc}    
\usepackage{hyperref}       
\usepackage{url}            
\usepackage{booktabs}       
\usepackage{amsfonts}       
\usepackage{nicefrac}       
\usepackage{microtype}      
\usepackage{lipsum}		
\usepackage{float}
\usepackage{capt-of}
\usepackage{multirow}
\usepackage{xcolor}
\usepackage{bm}
\usepackage[utf8]{inputenc}
\usepackage[ruled]{algorithm2e}
\usepackage{extarrows}
\usepackage{indentfirst}
\usepackage{listings}
\usepackage{pdfpages}
\usepackage{subfig}
\usepackage{enumitem} 
\usepackage{mathtools}
\usepackage{tikz}
\usepackage{xfp}
\usepackage{appendix}

\DeclareMathOperator*{\argmin}{arg\,min}
\DeclareMathOperator*{\argmax}{arg\,max}

\DeclareMathOperator{\Ae}{AE}
\DeclareMathOperator{\ET}{ET}
\DeclareMathOperator{\si}{si}
\DeclareMathOperator{\SI}{SI}
\DeclareMathOperator{\DB}{DB}

\startlocaldefs
\theoremstyle{plain}

\newtheorem{theorem}{Theorem}[section]

\theoremstyle{definition}
\newtheorem{definition}[theorem]{Definition}
\newtheorem{remark}[theorem]{Remark}

\endlocaldefs

\begin{document}

\begin{frontmatter}
\title{A Principal Submanifold-based Approach for Clustering and Multiscale RNA Correction}
\runtitle{Submanifold-based Clustering and RNA correction}

\begin{aug}
\author[A]{\fnms{Menghao}~\snm{Wu}\ead[label=e1]{menghao.wu@u.nus.edu}},
\author[B]{\fnms{Zhigang}~\snm{Yao}\ead[label=e2]{zhigang.yao@nus.edu.sg}} 
\address[A]{Department of Statistics and Data Science,
National University of Singapore, 117546 Singapore\printead[presep={ ,\ }]{e1}}

\address[B]{Department of Statistics and Data Science,
National University of Singapore, 117546 Singapore\printead[presep={,\ }]{e2}}

\end{aug}

\begin{abstract}
RNA structure determination is essential for understanding its biological functions. However, the reconstruction process often faces challenges, such as atomic clashes, which can lead to inaccurate models. To address these challenges, we introduce the principal submanifold (PSM) approach for analyzing RNA data on a torus. This method provides an accurate, low-dimensional feature representation, overcoming the limitations of previous torus-based methods. By combining PSM with DBSCAN, we propose a novel clustering technique, the principal submanifold-based DBSCAN (PSM-DBSCAN). Our approach achieves superior clustering accuracy and increased robustness to noise. Additionally, we apply this new method for multiscale corrections, effectively resolving RNA backbone clashes at both microscopic and mesoscopic scales. Extensive simulations and comparative studies highlight the enhanced precision and scalability of our method, demonstrating significant improvements over existing approaches. The proposed methodology offers a robust foundation for correcting complex RNA structures and has broad implications for applications in structural biology and bioinformatics.
\end{abstract}

\begin{keyword}
\kwd{Geometric statistics}
\kwd{Principal submanifold}
\kwd{Dimensionality reduction}
\kwd{High-dimensional clustering}
\kwd{Clash correction}
\end{keyword}

\end{frontmatter}


\section{Introduction}

\subsection{Background}
In the realm of life sciences, ribonucleic acid (RNA) is a pivotal molecule that undertakes various biological functions. Beyond its role as a genetic information conveyor and participant in protein synthesis, RNA is instrumental in regulating gene expression, catalyzing biochemical reactions, and other critical biological processes \citep{kieft2002crystal,moore1998three, jinek2014structures, chen2020expanding}. Research has revealed that RNA’s functionality is largely dictated by its $3$-dimensional structure rather than merely its nucleotide sequence. Consequently, a profound understanding of RNA’s spatial configuration is required to comprehend its functional mechanisms and decode its multifaceted roles in cellular processes \citep{doudna2002chemical,lilley2000structures}.

In recent years, the development of high-resolution structural determination techniques such as X-ray crystallography, nuclear magnetic resonance (NMR), and cryo-electron microscopy (Cryo-EM) has significantly enhanced our understanding of the interplay between RNA structure and function \citep{shen1995rna,jain2015computational}. However, reconstructing RNA’s backbone conformation remains challenging, often leading to RNA clashes, where atoms are too close based on van der Waals radii, resulting in physically implausible structures \citep{batsanov2001van,sponer2018rna}. These clashes highlight potential inaccuracies in structural prediction and require advanced computational correction methods. Effective clash correction is essential for accurate RNA modeling, as unresolved clashes can distort structural interpretations and hinder insights into RNA’s roles in catalysis, binding, and regulation. Improving RNA structural reliability through clash correction enables more precise models, crucial for studying RNA’s functions in biological systems and diseases \citep{serganov2013decade}. As RNA research advances, computational refinement tools are increasingly vital, not only in academia but also in medical and biotechnological applications, aiding the development of RNA-based therapeutics and diagnostics \citep{zhu2022rna,lu2024rna}.

For RNA clashes, most attention is focused on suites, which are segments of an RNA molecule spanning from one sugar ring to the next \citep{murray2003rna}. A comprehensive understanding of suites is achieved by adopting a dual-scale perspective that encompasses both microscopic and mesoscopic scales. At the microscopic scale, analysis concentrates on the atomic-level details of a suite’s RNA backbone, characterized by its dihedral angles that capture the conformational intricacies of the alternating phosphate and ribose sugar units with attached nitrogenous bases. Importantly, RNA dihedral angle data are not distributed uniformly on the torus. Empirical analyses demonstrate that RNA dihedral angle data are highly concentrated in low-dimensional substructures within the torus, rather than being uniformly distributed across the high-dimensional space. Biologically, this is consistent with steric hindrance, hydrogen bonding, and electrostatic interactions, which severely restrict the feasible combinations of torsion angles. These physical constraints effectively confine RNA suites to low-dimensional geometric structures within torus. This observation implies moving a suite toward a global average configuration would mix multiple conformational energy wells and lead to nonphysical structures. Instead, meaningful correction must be carried out within the correct conformational family to which the suite biologically belongs. This biological requirement naturally determines a statistical workflow in which (i) dimension reduction is used to separate conformational families geometrically, (ii) clustering identifies the appropriate family membership of each suite, and (iii) clash correction is then applied within that cluster. In contrast, the mesoscopic scale expands the analysis to a sequence of suites, emphasizing the geometric positioning of sugar ring centers along the RNA strand to capture the overall shape and form of the molecule. At this level, researchers examine interactions among multiple nucleotides, which give rise to structural features such as helices, loops, and bulges \citep{shivashankar2002mesoscopic}. Size-and-shape analysis at the mesoscopic scale thus provides a framework for understanding the geometric configuration of RNA molecules beyond the atomic details \citep{dryden2016statistical}.

\subsection{Existing methods}
Currently, there are two main methods to correct RNA clashes. One method is based on molecular dynamics. According to approximations based on biophysical and chemical laws, simulated atoms can move toward the position of lowest energy, as exemplified by ERRASER \citep{chou2013correcting}, which exhaustively samples nucleotide conformations and refines them using an energy function and electron-density correlation. ERRASER effectively corrects the majority of geometric errors while maintaining or improving agreement with diffraction data. However, it has limitations: the process is time-consuming and, despite its precision, this method often fails to resolve all clashes. 

Another method is data-driven by considering the geometric structure of RNA. The main challenge is performing a high-dimensional clustering of RNA suites on the torus. Traditional methods include linear and non-linear methods. Linear methods, such as PCA, efficiently capture variance but fail to fit non-linear structures such as a torus, leading to information loss. non-linear methods, including Isomap \citep{tenenbaum2000global}, LLE \citep{roweis2000nonlinear,donoho2003hessian}, t-SNE \citep{van2008visualizing}, and UMAP \citep{mcinnes2018umap}, better capture complex geometries by preserving local manifold structures. However, they lack strong theoretical guarantees, are computationally expensive at scale, and are sensitive to parameter choices, often failing to retain global structures. 

 To handle the RNA data on the high-dimensional torus, \cite{wiechers2023learning} proposed an RNA correction method MINT-AGE-CLEAN, a data-driven method utilizing MINT-AGE \citep{mardia2022principal} to categorize RNA data into different classes, before correcting RNA clashes across both microscopic and mesoscopic scales. It overcomes the limitations of traditional dimensionality reduction methods by mapping RNA data on the torus. The central idea is Torus PCA (tPCA) \citep{eltzner2018torus}, which involves a transformation — known as Torus-to-Stratified-Sphere (TOSS) mapping — that converts the data from a torus to a sphere. Although this approach allows the adaptation of principal nested spheres analysis (PNS) \citep{jung2012analysis}, a limitation still exists: tPCA fails to accurately capture the geometric structure of RNA data in its native space (e.g., the torus). Therefore, there is still room for improvement by exploring the hidden structure of its intrinsic space. 
Moreover, resolving clashes requires distinguishing between heterogeneous conformations. Different RNA suites correspond to distinct geometric patterns of torsion angles, and clash correction must consider these differences. A single global correction across all suites would obscure such heterogeneity and risk biologically implausible adjustments. Therefore, clustering RNA data into coherent groups along their geometric structures provides the essential basis for applying meaningful, suite-specific corrections.

\subsection{Main Contribution}

For high-dimensional clustering, the general approach is to encode features into a low-dimensional space prior to clustering. The quality of low-dimensional representation is crucial, as it directly impacts the preservation of the data's intrinsic structure and, consequently, the overall clustering performance. As mentioned above, RNA dihedral-angle data themselves are constrained by biophysics and concentrate on a low-dimensional structure on the torus, which is consistent with the manifold hypothesis \citep{fefferman2016testing}, suggesting that high-dimensional data often concentrate near a low-dimensional manifold. We note that RNA dihedral angle data are intrinsically defined on the torus $\mathbb{T}^D$. Therefore, the relevant task is not to embed arbitrary Euclidean data, but to explicitly estimate meaningful low-dimensional submanifolds within the torus that capture the geometric structure of RNA conformations. We adopt the Principal Submanifold (PSM) approach \citep{yao2024principalsubmanifolds,yao2023submanifold} to handle these high-dimensional data on the torus. Our work makes the following key contributions:

\begin{itemize}
    \item 
    PSM can accurately fit low-dimensional principal submanifolds with more general structures. Unlike existing methods like tPCA, which rely on spherical transformations and cause unignorable information loss, PSM preserves both local and global geometric features directly on the torus. 

    \item
    We introduce PSM-DBSCAN, a clustering method that combines PSM with the density-based DBSCAN algorithm \citep{ester1996density}. PSM provides a precise low-dimensional representation, enabling DBSCAN to detect clusters more accurately, even in complex and noisy data. Compared to tPCA-DBSCAN, MINT-AGE and traditional methods such as hierarchical and spectral clustering, PSM-DBSCAN consistently achieves higher accuracy and stability. 
    
    \item 
    We evaluate PSM-DBSCAN for RNA correction, demonstrating its practical utility in real-world scenarios. The enhanced clustering performance of PSM-DBSCAN leads to more accurate RNA correction by ensuring more reliable categorization of RNA data. Comparative studies show that our method outperforms MINT-AGE-CLEAN, achieving superior results in resolving atomic clashes and modeling complex RNA structures.

    \item 
      The proposed PSM is a general framework for data on the torus,which is geometry-oriented and task-independent. Unlike many modern machine-learning approaches that are driven by a specific downstream task, the principal submanifold is constructed purely from the geometric properties of the data on the torus. That is, PSM uncovers the intrinsic geometric structure directly from the data itself, without relying on biological supervision or any task-specific objective. This intrinsic representation serves as a universal geometric foundation that downstream pipelines may subsequently leverage for their own objectives.
yao2025mf      
\end{itemize}

\begin{figure}[H]

\includegraphics[width=1.0\textwidth]{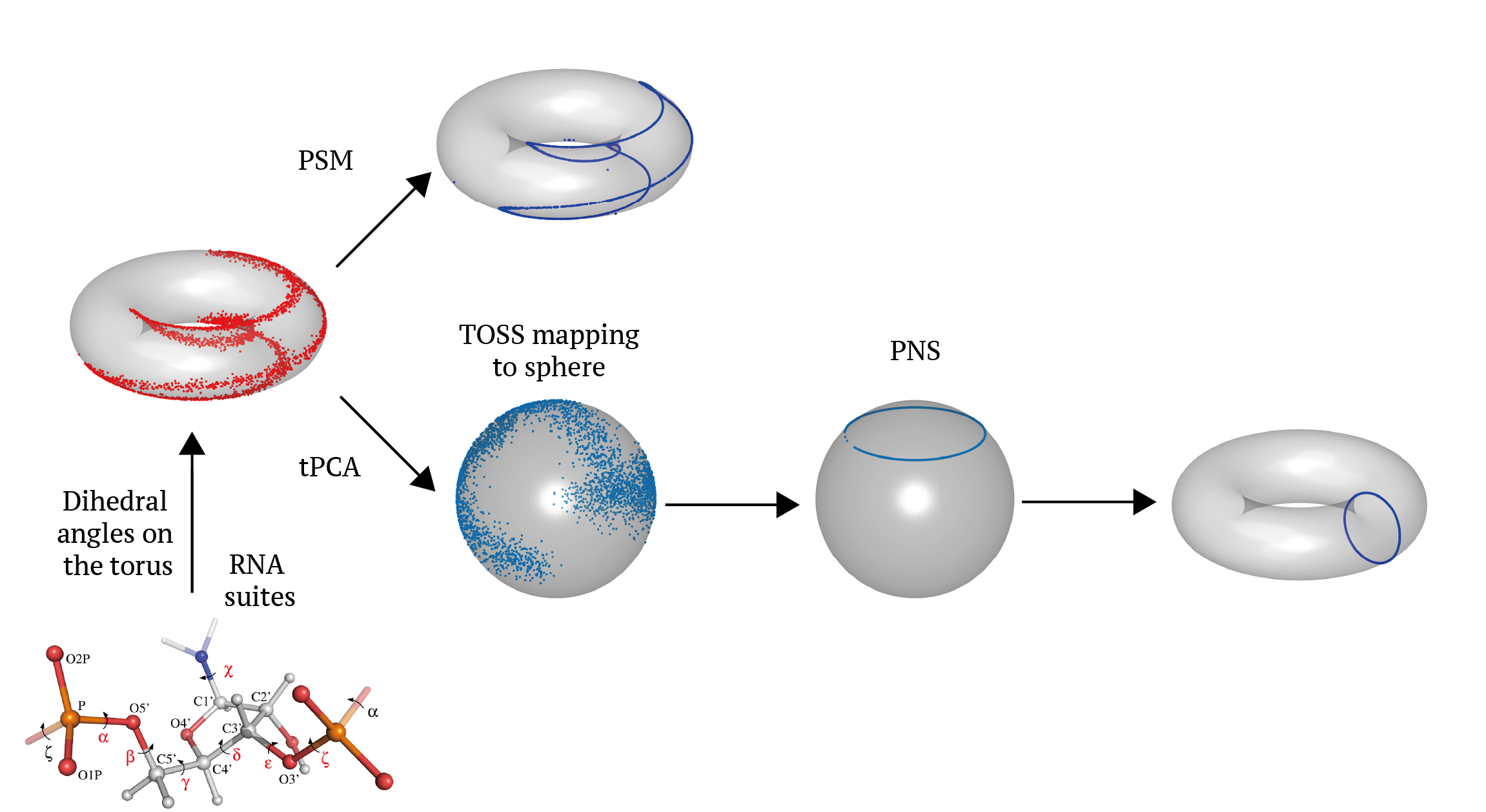} 
\caption{\textit{Visualization of the PSM and tPCA. RNA backbone dihedral-angle data are intrinsically torus-valued. The figure contrasts two low-dimensional fitting strategies: PSM directly fits a low-dimensional principal submanifold on the torus, yielding an intrinsic low-dimensional structure; tPCA maps the torus data (e.g., via a TOSS mapping) to the sphere for low-dimensional fitting (e.g., PNS) and maps back, extracting a corresponding low-dimensional fitted structure on the torus.
}}
\label{fig: 1}
\end{figure}

Without delving into details, we illustrate the improvement in two steps: dimensionality reduction and clustering performance. Figure \ref{fig: 1} illustrates samples on a $2$-dimensional torus in a $3$-dimensional space, situated around a $1$-dimensional curve. We can see that PSM accurately fits samples onto a $1$-dimensional curve, while tPCA fits samples onto a circle, thereby losing critical information of data. This highlights PSM's advantage in aligning the samples more precisely with the underlying low-dimensional structure, preserving the intrinsic geometric information in the data and retaining both local and global geometric features. 

\begin{figure}[H]
\centering
\subfloat[]{
  \includegraphics[width=0.3\textwidth]{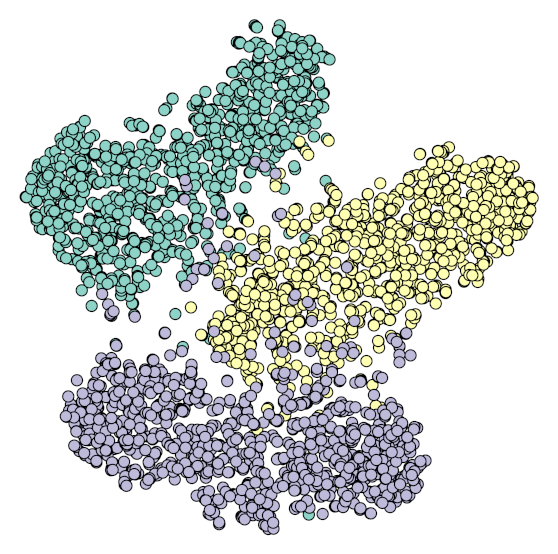}
}
\hfill
\subfloat[]{
  \includegraphics[width=0.3\textwidth]{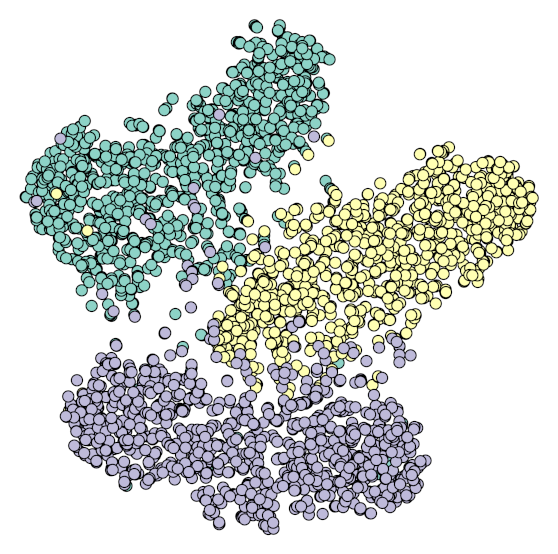}
}
\hfill
\subfloat[]{
  \includegraphics[width=0.3\textwidth]{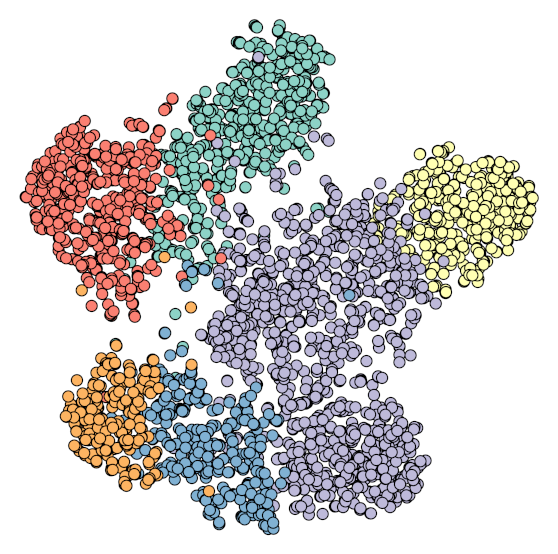}
}
\newline
\caption{ \textit{An intuitive illustration of clustering performance on the high-dimensional data using t-SNE. (a) shows the visualization of three different categories of points, (b) shows the clustering performance of PSM-DBSCAN and (c) shows MINT-AGE.}}
\label{fig: 2}
\end{figure}

For clustering, we consider a high-dimensional scenario where samples are situated around three $1$-dimensional curves on a $7$-dimensional torus, with added Gaussian noise (Figure \ref{fig: 2}(a)). Figure \ref{fig: 2}(b) demonstrates that PSM can effectively distinguish samples from different curves, forming three well-defined and compact clusters. In contrast, in Figure \ref{fig: 2}(c), MINT-AGE fails to differentiate between samples from distinct curves, resulting in a larger number of clusters and mistakenly grouping samples from two separate curves into the same cluster.

\subsection{Outline}
In Section \ref{Methodology}, we introduce the framework of the principal submanifold. We first describe the basic setup, then present PSM for dimensionality reduction and PSM-DBSCAN for clustering, followed by our method for RNA multiscale correction. In Section \ref{Simulations}, we provide numerical simulations that demonstrate the superior performance of PSM and the associated clustering techniques. Section \ref{Application for RNA data} details the RNA correction in real data. Finally, Section \ref{Disccusion} summarizes the main findings and contributions of our study and discusses several promising future works.

\section{Methodology}\label{Methodology}

\subsection{Preliminary}\label{Preliminary}

\begin{figure}[H]
\centering

\tikzset{every picture/.style={line width=0.75pt}} 

\begin{tikzpicture}[x=0.75pt,y=0.75pt,yscale=-1,xscale=1, scale=0.7]

\draw  [color={rgb, 255:red, 0; green, 0; blue, 0 }  ,draw opacity=1 ][fill={rgb, 255:red, 74; green, 74; blue, 74 }  ,fill opacity=1 ] (361,135.02) .. controls (361,127.28) and (367.28,121) .. (375.02,121) .. controls (382.77,121) and (389.05,127.28) .. (389.05,135.02) .. controls (389.05,142.77) and (382.77,149.05) .. (375.02,149.05) .. controls (367.28,149.05) and (361,142.77) .. (361,135.02) -- cycle ;
\draw    (225.47,203.73) -- (310,202.73) ;
\draw [line width=1.5]    (170.02,118.05) -- (183.2,145.6) ;
\draw    (191.05,135.71) -- (361,135.02) ;
\draw   (156,104.02) .. controls (156,96.28) and (162.28,90) .. (170.02,90) .. controls (177.77,90) and (184.05,96.28) .. (184.05,104.02) .. controls (184.05,111.77) and (177.77,118.05) .. (170.02,118.05) .. controls (162.28,118.05) and (156,111.77) .. (156,104.02) -- cycle ;
\draw  [fill={rgb, 255:red, 155; green, 155; blue, 155 }  ,fill opacity=1 ] (197.42,201.73) .. controls (197.42,193.99) and (203.7,187.71) .. (211.44,187.71) .. controls (219.19,187.71) and (225.47,193.99) .. (225.47,201.73) .. controls (225.47,209.48) and (219.19,215.76) .. (211.44,215.76) .. controls (203.7,215.76) and (197.42,209.48) .. (197.42,201.73) -- cycle ;
\draw  [fill={rgb, 255:red, 255; green, 255; blue, 255 }  ,fill opacity=1 ] (310,202.73) .. controls (310,194.99) and (316.28,188.71) .. (324.02,188.71) .. controls (331.77,188.71) and (338.05,194.99) .. (338.05,202.73) .. controls (338.05,210.48) and (331.77,216.76) .. (324.02,216.76) .. controls (316.28,216.76) and (310,210.48) .. (310,202.73) -- cycle ;
\draw    (389.05,135.02) -- (413.2,135.6) ;
\draw    (413.2,135.6) -- (362.2,203.6) ;
\draw    (338.05,203.73) -- (362.2,203.31) ;
\draw    (191.05,135.71) -- (140.05,203.71) ;
\draw    (140.05,203.71) -- (197.42,203.73) ;
\draw [line width=1.5]    (185.2,149.6) -- (189.38,159.15) ;
\draw [line width=1.5]    (192.2,164.6) -- (196.38,173.15) ;
\draw [line width=1.5]    (199.2,179.6) -- (204.38,190.15) ;
\draw    (184.05,104.02) -- (317.2,103.6) ;
\draw    (95.2,104.6) -- (156,104.02) ;
\draw    (95.2,104.6) -- (140.05,203.71) ;
\draw    (317.35,103.2) -- (362.2,203.31) ;
\draw  [draw opacity=0] (125.43,172.41) .. controls (125.51,171.83) and (125.65,171.26) .. (125.85,170.7) .. controls (128.14,164.26) and (137.62,161.75) .. (147.01,165.1) .. controls (153.49,167.42) and (158.24,171.91) .. (159.8,176.6) -- (142.85,176.76) -- cycle ; \draw   (125.43,172.41) .. controls (125.51,171.83) and (125.65,171.26) .. (125.85,170.7) .. controls (128.14,164.26) and (137.62,161.75) .. (147.01,165.1) .. controls (153.49,167.42) and (158.24,171.91) .. (159.8,176.6) ;  
\draw   (443,108.02) .. controls (443,100.28) and (449.28,94) .. (457.02,94) .. controls (464.77,94) and (471.05,100.28) .. (471.05,108.02) .. controls (471.05,115.77) and (464.77,122.05) .. (457.02,122.05) .. controls (449.28,122.05) and (443,115.77) .. (443,108.02) -- cycle ;
\draw  [fill={rgb, 255:red, 255; green, 255; blue, 255 }  ,fill opacity=1 ] (481,189.73) .. controls (481,181.99) and (487.28,175.71) .. (495.02,175.71) .. controls (502.77,175.71) and (509.05,181.99) .. (509.05,189.73) .. controls (509.05,197.48) and (502.77,203.76) .. (495.02,203.76) .. controls (487.28,203.76) and (481,197.48) .. (481,189.73) -- cycle ;
\draw  [fill={rgb, 255:red, 155; green, 155; blue, 155 }  ,fill opacity=1 ] (470.42,197.73) .. controls (470.42,189.99) and (476.7,183.71) .. (484.44,183.71) .. controls (492.19,183.71) and (498.47,189.99) .. (498.47,197.73) .. controls (498.47,205.48) and (492.19,211.76) .. (484.44,211.76) .. controls (476.7,211.76) and (470.42,205.48) .. (470.42,197.73) -- cycle ;
\draw  [color={rgb, 255:red, 0; green, 0; blue, 0 }  ,draw opacity=1 ][fill={rgb, 255:red, 74; green, 74; blue, 74 }  ,fill opacity=1 ] (528,118.02) .. controls (528,110.28) and (534.28,104) .. (542.02,104) .. controls (549.77,104) and (556.05,110.28) .. (556.05,118.02) .. controls (556.05,125.77) and (549.77,132.05) .. (542.02,132.05) .. controls (534.28,132.05) and (528,125.77) .. (528,118.02) -- cycle ;
\draw [line width=1.5]    (462.02,121.05) -- (478,185.67) ;
\draw [line width=1.5]    (535.44,129.73) -- (503,177.67) ;
\draw [line width=1.5]    (367.02,146.05) -- (332,191.67) ;
\draw  [draw opacity=0] (474.43,166.41) .. controls (474.51,165.83) and (474.65,165.26) .. (474.85,164.7) .. controls (477.14,158.26) and (486.62,155.75) .. (496.01,159.1) .. controls (502.49,161.42) and (507.24,165.91) .. (508.8,170.6) -- (491.85,170.76) -- cycle ; \draw   (474.43,166.41) .. controls (474.51,165.83) and (474.65,165.26) .. (474.85,164.7) .. controls (477.14,158.26) and (486.62,155.75) .. (496.01,159.1) .. controls (502.49,161.42) and (507.24,165.91) .. (508.8,170.6) ;  

\draw (163,66) node [anchor=north west][inner sep=0.75pt]  [font=\normalsize] [align=left] {O};
\draw (318,167) node [anchor=north west][inner sep=0.75pt]  [font=\normalsize] [align=left] {O};
\draw (205,168) node [anchor=north west][inner sep=0.75pt]  [font=\normalsize] [align=left] {P};
\draw (369,101) node [anchor=north west][inner sep=0.75pt]  [font=\normalsize] [align=left] {C};
\draw (138,139.4) node [anchor=north west][inner sep=0.75pt]  [font=\normalsize]  {$\alpha $};
\draw (487,130.4) node [anchor=north west][inner sep=0.75pt]  [font=\normalsize]  {$\alpha $};
\draw (450,70) node [anchor=north west][inner sep=0.75pt]  [font=\normalsize] [align=left] {O};
\draw (537,83) node [anchor=north west][inner sep=0.75pt]  [font=\normalsize] [align=left] {C};
\draw (461,211) node [anchor=north west][inner sep=0.75pt]  [font=\normalsize] [align=left] {P};
\draw (509,198) node [anchor=north west][inner sep=0.75pt]  [font=\normalsize] [align=left] {O};

\end{tikzpicture}

\caption{\textit{Illustration of the dihedral angles in the RNA backbone between the atom. Reproduced from \cite{mardia2013statistical}.}}
\label{fig:dihedral angles}
\end{figure}
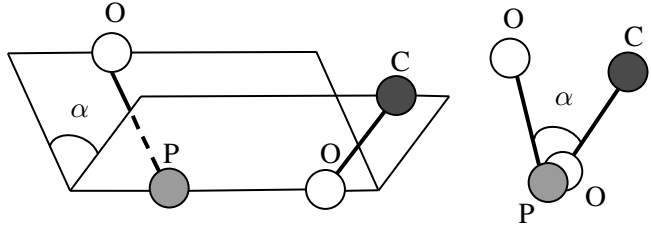

At the microscopic scale, the geometric relationship between atoms can be characterized by dihedral angles. As shown in Figure \ref{fig:dihedral angles}, four atoms within an RNA suite (specifically, $O-P-O-C$) can form a dihedral angle, denoted by $\alpha$, in a $3$-dimensional space. A dihedral angle is inherently periodic, ranging from $0$ to $2\pi$, which aligns naturally with the circular structure of $\mathbb{S}^1$. When considering multiple dihedral angles—each corresponding to a degree of rotational freedom along the RNA backbone—their collective multidimensional nature suggests a higher-dimensional torus. As the Cartesian product of circles, the torus elegantly accounts for angular periodicity, thereby enabling seamless transitions between rotations and enhancing computational simulations and structural predictions. At the microscopic scale, we focus on seven dihedral angles—$(\delta_i, \epsilon_i, \zeta_i, \alpha_{i+1}, \beta_{i+1}, \gamma_{i+1}, \epsilon_{i+1})$—for RNA suite $i$, as illustrated in Figure \ref{fig:RNA mesoscopic scale}. The details of RNA dihedral angles are shown in Appendix \ref{appendix: RNA dihedral angles}. 

For convenience, seven dihedral angles will be generally denoted as $x \in \mathbb{R}^D$ (i.e. $D=7$ in the RNA case), where $x = (x^{(1)}, \ldots, x^{(D)})^T$ with each $x^{(j)} \in [0,2\pi]$ representing the $j$-th angle.  Consequently, $x$ can be considered as embedded on a $D$-dimensional torus $\mathbb{T}^D$. Accordingly, we can define the samples $\mathcal{X} = \{x_i\}_{i=1}^N$ on the torus $\mathbb{T}^D$.  $\mathbb{T}^D$ is defined as the product of $D$ circles: 
    \begin{equation*}
        \mathbb{T}^D := [0,2\pi]^{D}/ \sim,
    \end{equation*}
where $\sim$ denotes that 0 and $2\pi$ are identified. We can define the geodesic distance between two samples $x_1=(x_1^{(1)}, \ldots, x_1^{(D)})^T$ and $x_2=(x_2^{(1)}, \ldots, x_2^{(D)})^T$ on the torus $\mathbb{T}^D$, as the canonical product distance
\begin{equation*}
    \operatorname{d}_{\mathbb{T}^D}(x_1, x_2) = \sqrt{\sum_{j=1}^D \operatorname{d}_{\mathbb{T}}(x_1^{(j)},x_2^{(j)})^2},
\end{equation*}
where the canonical distance is given by:
\begin{equation*}
    \operatorname{d}_{\mathbb{T}}(x_1^{(j)},x_2^{(j)}) = \min(|x_1^{(j)}-x_2^{(j)}|,2\pi-|x_1^{(j)}-x_2^{(j)}|).
\end{equation*}

Moreover, at the mesoscopic scale, RNA involves additional landmark positions of sugar rings in $3$-dimensional space and considering their relative configurations. Let \( \mathcal{S} = \{s_i\}_{i=1}^N \), where \( s_i \in \mathbb{R}^3 \) represents the position of the \( i \)-th sugar ring center. The size-and-shape space \( \Lambda_3^N \) is defined as:

\begin{equation*}
\Lambda_3^N = \{ [S] \mid S=(s_1, \ldots, s_N)^T \in \mathbb{R}^{N \times 3} \},
\end{equation*}
where \([S]\) denotes the equivalence class of \(S\) under proper Euclidean transformations ($\ET$), including rotations \(R \in \Xi(3)\) and translations \(v \in \mathbb{R}^3\):

\begin{equation} \label{mesoscopic} 
\ET.S = (Rs_1 + v, Rs_2 + v, \ldots, Rs_N + v)^T, \
[S] = \{\ET.S \mid \ET \in \Xi(3) \times \mathbb{R}^3\}.
\end{equation}

The Procrustes distance \( \operatorname{d}_{\Lambda} \) between two configurations \( [S] \) and \( [W] \) in size-and-shape space is given by:

\begin{equation} \label{shape distance}
\operatorname{d}_{\Lambda}([S], [W]) = \min_{\ET \in \Xi(3) \times \mathbb{R}^3} \| S - \ET.W \|_2,
\end{equation}
with the Euclidean norm \( \| \cdot \|_2 \) on \( \mathbb{R}^{N \times 3} \). This provides comprehensive insights into mesoscopic geometry, revealing structural variations, similarities, and potential functional motifs in RNA. At the mesoscopic scale, for RNA suite $i$, we focus on the $k$ suites before and $k$ suites after this central suites, as show in Figure \ref{fig:RNA mesoscopic scale}.

Since we have defined the geodesic distance in microscopic and mesoscopic scale spaces respectively, we can define the Fr\'echet mean $\mu$ at both two scale space $\mathbb{T}^D$ and $\Lambda_3^N$: 
\begin{equation*} \label{Fr\'echet mean}
\mu_{\mathbb{T}} \in \argmin_{x \in {\mathbb{T}^D}} \sum_{i=1}^N \operatorname{d}_{{\mathbb{T}^D}}(x,x_i)^2, \quad \mu_{\Lambda} \in \argmin_{[S] \in {\Lambda_3^N}} \sum_{i=1}^N \operatorname{d}_{{\Lambda_3^N}}([S],[S_i])^2.
\end{equation*}

\begin{figure}[ht]
\centering

\tikzset{every picture/.style={line width=0.75pt}} 

\begin{tikzpicture}[x=0.75pt,y=0.75pt,yscale=-1,xscale=1]

\draw  [color={rgb, 255:red, 74; green, 144; blue, 226 }  ,draw opacity=1 ] (171.71,210.73) -- (162.87,238.76) -- (133.49,239.02) -- (124.16,211.15) -- (147.78,193.67) -- cycle ;
\draw  [color={rgb, 255:red, 74; green, 144; blue, 226 }  ,draw opacity=1 ] (245.71,210.73) -- (236.87,238.76) -- (207.49,239.02) -- (198.16,211.15) -- (221.78,193.67) -- cycle ;
\draw  [color={rgb, 255:red, 74; green, 144; blue, 226 }  ,draw opacity=1 ] (319.71,210.73) -- (310.87,238.76) -- (281.49,239.02) -- (272.16,211.15) -- (295.78,193.67) -- cycle ;
\draw  [color={rgb, 255:red, 74; green, 144; blue, 226 }  ,draw opacity=1 ] (393.71,210.73) -- (384.87,238.76) -- (355.49,239.02) -- (346.16,211.15) -- (369.78,193.67) -- cycle ;
\draw  [color={rgb, 255:red, 74; green, 144; blue, 226 }  ,draw opacity=1 ] (467.71,210.73) -- (458.87,238.76) -- (429.49,239.02) -- (420.16,211.15) -- (443.78,193.67) -- cycle ;
\draw  [color={rgb, 255:red, 74; green, 144; blue, 226 }  ,draw opacity=1 ] (541.71,210.73) -- (532.87,238.76) -- (503.49,239.02) -- (494.16,211.15) -- (517.78,193.67) -- cycle ;
\draw [color={rgb, 255:red, 117; green, 186; blue, 45 }  ,draw opacity=1 ]   (162.87,238.76) -- (207.49,239.02) ;
\draw [color={rgb, 255:red, 117; green, 186; blue, 45 }  ,draw opacity=1 ]   (236.87,238.76) -- (281.49,239.02) ;
\draw [color={rgb, 255:red, 117; green, 186; blue, 45 }  ,draw opacity=1 ]   (310.87,238.76) -- (355.49,239.02) ;
\draw [color={rgb, 255:red, 117; green, 186; blue, 45 }  ,draw opacity=1 ]   (384.87,238.76) -- (429.49,239.02) ;
\draw [color={rgb, 255:red, 117; green, 186; blue, 45 }  ,draw opacity=1 ]   (458.87,238.76) -- (503.49,239.02) ;
\draw  [color={rgb, 255:red, 208; green, 2; blue, 27 }  ,draw opacity=1 ][fill={rgb, 255:red, 208; green, 2; blue, 27 }  ,fill opacity=1 ] (147,218.5) .. controls (147,217.67) and (147.67,217) .. (148.5,217) .. controls (149.33,217) and (150,217.67) .. (150,218.5) .. controls (150,219.33) and (149.33,220) .. (148.5,220) .. controls (147.67,220) and (147,219.33) .. (147,218.5) -- cycle ;
\draw  [color={rgb, 255:red, 208; green, 2; blue, 27 }  ,draw opacity=1 ][fill={rgb, 255:red, 208; green, 2; blue, 27 }  ,fill opacity=1 ] (221,218.5) .. controls (221,217.67) and (221.67,217) .. (222.5,217) .. controls (223.33,217) and (224,217.67) .. (224,218.5) .. controls (224,219.33) and (223.33,220) .. (222.5,220) .. controls (221.67,220) and (221,219.33) .. (221,218.5) -- cycle ;
\draw  [color={rgb, 255:red, 208; green, 2; blue, 27 }  ,draw opacity=1 ][fill={rgb, 255:red, 208; green, 2; blue, 27 }  ,fill opacity=1 ] (295,218.5) .. controls (295,217.67) and (295.67,217) .. (296.5,217) .. controls (297.33,217) and (298,217.67) .. (298,218.5) .. controls (298,219.33) and (297.33,220) .. (296.5,220) .. controls (295.67,220) and (295,219.33) .. (295,218.5) -- cycle ;
\draw  [color={rgb, 255:red, 208; green, 2; blue, 27 }  ,draw opacity=1 ][fill={rgb, 255:red, 208; green, 2; blue, 27 }  ,fill opacity=1 ] (369,218.5) .. controls (369,217.67) and (369.67,217) .. (370.5,217) .. controls (371.33,217) and (372,217.67) .. (372,218.5) .. controls (372,219.33) and (371.33,220) .. (370.5,220) .. controls (369.67,220) and (369,219.33) .. (369,218.5) -- cycle ;
\draw  [color={rgb, 255:red, 208; green, 2; blue, 27 }  ,draw opacity=1 ][fill={rgb, 255:red, 208; green, 2; blue, 27 }  ,fill opacity=1 ] (442,218.5) .. controls (442,217.67) and (442.67,217) .. (443.5,217) .. controls (444.33,217) and (445,217.67) .. (445,218.5) .. controls (445,219.33) and (444.33,220) .. (443.5,220) .. controls (442.67,220) and (442,219.33) .. (442,218.5) -- cycle ;
\draw  [color={rgb, 255:red, 208; green, 2; blue, 27 }  ,draw opacity=1 ][fill={rgb, 255:red, 208; green, 2; blue, 27 }  ,fill opacity=1 ] (516,218.5) .. controls (516,217.67) and (516.67,217) .. (517.5,217) .. controls (518.33,217) and (519,217.67) .. (519,218.5) .. controls (519,219.33) and (518.33,220) .. (517.5,220) .. controls (516.67,220) and (516,219.33) .. (516,218.5) -- cycle ;
\draw  [color={rgb, 255:red, 189; green, 16; blue, 224 }  ,draw opacity=1 ] (257.71,197.47) .. controls (257.71,190.4) and (263.44,184.67) .. (270.51,184.67) -- (395.2,184.67) .. controls (402.27,184.67) and (408,190.4) .. (408,197.47) -- (408,235.87) .. controls (408,242.94) and (402.27,248.67) .. (395.2,248.67) -- (270.51,248.67) .. controls (263.44,248.67) and (257.71,242.94) .. (257.71,235.87) -- cycle ;
\draw [color={rgb, 255:red, 189; green, 16; blue, 224 }  ,draw opacity=1 ]   (331,185) -- (331,155.67) ;
\draw [shift={(331,153.67)}, rotate = 90] [color={rgb, 255:red, 189; green, 16; blue, 224 }  ,draw opacity=1 ][line width=0.75]    (10.93,-3.29) .. controls (6.95,-1.4) and (3.31,-0.3) .. (0,0) .. controls (3.31,0.3) and (6.95,1.4) .. (10.93,3.29)   ;
\draw [color={rgb, 255:red, 208; green, 2; blue, 27 }  ,draw opacity=1 ]   (150,218.5) -- (221,218.5) ;
\draw [color={rgb, 255:red, 208; green, 2; blue, 27 }  ,draw opacity=1 ]   (224,218.5) -- (295,218.5) ;
\draw [color={rgb, 255:red, 208; green, 2; blue, 27 }  ,draw opacity=1 ]   (298,218.5) -- (369,218.5) ;
\draw [color={rgb, 255:red, 208; green, 2; blue, 27 }  ,draw opacity=1 ]   (371,218.5) -- (442,218.5) ;
\draw [color={rgb, 255:red, 208; green, 2; blue, 27 }  ,draw opacity=1 ]   (445,218.5) -- (516,218.5) ;
\draw [color={rgb, 255:red, 117; green, 186; blue, 45 }  ,draw opacity=1 ]   (88.87,238.76) -- (133.49,239.02) ;
\draw [color={rgb, 255:red, 117; green, 186; blue, 45 }  ,draw opacity=1 ]   (532.87,238.76) -- (577.49,239.02) ;
\draw    (310.87,258.76) -- (355.49,259.02) ;
\draw    (385.87,258.76) -- (430.49,259.02) ;
\draw    (458.87,257.76) -- (503.49,258.02) ;
\draw    (162.87,258.76) -- (207.49,259.02) ;
\draw    (235.87,258.76) -- (280.49,259.02) ;
\draw [color={rgb, 255:red, 74; green, 144; blue, 226 }  ,draw opacity=1 ]   (237,55) -- (222,68) ;
\draw [color={rgb, 255:red, 74; green, 144; blue, 226 }  ,draw opacity=1 ]   (226,100) -- (219,85) ;
\draw [color={rgb, 255:red, 74; green, 144; blue, 226 }  ,draw opacity=1 ]   (266,68) -- (250,55) ;
\draw [color={rgb, 255:red, 74; green, 144; blue, 226 }  ,draw opacity=1 ]   (262,100) -- (269,84) ;
\draw [color={rgb, 255:red, 74; green, 144; blue, 226 }  ,draw opacity=1 ]   (234,109) -- (253,109) ;
\draw [color={rgb, 255:red, 117; green, 186; blue, 45 }  ,draw opacity=1 ]   (268,109) -- (287,109) ;
\draw [color={rgb, 255:red, 117; green, 186; blue, 45 }  ,draw opacity=1 ]   (199,109) -- (218,109) ;
\draw [color={rgb, 255:red, 117; green, 186; blue, 45 }  ,draw opacity=1 ]   (303,109) -- (322,109) ;
\draw    (329.44,86) -- (329.56,99) ;
\draw    (331.44,86) -- (331.56,99) ;
\draw [color={rgb, 255:red, 117; green, 186; blue, 45 }  ,draw opacity=1 ]   (338,109) -- (357,109) ;
\draw [color={rgb, 255:red, 117; green, 186; blue, 45 }  ,draw opacity=1 ]   (164,109) -- (183,109) ;
\draw [color={rgb, 255:red, 74; green, 144; blue, 226 }  ,draw opacity=1 ]   (446,55) -- (431,68) ;
\draw [color={rgb, 255:red, 74; green, 144; blue, 226 }  ,draw opacity=1 ]   (435,100) -- (428,85) ;
\draw [color={rgb, 255:red, 74; green, 144; blue, 226 }  ,draw opacity=1 ]   (475,68) -- (459,55) ;
\draw [color={rgb, 255:red, 74; green, 144; blue, 226 }  ,draw opacity=1 ]   (471,100) -- (478,84) ;
\draw [color={rgb, 255:red, 74; green, 144; blue, 226 }  ,draw opacity=1 ]   (443,109) -- (462,109) ;
\draw [color={rgb, 255:red, 117; green, 186; blue, 45 }  ,draw opacity=1 ]   (477,109) -- (496,109) ;
\draw [color={rgb, 255:red, 117; green, 186; blue, 45 }  ,draw opacity=1 ]   (408,109) -- (427,109) ;
\draw [color={rgb, 255:red, 117; green, 186; blue, 45 }  ,draw opacity=1 ]   (512,109) -- (531,109) ;
\draw [color={rgb, 255:red, 117; green, 186; blue, 45 }  ,draw opacity=1 ]   (373,109) -- (392,109) ;
\draw    (186.22,117) -- (176.33,131) ;
\draw    (195.22,117) -- (203.33,131) ;
\draw    (330.44,118) -- (330.56,131) ;
\draw    (226.44,118) -- (226.56,131) ;
\draw    (260.44,118) -- (260.56,131) ;
\draw    (396.22,117) -- (386.33,131) ;
\draw    (405.22,117) -- (413.33,131) ;
\draw    (436.44,118) -- (436.56,131) ;
\draw    (470.44,118) -- (470.56,131) ;
\draw    (243.44,29) -- (243.56,42) ;
\draw    (452.44,29) -- (452.56,42) ;

\draw (138,202) node [anchor=north west][inner sep=0.75pt]  [font=\footnotesize] [align=left] {$\displaystyle s_{i-2}$};
\draw (211,202) node [anchor=north west][inner sep=0.75pt]  [font=\footnotesize] [align=left] {$\displaystyle s_{i-1}$};
\draw (291,201) node [anchor=north west][inner sep=0.75pt]  [font=\footnotesize] [align=left] {$\displaystyle s_{i}$};
\draw (360,201) node [anchor=north west][inner sep=0.75pt]  [font=\footnotesize] [align=left] {$\displaystyle s_{i+1}$};
\draw (434,202) node [anchor=north west][inner sep=0.75pt]  [font=\footnotesize] [align=left] {$\displaystyle s_{i+2}$};
\draw (507,202) node [anchor=north west][inner sep=0.75pt]  [font=\footnotesize] [align=left] {$\displaystyle s_{i+3}$};
\draw (315,188) node [anchor=north west][inner sep=0.75pt]   [align=left] {Suite i};
\draw (324,261) node [anchor=north west][inner sep=0.75pt]   [align=left] {$\displaystyle x_{i}$};
\draw (394,261) node [anchor=north west][inner sep=0.75pt]   [align=left] {$\displaystyle x_{i+1}$};
\draw (467,260) node [anchor=north west][inner sep=0.75pt]   [align=left] {$\displaystyle x_{i+2}$};
     \draw (171,261) node [anchor=north west][inner sep=0.75pt]   [align=left] {$\displaystyle x_{i-2}$};
\draw (244,261) node [anchor=north west][inner sep=0.75pt]   [align=left] {$\displaystyle x_{i-1}$};
\draw (119,260) node [anchor=north west][inner sep=0.75pt]   [align=left] {$\displaystyle \dotsc $};
\draw (516,259) node [anchor=north west][inner sep=0.75pt]   [align=left] {$\displaystyle \dotsc $};
\draw (237,43) node [anchor=north west][inner sep=0.75pt]   [align=left] {C};
\draw (325,103) node [anchor=north west][inner sep=0.75pt]   [align=left] {P};
\draw (211,70) node [anchor=north west][inner sep=0.75pt]   [align=left] {O};
\draw (264,70) node [anchor=north west][inner sep=0.75pt]   [align=left] {CH$\displaystyle _{2}$OH};
\draw (221,103) node [anchor=north west][inner sep=0.75pt]   [align=left] {C};
\draw (253,103) node [anchor=north west][inner sep=0.75pt]   [align=left] {C};
\draw (185,103) node [anchor=north west][inner sep=0.75pt]   [align=left] {C};
\draw (288,103) node [anchor=north west][inner sep=0.75pt]   [align=left] {O};
\draw (324,71) node [anchor=north west][inner sep=0.75pt]   [align=left] {O};
\draw (324,132) node [anchor=north west][inner sep=0.75pt]   [align=left] {O};
\draw (358,103) node [anchor=north west][inner sep=0.75pt]   [align=left] {O};
\draw (446,43) node [anchor=north west][inner sep=0.75pt]   [align=left] {C};
\draw (420,70) node [anchor=north west][inner sep=0.75pt]   [align=left] {O};
\draw (473,70) node [anchor=north west][inner sep=0.75pt]   [align=left] {CH$\displaystyle _{2}$OH};
\draw (430,103) node [anchor=north west][inner sep=0.75pt]   [align=left] {C};
\draw (464,103) node [anchor=north west][inner sep=0.75pt]   [align=left] {C};
\draw (394,103) node [anchor=north west][inner sep=0.75pt]   [align=left] {C};
\draw (497,103) node [anchor=north west][inner sep=0.75pt]   [align=left] {O};
\draw (167.22,132) node [anchor=north west][inner sep=0.75pt]   [align=left] {H};
\draw (202.22,132) node [anchor=north west][inner sep=0.75pt]   [align=left] {H};
\draw (220,132) node [anchor=north west][inner sep=0.75pt]   [align=left] {H};
\draw (254,132) node [anchor=north west][inner sep=0.75pt]   [align=left] {H};
\draw (377.22,132) node [anchor=north west][inner sep=0.75pt]   [align=left] {H};
\draw (412.22,132) node [anchor=north west][inner sep=0.75pt]   [align=left] {H};
\draw (430,132) node [anchor=north west][inner sep=0.75pt]   [align=left] {H};
\draw (464,132) node [anchor=north west][inner sep=0.75pt]   [align=left] {H};
\draw (336.22,92.4) node [anchor=north west][inner sep=0.75pt]  [font=\scriptsize,color={rgb, 255:red, 245; green, 166; blue, 35 }  ,opacity=1 ]  {$\alpha _{i+1}$};
\draw (370.22,90.4) node [anchor=north west][inner sep=0.75pt]  [font=\scriptsize,color={rgb, 255:red, 245; green, 166; blue, 35 }  ,opacity=1 ]  {$\beta _{i+1}$};
\draw (405.22,92.4) node [anchor=north west][inner sep=0.75pt]  [font=\scriptsize,color={rgb, 255:red, 245; green, 166; blue, 35 }  ,opacity=1 ]  {$\gamma _{i+1}$};
\draw (440.22,90.4) node [anchor=north west][inner sep=0.75pt]  [font=\scriptsize,color={rgb, 255:red, 245; green, 166; blue, 35 }  ,opacity=1 ]  {$\delta _{i+1}$};
\draw (476.22,92.4) node [anchor=north west][inner sep=0.75pt]  [font=\scriptsize,color={rgb, 255:red, 0; green, 0; blue, 0 }  ,opacity=1 ]  {$\epsilon _{i+1}$};
\draw (203.22,92.4) node [anchor=north west][inner sep=0.75pt]  [font=\scriptsize,color={rgb, 255:red, 0; green, 0; blue, 0 }  ,opacity=1 ]  {$\gamma _{i}$};

\draw (236.22,90.4) node [anchor=north west][inner sep=0.75pt]  [font=\scriptsize,color={rgb, 255:red, 245; green, 166; blue, 35 }  ,opacity=1 ]  {$\delta _{i}$};
\draw (273,92.4) node [anchor=north west][inner sep=0.75pt]  [font=\scriptsize,color={rgb, 255:red, 245; green, 166; blue, 35 }  ,opacity=1 ]  {$\epsilon _{i}$};
\draw (307,90.4) node [anchor=north west][inner sep=0.75pt]  [font=\scriptsize,color={rgb, 255:red, 245; green, 166; blue, 35 }  ,opacity=1 ]  {$\zeta _{i}$};
\draw (231.2,13) node [anchor=north west][inner sep=0.75pt]   [align=left] {base};
\draw (440.2,13) node [anchor=north west][inner sep=0.75pt]   [align=left] {base};
\draw (236.22,59.4) node [anchor=north west][inner sep=0.75pt]  [font=\scriptsize,color={rgb, 255:red, 0; green, 0; blue, 0 }  ,opacity=1 ]  {$\chi _{i}$};
\draw (442,59.4) node [anchor=north west][inner sep=0.75pt]  [font=\scriptsize,color={rgb, 255:red, 0; green, 0; blue, 0 }  ,opacity=1 ]  {$\chi _{i+1}$};
\draw (545.2,86) node [anchor=north west][inner sep=0.75pt]   [align=left] {\begin{minipage}[lt]{59.4pt}\setlength\topsep{0pt}
Microscopic 
\begin{center}
scale
\end{center}

\end{minipage}};
\draw (547.2,186) node [anchor=north west][inner sep=0.75pt]   [align=left] {\begin{minipage}[lt]{56.58pt}\setlength\topsep{0pt}
Mesoscopic
\begin{center}
scale
\end{center}

\end{minipage}};

\end{tikzpicture}

\caption{\textit{Visualization of an RNA backbone at microscopic and mesoscopic scale respectively, illustrating seven dihedral angles of $i$-$th$ suite at microscopic scale and six centers $(k=2)$ of the sugar rings (from $s_{i-2}$ to $s_{i+3}$) defining shape of $i$-$th$ suite at mesoscopic scale.}}

\label{fig:RNA mesoscopic scale} 
\end{figure}
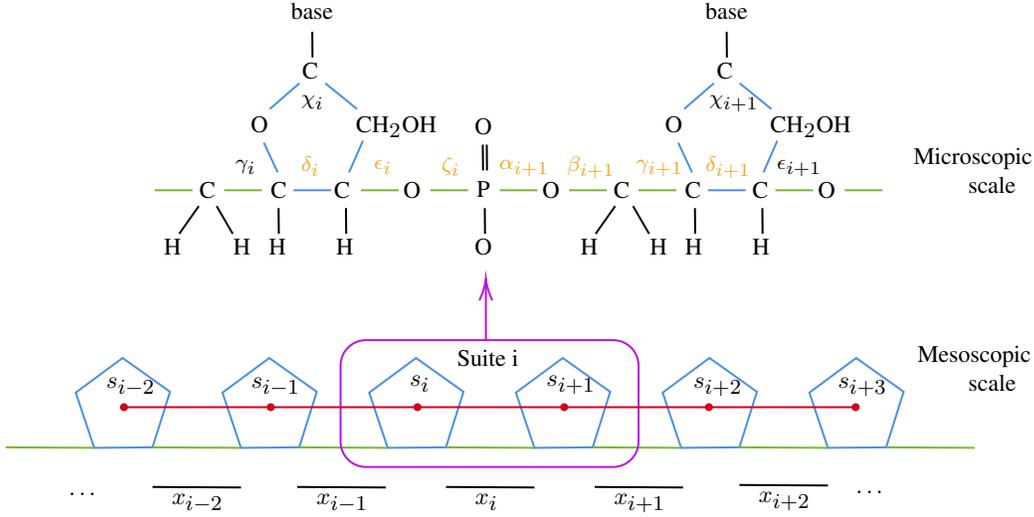

\subsection{Statistical model for PSM}
To analyze \( \mathcal{X} \) on the torus \( \mathbb{T}^D \), tPCA was introduced as a two-step procedure: first mapping the torus data onto a sphere via TOSS, and then applying PNS to obtain a lower-dimensional representation. Compared with linear methods, tPCA takes into account the periodicity of angular data; however, it still has inherent limitations. The TOSS projection inevitably distorts the intrinsic product geometry of the torus, while the subsequent PNS step constrains the reduced structure to lie on a lower-dimensional sphere rather than a flexible submanifold adapted to the data. These restrictions lead to information loss and a mismatch with the true torus geometry. Mathematical details of tPCA are provided in Appendix \ref{appendix: tPCA}.

To overcome the challenges posed by torus setting and the weakness of tPCA, we need a new method which can accurately fit low-dimensional representation on the high-dimensional torus. Here, we adopt an advanced method, which comes from a series of works represented by principal submanifold \citep{yao2023submanifold, yao2024principalsubmanifolds}, a
generalization of the principal flow \citep{panaretos2014principal}. This paper adopt the theories and methodologies and aims to fit a smooth low-dimensional principal submanifold within data situated on the manifold $\mathcal{M}$, which can capture the main geometric structures of data and minimize the information loss. 

 To accommodate our RNA setting, we set $\mathcal{M} = \mathbb{T}^D$. The statistical problem is to recover an accurate underlying low-dimensional geometric structure of data on the $\mathbb{T}^D$ while minimizing information loss. We assume $d^*$-dimensional submanifolds class $\Psi= \{\psi: \psi \subset \mathbb{T}^D , \dim(\psi)=d^*, \psi \in C^2(\mathbb{T}^D)\}$, where $d^* < D$ and $C^2(\mathbb{T}^D)$ denotes the class of twice continuously differentiable submanifolds embedded in $\mathbb{T}^D$. Then, the \emph{population principal submanifold} \( \psi^* \) is conceptually idealized to summarize the data by optimizing the expected squared geodesic distance with respect to the distribution $\tau$. Formally, \( \psi^* \) is defined as: 
\begin{equation}
\psi^* = \argmin_{\psi' \in \Psi} \int \operatorname{d}_{\mathbb{T}^D}(x, \psi')^2 \tau(dx).
\label{eq:population_PSM}
\end{equation}

\begin{remark}
Equation~\eqref{eq:population_PSM} is formulated under the assumption that the dominant low-dimensional structure of the data is a smooth embedded submanifold of $\mathbb{T}^D$. This excludes singular sets such as trees with branching points, since such sets are not locally Euclidean and do not admit a uniquely defined tangent space at the branch points. Under this assumption, the projection distance and the optimization criterion in Equation~\eqref{eq:population_PSM} are well-defined.
\end{remark}

\begin{remark}
In our setting, the RNA dihedral-angle representation lies on a torus, and the observed conformational variation is known to change continuously with respect to these angles. This makes a smooth low-dimensional submanifold on the torus a natural model for the underlying structure: it preserves the intrinsic periodic topology of the torus and reflects the smooth, non-branching nature of RNA conformational changes, whereas tree-type structures would introduce artificial cuts and branching singularities that are not physically meaningful in this space. In Equation~\eqref{eq:population_PSM}, at the population level, the principal submanifold is defined as the minimizer of the expected squared geodesic distance on the torus, which is the natural nonlinear generalization of PCA in this Riemannian setting. The empirical procedure in Equation~\eqref{eq:population_PSM} together with Algorithm \ref{Principal Submanifold Algorithm} introduced below, should be understood as a discrete approximation of this object: it produces fitted samples that are locally aligned with estimated tangent spaces and therefore concentrate around a smooth low-dimensional structure on the torus. In this sense, the output is not an explicit parametric manifold, but a consistent discrete representation of the principal submanifold, which is appropriate for the data and downstream clustering tasks considered in this work.
\end{remark}

In practice, $\tau$ is unknown and only sample $\mathcal{X}$ is observed. Replacing the expectation in \eqref{eq:population_PSM} by the empirical average, we define the \emph{empirical principal submanifold} as follows:
\begin{equation}
    \widehat{\psi}_N = \argmin_{\psi' \in \Psi} \frac{1}{N}\sum_{i=1}^N\operatorname{d}_{\mathbb{T}^D}(x_i, \psi')^2.
    \label{eq:empirical_PSM}
\end{equation} 
Our principal submanifold framework seeks to use $\widehat{\psi}_N$ as an estimator of $\psi^*$ and to ensure that $\widehat{\psi}_N$ is very close to $\psi^*$ on the torus $\mathbb{T}^D$. Theoretical convergence results have been established in earlier works \citep{yao2024principalsubmanifolds, yao2025mf,yao2023submanifold}, which show that, as $N \to \infty$, the Hausdorff distance between $\widehat{\psi}_N$ and $\psi^*$ converges to zero. Building on this theoretical foundation, our paper extends the framework to specifically investigate the case of submanifolds on the torus.

\subsection{Algorithm for PSM}
Directly solving the minimization problem in \eqref{eq:empirical_PSM} on the torus $\mathbb{T}^D$ is in general computationally intractable due to the manifold’s periodic structure and the nonlinearity of its geodesic distance. To address this, we propose an iterative PSM algorithm, outlined in Algorithm \ref{Principal Submanifold Algorithm}, which operates in an augmented Euclidean embedding space. Here, we elaborate the method to calculate the principal submanifold $\widehat{\psi}_N^{{\mathrm{PSM}}}$ based on the samples $\mathcal{X}$, visualized in Figure \ref{fig: Algorithm for PSM}.

\begin{algorithm}[H]
\caption{PSM Algorithm}
\SetAlgoLined
\KwResult{Compute $\widehat{\psi}_N^{\mathrm{PSM}}$ based on the sample set $\mathcal{X}$.}
\textbf{Input:} Sample set $\mathcal{X}$, maximum number of iterations $maxiter$, learning rate $\lambda$, initial neighbourhood radius $r_s$, submanifold dimension $d$, convergence threshold $\rho$.

\textbf{Initialization:} Augment each $x_i \in \mathcal{X}$ to $y_i$ by (\ref{augmentation});

\For{$i\leftarrow 1$ \KwTo $N$}{
    Init $y_i^0 = y_i$;
    
    \For{$t\leftarrow 1$ \KwTo $maxiter$}{
        Calculate $\mathcal{N}-U$ with the neighbourhood $\mathcal{N}_{r_s}(y^{t-1}_i)$ of $y^{t-1}_i$ by (\ref{neighborhood matrix}) and (\ref{mean matrix});
        
        Calculate the singular vector matrix $\Omega_d$ by (\ref{omega}) and the projection direction $v_d$ by (\ref{v_d});
        
        Update $y^{t}_i$ by $y^{t-1}_i - \lambda v_d$;
        
        \If{$\|y^{t}_i - y^{t-1}_i\|_2 \leq \rho$}{            
            break;
        }
    }
    $\widetilde{y}_{i} = y_i^{t}$;
    
    Normalize $\widetilde{y}_{i}$ by (\ref{normalization}) and map $\widetilde{y}_{i}$ from $\mathbb{R}^{2D}$ back to the torus $\mathbb{T}^D$ as $\widetilde{x}_{i}$ by (\ref{mapping back}).
}

\textbf{Output:} The fitted samples $\widetilde{\mathcal{X}} = \{\widetilde{x}_i\}_{i=1}^N$ as the $d$-dimentional  $\widehat{\psi}_N^{\mathrm{PSM}}$.

\label{Principal Submanifold Algorithm}

\end{algorithm}

    \begin{enumerate}
      \item \textit{Augmentation}: To account for the periodicity of the angular coordinates on the torus, we embed each original sample $x_i \in \mathcal{X} \subset \mathbb{T}^D$ into a $2D$-dimensional Euclidean space by lifting each coordinate onto the unit circle. Specifically, we define:
        \begin{equation} \label{augmentation}
        x_i^{(j)} \mapsto y_i^{(2j-1)} = \cos(x_i^{(j)}), \quad y_i^{(2j)} = \sin(x_i^{(j)}), \quad j = 1, \ldots, D.
        \end{equation}
        The resulting vector $y_i = (y_i^{(1)}, y_i^{(2)}, \ldots, y_i^{(2D)})^\top \in \mathbb{R}^{2D}$ represents the angular lifting of $x_i$, and we denote the lifted dataset as $\mathcal{Y} = \{y_i\}_{i=1}^N$.
             
        \item \textit{Iteration}: In the new $2D$-dimensional space, the algorithm iteratively updates $y_i^{t}$ to $y_i^{t+1}$ such that $y_i^{t+1}$ is closer to $\psi^*$. At stage $t$, the neighbours $\mathcal{N}_{r_s}(y_i^{t}) = \mathbb{B}(y_i^{t},r_s) \cap \mathcal{Y}$ within the neighbourhood radius $r_s$ and the neighborhood sample matrix $\mathcal{N}$ of $y_i^t$ as:
            \begin{equation} \label{neighborhood matrix}
                \mathcal{N} = (y_1, \ldots, y_{n_{\mathcal{N}}})^T,\ y_j \in \mathcal{N}_{r_s}(y_i^{t}), \ j = 1, \ldots, n_{\mathcal{N}},
            \end{equation}
        where $n_{\mathcal{N}}$ is the number of elements in $\mathcal{N}_{r_s}(y_i^{t})$.
        
         If there exists a sufficient number of neighbors $\mathcal{N}_{r_s}(y_i^{t})$, we can calculate the local mean $u_{y_i^{t}}$:
                 \begin{equation*}
                    u_{y_i^{t}} = \frac{1}{n_{\mathcal{N}}} \sum_{y_j \in \mathcal{N}_{r_s}(y_i^{t})} y_j,
                \end{equation*}
        and define the local mean matrix $U$:
                \begin{equation} \label{mean matrix}
                    U = \mathbf{1} u_{y_i^t}^T,
                \end{equation}
        where $\mathbf{1} = (1, 1, \dots, 1)^T \in \mathbb{R}^{n_{\mathcal{N}}}.$
        
        Then, given the dimension $d$ of data to fit, where $1 \leq d \leq d^*$, we calculate singular vector matrix $\Omega_d$, defined as:
            \begin{equation} \label{omega}
            \Omega_d = (\omega_1, \omega_2, \ldots, \omega_{2D-d})^T,
            \end{equation}
        where $\omega_i$ is the singular vector corresponding to the smallest $i$ singular values of the local zero-centered neighborhood matrix $\mathcal{N}-U$.
            
        Based on $\Omega_d$, we can determine the projection direction towards to the empirical principal submanifold $\widehat{\psi}_N$: 
        \begin{equation} \label{v_d}
        v_d = \Omega_d^T \Omega_d(y_i^{t} - u_{y_i^{t}}).
        \end{equation}
        
        Update $y_i^{t+1}$ by $y_i^{t} - \lambda v_d$ using a learning rate $\lambda$ and continue until movement falls below a predefined convergence threshold $\rho$: $\|y_i^{t+1}-y_i^{t} \|_2 \leq \rho$. 
    
       \item \textit{Output}: Let $\widetilde{y}_i$ denote the final iteration $y_i^{t}$. After the iterative procedure, each fitted point $\widetilde{y}_i \in \mathbb{R}^{2D}$ is projected back onto the product of unit circles to ensure the result lies on the torus $\mathbb{T}^D$. Specifically, for each $j = 1, \ldots, D$, we normalize the $j$-th angular pair $\left(\widetilde{y}_i^{(2j-1)}, \widetilde{y}_i^{(2j)}\right)$ onto the unit circle:
        \begin{equation} \label{normalization}
        \left(\widetilde{y}_i^{(2j-1)}, \widetilde{y}_i^{(2j)}\right) \leftarrow \frac{\left(\widetilde{y}_i^{(2j-1)}, \widetilde{y}_i^{(2j)}\right)}{\left\| \left(\widetilde{y}_i^{(2j-1)}, \widetilde{y}_i^{(2j)} \right) \right\|_2}.
        \end{equation}

        Finally, we map each normalized pair back to its corresponding angular coordinate on the torus $\mathbb{T}^D$ via inverse angular lifting, where the normalization is applied only once at the end to improve efficiency and avoid distorting the update trajectory, while still ensuring that the fitted data lies on the torus. Specifically, for each $j = 1, \ldots, D$, we recover the $j$-th angular component by:
        \begin{equation} \label{mapping back}
        \widetilde{x}_i^{(j)} = \left( \mathrm{atan2} \left( \widetilde{y}_i^{(2j)}, \widetilde{y}_i^{(2j-1)} \right) + 2\pi \right) \bmod 2\pi,
        \end{equation}
        where $\mathrm{atan2}(y, x)$ denotes the two-argument inverse tangent function that returns the angle of the vector $(x, y)$ in $(-\pi, \pi]$. 
            
        The algorithm outputs fitted samples $\widetilde{\mathcal{X}}^d=\{\widetilde{x}_i^d\}_{i=1}^N$, where $\widetilde{x}_i^d \in \widehat{\psi}_N^{\mathrm{PSM}} \subset \mathbb{R}^D$. For notational simplicity, we omit the superscript $d$ hereafter, including in Algorithm~\ref{Principal Submanifold Algorithm}, and write $\widetilde{\mathcal{X}}$ and $\widetilde{x}_i$ unless the dimension needs to be emphasized.
        
        \begin{figure}[H]
        \centering
        \includegraphics[width=0.75\textwidth]{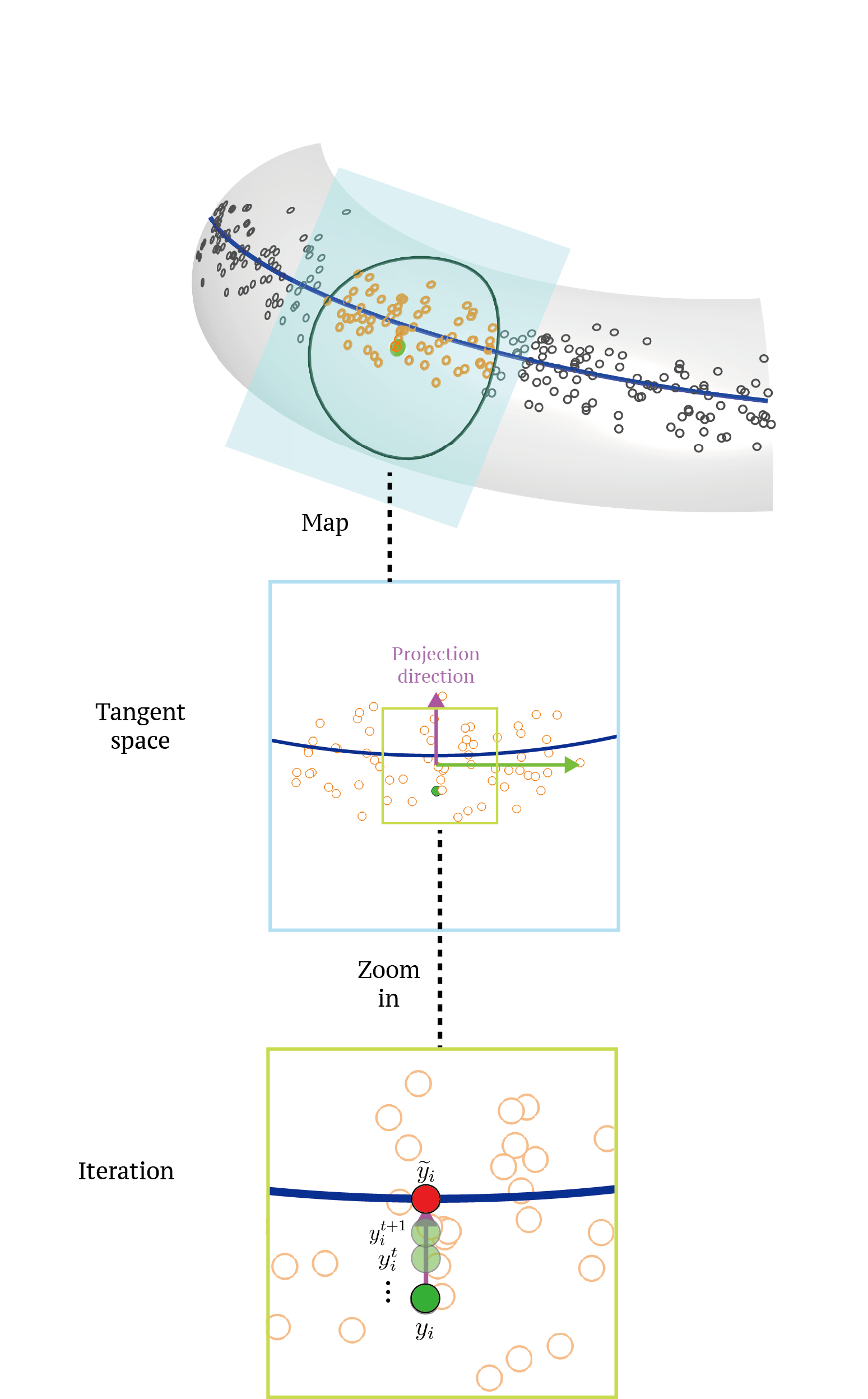} 
        \caption{\textit{Illustration of principal submanifold estimation. \textbf{(Top)} A neighborhood of $y_i$ on $\mathcal{M}$ is used to approximate the tangent space. \textbf{(Middle)} SVD in this local tangent space determines the projection direction from the smallest $2D-d$ singular vectors. \textbf{(Bottom)} Iterative updates from $y_i$ converge to the fitted point $\widetilde{y}_i$.}}
        \label{fig: Algorithm for PSM}
        \end{figure}
    \end{enumerate}

\subsection{Clustering based on PSM}
Based on the fitted $\widetilde{\mathcal{X}}$, DBSCAN is applied for clustering. We note that any clustering method can be suitable for $\widetilde{\mathcal{X}}$. Here, we select DBSCAN because, combined with PSM’s accurate low-dimensional representation, it is robust to noise and capable of detecting clusters with irregular shapes, making it well-suited for the heterogeneous geometric patterns in RNA data. We refer to this combined approach as PSM-DBSCAN.

There are two parameters associated with DBSCAN: \( r_C \), which defines the neighborhood radius; and \( \eta \), which is the threshold for density.

PSM-DBSCAN starts with an initial assignment for each sample from the samples \( \widetilde{\mathcal{X}} \) to any of three sets: $\mathscr{C}$, $\mathscr{B}$, or $\mathscr{O}$ using the following rule:

\begin{equation*}
\widetilde{x}_{i} \in 
\begin{cases} 
\mathscr{C}(r_C,\eta), & \text{if } \#(\mathcal{N}_{r_C}(\widetilde{x}_{i})) \geq \eta, \\
\mathscr{B}(r_C,\eta), & \text{if } \#(\mathcal{N}_{r_C}(\widetilde{x}_{i})) < \eta \ ; \ \exists\ \widetilde{x}_j \in \mathcal{N}_{r_C}(\widetilde{x}_{i}),\ \text{s.t.}\ \#(\mathcal{N}_{r_C}(\widetilde{x}_j)) \geq \eta, \\
\mathscr{O}(r_C,\eta), & \text{otherwise}.
\end{cases}
\end{equation*}

Based on the sets assigned to $ \widetilde{\mathcal{X}}$, DBSCAN forms clusters $C_1, \ldots, C_r$ starting from each sample $\widetilde{x}_{i} \in \mathscr{C}$. Then, expanding cluster set $\mathcal{C}(\widetilde{x}_{i})$ is defined as follows:
\begin{equation*}
\mathcal{C}(\widetilde{x}_{i}) = \{\widetilde{x}_j \mid \widetilde{x}_j \in \mathcal{D}(\widetilde{x}_{i}), j = 1, \ldots, N\},  
\end{equation*}
where we define the density-reachable set $\mathcal{D}(\widetilde{x}_{i})$ of \( \widetilde{x}_{i} \) as:

\begin{equation*}
\begin{aligned}
\mathcal{D}(\widetilde{x}_{i}) = \{&\widetilde{x}_j \in \widetilde{\mathcal{X}} \mid \exists p_1, \ldots, p_n,\\
&\text{s.t.}\ \ p_1=\widetilde{x}_{i},\ p_n=\widetilde{x}_j,\ p_{m+1} \in \mathcal{N}_{r_C}(p_m),\ p_{m+1} \in \mathscr{C},\ for\ m = 1, \ldots, n \}.
\end{aligned}
\end{equation*}

Once the cluster sets are constructed through above density-based expansion process, each $\widetilde{x}_i$ is then assigned a label $l_i$, and cluster set $\mathcal{C}_h$ for cluster $C_h$ can be defined as:
\begin{equation*}
 \mathcal{C}_h = \{\widetilde{x}_i \mid l_i = C_h \},\ h = 1, \ldots, r.
\end{equation*}

We can identify the outliers set $\mathcal{C}_{o}$, which defined as:
\begin{equation*}
\mathcal{C}_{o} = \left\{ \widetilde{x}_i \in \widetilde{\mathcal{X}} \,\middle|\, \widetilde{x}_i \notin \mathscr{C} \text{ and } \nexists\, \widetilde{x}_j \in \mathcal{N}_{r_C}(\widetilde{x}_i) \text{, s.t. } \widetilde{x}_j \in \mathscr{C} \right\}.
\end{equation*}

\subsection{Multiscale correction}
\begin{figure}[ht]
    \centering

\tikzset{every picture/.style={line width=0.75pt}} 

\tikzset{every picture/.style={line width=0.75pt}} 

\tikzset{every picture/.style={line width=0.75pt}} 

\begin{tikzpicture}[x=0.75pt,y=0.75pt,yscale=-1,xscale=1]

\draw   (154.71,64.73) -- (145.87,92.76) -- (116.49,93.02) -- (107.16,65.15) -- (130.78,47.67) -- cycle ;
\draw   (228.71,64.73) -- (219.87,92.76) -- (190.49,93.02) -- (181.16,65.15) -- (204.78,47.67) -- cycle ;
\draw   (302.71,64.73) -- (293.87,92.76) -- (264.49,93.02) -- (255.16,65.15) -- (278.78,47.67) -- cycle ;
\draw   (376.71,64.73) -- (367.87,92.76) -- (338.49,93.02) -- (329.16,65.15) -- (352.78,47.67) -- cycle ;
\draw   (450.71,64.73) -- (441.87,92.76) -- (412.49,93.02) -- (403.16,65.15) -- (426.78,47.67) -- cycle ;
\draw   (524.71,64.73) -- (515.87,92.76) -- (486.49,93.02) -- (477.16,65.15) -- (500.78,47.67) -- cycle ;
\draw    (145.87,92.76) -- (190.49,93.02) ;
\draw    (219.87,92.76) -- (264.49,93.02) ;
\draw    (293.87,92.76) -- (338.49,93.02) ;
\draw    (367.87,92.76) -- (412.49,93.02) ;
\draw    (441.87,92.76) -- (486.49,93.02) ;
\draw  [color={rgb, 255:red, 208; green, 2; blue, 27 }  ,draw opacity=1 ][fill={rgb, 255:red, 208; green, 2; blue, 27 }  ,fill opacity=1 ] (130,72.5) .. controls (130,71.67) and (130.67,71) .. (131.5,71) .. controls (132.33,71) and (133,71.67) .. (133,72.5) .. controls (133,73.33) and (132.33,74) .. (131.5,74) .. controls (130.67,74) and (130,73.33) .. (130,72.5) -- cycle ;
\draw  [color={rgb, 255:red, 208; green, 2; blue, 27 }  ,draw opacity=1 ][fill={rgb, 255:red, 208; green, 2; blue, 27 }  ,fill opacity=1 ] (204,72.5) .. controls (204,71.67) and (204.67,71) .. (205.5,71) .. controls (206.33,71) and (207,71.67) .. (207,72.5) .. controls (207,73.33) and (206.33,74) .. (205.5,74) .. controls (204.67,74) and (204,73.33) .. (204,72.5) -- cycle ;
\draw  [color={rgb, 255:red, 208; green, 2; blue, 27 }  ,draw opacity=1 ][fill={rgb, 255:red, 208; green, 2; blue, 27 }  ,fill opacity=1 ] (278,72.5) .. controls (278,71.67) and (278.67,71) .. (279.5,71) .. controls (280.33,71) and (281,71.67) .. (281,72.5) .. controls (281,73.33) and (280.33,74) .. (279.5,74) .. controls (278.67,74) and (278,73.33) .. (278,72.5) -- cycle ;
\draw  [color={rgb, 255:red, 208; green, 2; blue, 27 }  ,draw opacity=1 ][fill={rgb, 255:red, 208; green, 2; blue, 27 }  ,fill opacity=1 ] (352,72.5) .. controls (352,71.67) and (352.67,71) .. (353.5,71) .. controls (354.33,71) and (355,71.67) .. (355,72.5) .. controls (355,73.33) and (354.33,74) .. (353.5,74) .. controls (352.67,74) and (352,73.33) .. (352,72.5) -- cycle ;
\draw  [color={rgb, 255:red, 208; green, 2; blue, 27 }  ,draw opacity=1 ][fill={rgb, 255:red, 208; green, 2; blue, 27 }  ,fill opacity=1 ] (425,72.5) .. controls (425,71.67) and (425.67,71) .. (426.5,71) .. controls (427.33,71) and (428,71.67) .. (428,72.5) .. controls (428,73.33) and (427.33,74) .. (426.5,74) .. controls (425.67,74) and (425,73.33) .. (425,72.5) -- cycle ;
\draw  [color={rgb, 255:red, 208; green, 2; blue, 27 }  ,draw opacity=1 ][fill={rgb, 255:red, 208; green, 2; blue, 27 }  ,fill opacity=1 ] (499,72.5) .. controls (499,71.67) and (499.67,71) .. (500.5,71) .. controls (501.33,71) and (502,71.67) .. (502,72.5) .. controls (502,73.33) and (501.33,74) .. (500.5,74) .. controls (499.67,74) and (499,73.33) .. (499,72.5) -- cycle ;
\draw [color={rgb, 255:red, 208; green, 2; blue, 27 }  ,draw opacity=1 ]   (133,72.5) -- (204,72.5) ;
\draw [color={rgb, 255:red, 208; green, 2; blue, 27 }  ,draw opacity=1 ]   (207,72.5) -- (278,72.5) ;
\draw [color={rgb, 255:red, 208; green, 2; blue, 27 }  ,draw opacity=1 ]   (281,72.5) -- (352,72.5) ;
\draw [color={rgb, 255:red, 208; green, 2; blue, 27 }  ,draw opacity=1 ]   (354,72.5) -- (425,72.5) ;
\draw [color={rgb, 255:red, 208; green, 2; blue, 27 }  ,draw opacity=1 ]   (428,72.5) -- (499,72.5) ;
\draw    (71.87,92.76) -- (116.49,93.02) ;
\draw    (515.87,92.76) -- (560.49,93.02) ;
\draw    (293.87,112.76) -- (338.49,113.02) ;
\draw    (368.87,112.76) -- (413.49,113.02) ;
\draw    (441.87,111.76) -- (486.49,112.02) ;
\draw    (145.87,112.76) -- (190.49,113.02) ;
\draw    (218.87,112.76) -- (263.49,113.02) ;
\draw [color={rgb, 255:red, 157; green, 198; blue, 114 }  ,draw opacity=1 ]   (167,135.67) -- (167,160.67) ;
\draw [shift={(167,162.67)}, rotate = 270] [color={rgb, 255:red, 157; green, 198; blue, 114 }  ,draw opacity=1 ][line width=0.75]    (10.93,-3.29) .. controls (6.95,-1.4) and (3.31,-0.3) .. (0,0) .. controls (3.31,0.3) and (6.95,1.4) .. (10.93,3.29)   ;
\draw [color={rgb, 255:red, 157; green, 198; blue, 114 }  ,draw opacity=1 ]   (239,136.67) -- (239,161.67) ;
\draw [shift={(239,163.67)}, rotate = 270] [color={rgb, 255:red, 157; green, 198; blue, 114 }  ,draw opacity=1 ][line width=0.75]    (10.93,-3.29) .. controls (6.95,-1.4) and (3.31,-0.3) .. (0,0) .. controls (3.31,0.3) and (6.95,1.4) .. (10.93,3.29)   ;
\draw [color={rgb, 255:red, 157; green, 198; blue, 114 }  ,draw opacity=1 ]   (315,137.67) -- (315,162.67) ;
\draw [shift={(315,164.67)}, rotate = 270] [color={rgb, 255:red, 157; green, 198; blue, 114 }  ,draw opacity=1 ][line width=0.75]    (10.93,-3.29) .. controls (6.95,-1.4) and (3.31,-0.3) .. (0,0) .. controls (3.31,0.3) and (6.95,1.4) .. (10.93,3.29)   ;
\draw [color={rgb, 255:red, 157; green, 198; blue, 114 }  ,draw opacity=1 ]   (391,136.67) -- (391,161.67) ;
\draw [shift={(391,163.67)}, rotate = 270] [color={rgb, 255:red, 157; green, 198; blue, 114 }  ,draw opacity=1 ][line width=0.75]    (10.93,-3.29) .. controls (6.95,-1.4) and (3.31,-0.3) .. (0,0) .. controls (3.31,0.3) and (6.95,1.4) .. (10.93,3.29)   ;
\draw [color={rgb, 255:red, 157; green, 198; blue, 114 }  ,draw opacity=1 ]   (465,137.67) -- (465,162.67) ;
\draw [shift={(465,164.67)}, rotate = 270] [color={rgb, 255:red, 157; green, 198; blue, 114 }  ,draw opacity=1 ][line width=0.75]    (10.93,-3.29) .. controls (6.95,-1.4) and (3.31,-0.3) .. (0,0) .. controls (3.31,0.3) and (6.95,1.4) .. (10.93,3.29)   ;
\draw [color={rgb, 255:red, 74; green, 144; blue, 226 }  ,draw opacity=1 ]   (166,198.67) -- (166,223.67) ;
\draw [shift={(166,225.67)}, rotate = 270] [color={rgb, 255:red, 74; green, 144; blue, 226 }  ,draw opacity=1 ][line width=0.75]    (10.93,-3.29) .. controls (6.95,-1.4) and (3.31,-0.3) .. (0,0) .. controls (3.31,0.3) and (6.95,1.4) .. (10.93,3.29)   ;
\draw [color={rgb, 255:red, 74; green, 144; blue, 226 }  ,draw opacity=1 ]   (238,199.67) -- (238,224.67) ;
\draw [shift={(238,226.67)}, rotate = 270] [color={rgb, 255:red, 74; green, 144; blue, 226 }  ,draw opacity=1 ][line width=0.75]    (10.93,-3.29) .. controls (6.95,-1.4) and (3.31,-0.3) .. (0,0) .. controls (3.31,0.3) and (6.95,1.4) .. (10.93,3.29)   ;
\draw [color={rgb, 255:red, 74; green, 144; blue, 226 }  ,draw opacity=1 ]   (314,200.67) -- (314,225.67) ;
\draw [shift={(314,227.67)}, rotate = 270] [color={rgb, 255:red, 74; green, 144; blue, 226 }  ,draw opacity=1 ][line width=0.75]    (10.93,-3.29) .. controls (6.95,-1.4) and (3.31,-0.3) .. (0,0) .. controls (3.31,0.3) and (6.95,1.4) .. (10.93,3.29)   ;
\draw [color={rgb, 255:red, 74; green, 144; blue, 226 }  ,draw opacity=1 ]   (390,199.67) -- (390,224.67) ;
\draw [shift={(390,226.67)}, rotate = 270] [color={rgb, 255:red, 74; green, 144; blue, 226 }  ,draw opacity=1 ][line width=0.75]    (10.93,-3.29) .. controls (6.95,-1.4) and (3.31,-0.3) .. (0,0) .. controls (3.31,0.3) and (6.95,1.4) .. (10.93,3.29)   ;
\draw [color={rgb, 255:red, 74; green, 144; blue, 226 }  ,draw opacity=1 ]   (464,200.67) -- (464,225.67) ;
\draw [shift={(464,227.67)}, rotate = 270] [color={rgb, 255:red, 74; green, 144; blue, 226 }  ,draw opacity=1 ][line width=0.75]    (10.93,-3.29) .. controls (6.95,-1.4) and (3.31,-0.3) .. (0,0) .. controls (3.31,0.3) and (6.95,1.4) .. (10.93,3.29)   ;
\draw [color={rgb, 255:red, 245; green, 166; blue, 35 }  ,draw opacity=1 ]   (484,246) -- (532.2,246) ;
\draw [color={rgb, 255:red, 245; green, 166; blue, 35 }  ,draw opacity=1 ]   (532.2,246) -- (532.2,141) ;
\draw [shift={(532.2,139)}, rotate = 90] [color={rgb, 255:red, 245; green, 166; blue, 35 }  ,draw opacity=1 ][line width=0.75]    (10.93,-3.29) .. controls (6.95,-1.4) and (3.31,-0.3) .. (0,0) .. controls (3.31,0.3) and (6.95,1.4) .. (10.93,3.29)   ;
\draw [color={rgb, 255:red, 245; green, 166; blue, 35 }  ,draw opacity=1 ]   (532.2,246) -- (580.4,246) ;
\draw [color={rgb, 255:red, 245; green, 166; blue, 35 }  ,draw opacity=1 ]   (580.4,246) -- (580.4,88) ;
\draw [shift={(580.4,86)}, rotate = 90] [color={rgb, 255:red, 245; green, 166; blue, 35 }  ,draw opacity=1 ][line width=0.75]    (10.93,-3.29) .. controls (6.95,-1.4) and (3.31,-0.3) .. (0,0) .. controls (3.31,0.3) and (6.95,1.4) .. (10.93,3.29)   ;

\draw (121,55) node [anchor=north west][inner sep=0.75pt]  [font=\footnotesize] [align=left] {$\displaystyle s_{i-2}$};
\draw (194,55) node [anchor=north west][inner sep=0.75pt]  [font=\footnotesize] [align=left] {$\displaystyle s_{i-1}$};
\draw (274,55) node [anchor=north west][inner sep=0.75pt]  [font=\footnotesize] [align=left] {$\displaystyle s_{i}$};
\draw (343,55) node [anchor=north west][inner sep=0.75pt]  [font=\footnotesize] [align=left] {$\displaystyle s_{i+1}$};
\draw (417,55) node [anchor=north west][inner sep=0.75pt]  [font=\footnotesize] [align=left] {$\displaystyle s_{i+2}$};
\draw (490,55) node [anchor=north west][inner sep=0.75pt]  [font=\footnotesize] [align=left] {$\displaystyle s_{i+3}$};
\draw (298,42) node [anchor=north west][inner sep=0.75pt]   [align=left] {Suite i};
\draw (307,115) node [anchor=north west][inner sep=0.75pt]   [align=left] {$\displaystyle x_{i}$};
\draw (377,115) node [anchor=north west][inner sep=0.75pt]   [align=left] {$\displaystyle x_{i+1}$};
\draw (450,114) node [anchor=north west][inner sep=0.75pt]   [align=left] {$\displaystyle x_{i+2}$};
\draw (154,115) node [anchor=north west][inner sep=0.75pt]   [align=left] {$\displaystyle x_{i-2}$};
\draw (227,115) node [anchor=north west][inner sep=0.75pt]   [align=left] {$\displaystyle x_{i-1}$};
\draw (154,238) node [anchor=north west][inner sep=0.75pt]   [align=left] {$\displaystyle l_{i-2}$};
\draw (153,175) node [anchor=north west][inner sep=0.75pt]   [align=left] {$\displaystyle \widetilde{x}_{i-2}$};
\draw (226,175) node [anchor=north west][inner sep=0.75pt]   [align=left] {$\displaystyle \widetilde{x}_{i-1}$};
\draw (309,175) node [anchor=north west][inner sep=0.75pt]   [align=left] {$\displaystyle \widetilde{x}_{i}$};
\draw (378,175) node [anchor=north west][inner sep=0.75pt]   [align=left] {$\displaystyle \widetilde{x}_{i+1}$};
\draw (450,175) node [anchor=north west][inner sep=0.75pt]   [align=left] {$\displaystyle \widetilde{x}_{i+2}$};
\draw (88,203) node [anchor=north west][inner sep=0.75pt]   [align=left] {\textcolor[rgb]{0.29,0.56,0.89}{DBSCAN}};
\draw (101,141) node [anchor=north west][inner sep=0.75pt]   [align=left] {\textcolor[rgb]{0.62,0.78,0.45}{PSM}};
\draw (229,238) node [anchor=north west][inner sep=0.75pt]   [align=left] {$\displaystyle l_{i-1}$};
\draw (310,239) node [anchor=north west][inner sep=0.75pt]   [align=left] {$\displaystyle l_{i}$};
\draw (380,239) node [anchor=north west][inner sep=0.75pt]   [align=left] {$\displaystyle l_{i+1}$};
\draw (454,239) node [anchor=north west][inner sep=0.75pt]   [align=left] {$\displaystyle l_{i+2}$};
\draw (524,115) node [anchor=north west][inner sep=0.75pt]    {$\mu ^{c}_{\mathbb{T}}$};
\draw (570,59.4) node [anchor=north west][inner sep=0.75pt]    {$\mu ^{c}_{\Lambda}$};
\draw (544,176) node [anchor=north west][inner sep=0.75pt]  [color={rgb, 255:red, 245; green, 166; blue, 35 }  ,opacity=1 ] [align=left] {MC};
\draw (510,250.4) node [anchor=north west][inner sep=0.75pt]    {$\mathcal{C}_1, \ldots, \mathcal{C}_r$};

\end{tikzpicture}

 \caption{\textit{Framework of PSM-DBSCASN-MC for RNA correction at microscopic and mesoscopic scale. }}

\label{fig:RNA multiscale correction} 
\end{figure}
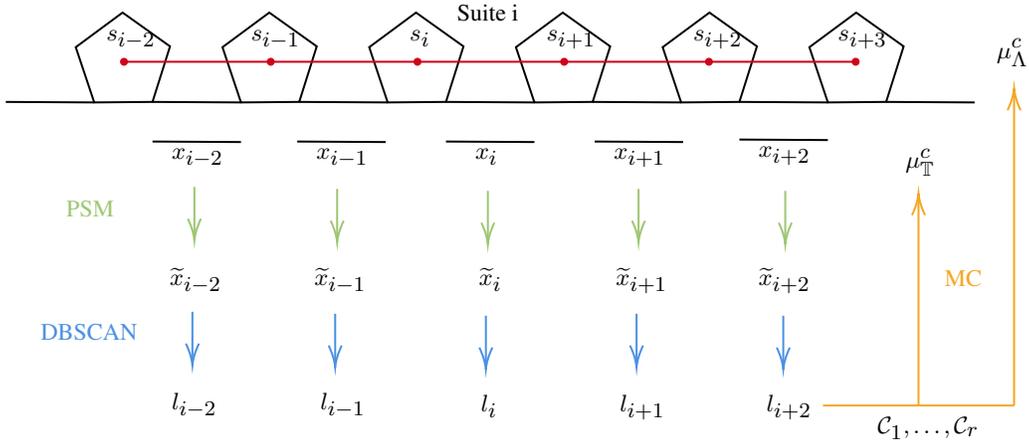

\cite{wiechers2023learning} introduce the multiscale correction method that integrates RNA correction at both microscopic and mesoscopic scales. Based on the cluster set $\mathcal{C}_1, \ldots, \mathcal{C}_r$ generated by PSM-DBSCAN, which efficiently identifies the different RNA suites, we propose a method named as PSM-DBSCAN-MC (Figure \ref{fig:RNA multiscale correction}). 

Multiscale RNA correction starts with a clash suites sample set \(\mathcal{X}^c\). For each clash suite $x_i^c$, its corresponding mesoscopic shape \([S^c]\) is defined as (\ref{mesoscopic}) by the landmark configuration matrix $S^c = (s_1, \ldots, s_{2k+2})^T$. Based on a clash-free suites sample set $\mathcal{X}$, we address each clash suite $x_i^c$ as follows. First, we define a neighborhood \(\mathcal{N}^c\) of $x_i^c$ consisting of the \( \kappa \)-nearest clash-free suites in $\mathcal{X}$. This neighborhood is determined by the mesoscopic shape space distance (\ref{shape distance}):

\begin{equation*}
\mathcal{N}^c := \{ x_i \in \mathcal{X} : \#\{ x_j \in \mathcal{X} : \operatorname{d}_{\Lambda}([S_j], [S^c]) \leq \operatorname{d}_{\Lambda}([S_i], [S^c])\} \leq \kappa \},
\end{equation*}
where $\#$ is the number of set elements and $[S_i]$ denotes corresponding mesoscopic shape $[S]$ of training suite $x_i$. Then, we identify the dominant class \(\mathcal{C}_{h^c}\) of $x_i^c$ within its neighborhood $\mathcal{N}^c$:

\begin{equation*}
h^c \in \argmax_{h=1, \dots, r} \#(\mathcal{C}_h \cap \mathcal{N}^c). 
\end{equation*}

After establishing \(\mathcal{C}_{h^c}\), Fr\'echet mean \(\mu_{\mathbb{T}}^c\) is used to correct the microscopic shape, representing the geometric center of the clash-free suites with the same cluster $\mathcal{C}_{h^c}$ within neighborhood \(\mathcal{N}^c\):

\begin{equation*}
\mu_{\mathbb{T}}^c \in \argmin_{x \in \mathbb{T}^D} \sum_{x_i \in \mathcal{C}_{h^c} \cap \mathcal{N}^c} \operatorname{d}_{\mathbb{T}^D}(x, x_i)^2.
\end{equation*}

Fr\'echet mean \(\mu_{\Lambda}^c = (\mu_1, \ldots, \mu_{2k+2})^T\) is used to correct the mesoscopic shape:

\begin{equation*}
\mu_{\Lambda}^c \in \argmin_{[S] \in \Lambda^{2k+2}_3} \sum_{i \in \{i \mid x_i \in \mathcal{C}_{h^c} \cap \mathcal{N}^c \}} \operatorname{d}_{\Lambda}([S], [S_i])^2.
\end{equation*}

The adjusted mesoscopic shape is then calculated by modifying the distances between the core sugar rings to the new length, determined through the Fr\'echet mean \(\mu_{\mathbb{T}}^c\), while satisfying the constrained condition set
\[
\left\{[S] = [s_1, \ldots, s_{2k+2}] \in \Lambda^{2k+2}_3 : \|s_1 - s_{2k+2}\|_2 = \theta_1,\ \|s_{k+2} - s_{k+1}\|_2 = \theta_2 \right\},
\]
where \(\theta_1\) denotes the Euclidean distance between the first and last sugar ring in the original mesoscopic shape, and \(\theta_2\) denotes the distance between the central two landmarks. The first constraint enforces compatibility with the global geometry of the RNA strand, while the second condition aligns the distance between the central landmarks with the length of the revised suite. This adjustment involves an orthogonal projection of the size-and-shape $W^*$ of the Fr\'echet mean $\mu_{\Lambda}^c$ for $i \in \{1, k+1, k+2, 2k+2\}$:

\begin{equation*}
w^*_1 = p \mu_1 + (1 - p) \mu_{2k+2}, \quad w^*_{2k+2} = p \mu_{2k+2} + (1 - p) \mu_1,
\end{equation*}
\begin{equation*}
w^*_{k+1} = q \mu_{k+1} + (1 - q) \mu_{k+2}, \quad w^*_{k+2} = q \mu_{k+2} + (1 - q) \mu_{k+1},
 \end{equation*}
where
\begin{equation*}
p = \frac{1}{2} \left(1 + \frac{\|s_{2k+2} - s_1\|_2}{\|\mu_{2k+2} - \mu_1\|_2}\right), \quad q = \frac{1}{2} \left(1 + \frac{\|s_{k+2} - s_{k+1}\|_2}{\|\mu_{k+2} - \mu_{k+1}\|_2}\right).
\end{equation*}
and $w^*_i = \mu_i$ for otherwise.
 
PSM-DBSCAN-MC outputs a corrected microscopic shape $\mu_{\mathbb{T}}^c$ and its corresponding corrected mesoscopic shape \([W^*]\) to reflect the new, clash-free configuration of the RNA suites.

\section{Simulations} \label{Simulations}
We conduct a series of simulation experiments divided into two parts. In the first part, we systematically compare the dimensionality reduction performance of PSM and tPCA for data situated on the torus $\mathbb{T}^D$. In the second part, we compare the clustering performance of PSM-DBSCAN with other clustering methods. 

First, we discuss how to set the tuning parameters. We adopted a grid search strategy for selecting tuning parameters across all methods, including our proposed approach and the baseline methods. In simulation studies where ground-truth labels are available, the dataset was randomly split into training and validation subsets. Parameters are then chosen to maximize the average performance on the validation set, measured by Adjusted Mutual Information (AMI) and Normalized Mutual Information (NMI). For data without labels, parameter selection is guided by internal clustering criteria.

For all baseline methods (e.g., DBSCAN, tPCA-DBSCAN, MINT-AGE, HC, and SC), we apply a consistent grid search procedure over parameter ranges recommended in the corresponding literature. Here, we outline the basic search procedure; implementation details can be found in the Appendix \ref{appendix:simulation_results}. For DBSCAN-based clustring methods, the two main parameters are $r_{C}$ and $\eta$. We consider a candidate set of $\eta \in \{4,5,8,10,15,20,30,40,50\}$, while $r_C$ is selected from quantiles $\{0.60,0.65,\dots,0.95\}$. For MINT-AGE, the tuning parameters are $d_{\max}$, $\kappa$, and $q$. We set $\kappa \in \{15,20,30\}$ and $q \in \{0.10,0.15,0.20\}$, while $d_{\max}$ is chosen to yield outlier fractions in $\{5\%,10\%,15\%,20\%\}$. For HC, the key parameters are the linkage criterion and the cut level. We consider linkage methods from $\in \{\text{average}, \text{complete}, \text{ward}\}$, and apply cuts either by specifying the number of clusters $K \in \{2,\dots,10\}$, or by thresholding at quantiles $\{0.60,0.65,\dots,0.95\}$ of the fusion distances. For SC, the parameters include the number of clusters $K$, the graph construction method, and the kernel scale. We vary $K \in \{2,\dots,10\}$; for the RBF graph we set the kernel scale $\sigma \in \{0.25,0.5,1,2,4\}$ times the median pairwise distance, and for the $k$-NN graph we use $k \in \{5,10,15,20,30,50\}$, with optional RBF weights at scales $\{0.5,1,2\}$.

\subsection{Experiment 1 (dimensionality reduction)}

A series of subexperiments $(a-l)$ is conducted to compare PSM with tPCA. In each subexperiment, we generate 3000 samples on the torus $\mathbb{T}^D$, ranging from $D=2, \ldots, 7$. Each subexperiment is repeated 10 times to assess the stability and robustness. For each $D$-dimensional torus, two subexperiments are conducted to generate samples $\mathcal{X}$ around a submanifold $\psi$ with intrinsic dimension $1$ and $D-1$. Two types of submanifolds are considered: linear and non-linear. For the linear case, the samples $\mathcal{X}$ are generated as follows: all components $x_i^{(j)}$ are set equal to a value drawn from a $\text{Uniform}(0, 2\pi)$ sample. For the non-linear case, a relationship among the different components is established via a sinusoidal function: $x_i^{(2j)} = \sin(2x_i^{(2j-1)})+\pi$. Gaussian noise \(\epsilon \overset{\text{i.i.d.}}{\sim} N(0, \Sigma)\) is added to $x_i$, with the covariance matrix \( \Sigma = \sigma^2 I \) ($\sigma = 0.2$).

Traditional metrics often fail to accurately assess the effectiveness of dimensionality reduction methods on the manifold. To address this challenge, we use two novel metrics: approximation error $\Ae$ and proportion of information retained $P_d$, which are provided in Appendix \ref{appendix: Definitions of evaluation metrics}.

Table \ref{tab:approximation error} reports the approximation error, reflecting how well the low-dimensional data aligns with the underlying submanifold \(\psi\). Overall, the approximation errors of the PSM method are consistently lower than those of tPCA for low-dimensional representation, especially when \(d=1\), which demonstrates its superior performance across most subexperimental scenarios. Specifically, the approximation errors of PSM range from \(0.01\) to \(0.29\), whereas tPCA errors range from about \(0.11\) to \(0.43\), clearly indicating that PSM significantly reduces the approximation error. This suggests that the low-dimensional representation fitted by PSM is closer to the intrinsic true structure of the samples than tPCA. Additionally, the standard deviations (shown in parentheses) for PSM are generally smaller than those for tPCA, indicating that PSM provides more stable and reliable approximations.

\begin{table}[H]
\centering
\caption{ \textit{Comparison of approximation error between PSM and tPCA. Standard deviations computed based on 10 repetitions of simulations are shown in parentheses, and all standard deviations are scaled by $10^{-4}$. Bold font indicates the best performance under each experimental setting.}}

\begin{tabular}{ccllllll}
\hline
\multicolumn{8}{c}{{PSM}} \\ \hline
${D}$ & Exp1 & \phantom{X}${d=1}$ & \phantom{X}${d=2}$ & \phantom{X}${d=3}$ & \phantom{X}${d=4}$ & \phantom{X}${d=5}$ & \phantom{X}${d=6}$ \\ \hline
$2$ & $a$ & $\mathbf{.01} (11)$ & & & & & \\
    & $b$ & $\mathbf{.05} (58)$ & & & & & \\ 
$3$ & $c$ & $\mathbf{.02} (43)$ & $\mathbf{.15} (24)$ & & & & \\
    & $d$ & $\mathbf{.07} (50)$ & $\mathbf{.09} (111)$ & & & & \\ 
$4$ & $e$ & $\mathbf{.03} (24)$ & $.17 (43)$ & $\mathbf{.24} (30)$ & & &\\
    & $f$ & $\mathbf{.11} (75)$ & $\mathbf{.11} (48)$ & $.19 (56)$ & & &\\ 
$5$ & $g$ & $\mathbf{.03} (9)$ & $\mathbf{.18} (37)$ & $\mathbf{.24} (48)$ & $\mathbf{.29} (17)$ & &\\
    & $h$ & $\mathbf{.15} (158)$ & $\mathbf{.12} (101)$ & $.19 (26)$ & $\mathbf{.23} (21)$ & &\\ 
$6$ & $i$ & $\mathbf{.01} (14)$ & $\mathbf{.07} (31)$ & $\mathbf{.11} (26)$ & $\mathbf{.14} (17)$ & $\mathbf{.17} (9)$ &\\
    & $j$ & $\mathbf{.11} (29)$ & $.13 (31)$ & $.11 (39)$ & $.12 (7)$ & $\mathbf{.15} (12)$  &\\ 
$7$ & $k$ & $\mathbf{.02} (17)$ & $\mathbf{.07} (36)$ & $\mathbf{.12} (11)$ & $\mathbf{.15} (28)$ & $\mathbf{.17} (15)$ & $\mathbf{.19} (11)$ \\
    & $l$ & $\mathbf{.12} (32)$ & $.13 (37)$ & $.12 (34)$ & $.13 (36)$ & $\mathbf{.14} (15)$ & $\mathbf{.16} (11)$ \\ \hline
    
\multicolumn{8}{c}{{tPCA}} \\ \hline
${D}$ & Exp1 & \phantom{X}${d=1}$ & \phantom{X}${d=2}$ & \phantom{X}${d=3}$ & \phantom{X}${d=4}$ & \phantom{X}${d=5}$ & \phantom{X}${d=6}$  \\ \hline
$2$ & $a$ & $.29 (40)$ & & & & &\\
    & $b$ & $.43 (978)$ & & & & &\\ 
$3$ & $c$ & $.26 (21)$ & $.15 (28)$ & & & & \\
    & $d$ & $.22 (295)$ & $.14 (36)$ & & & & \\ 
$4$ & $e$ & $.30 (48)$ & $.17 (40)$ & $.25 (24)$ & & & \\
    & $f$ & $.23 (429)$ & $.15 (51)$ & $\mathbf{.17} (46)$ & & &  \\ 
$5$ & $g$ & $.29 (26)$ & $.22 (57)$ & $.25 (32)$ & $.32 (19)$ & &  \\
    & $h$ & $.16 (363)$ & $.15 (36)$ & $\mathbf{.17} (39)$ & $.25 (26)$ & & \\ 
$6$ & $i$ & $.28 (48)$ & $.22 (40)$ & $.13 (19)$ & $.16 (14)$ & $.19 (11)$ & \\
    & $j$ & $.17 (42)$ & $\mathbf{.12} (14)$ & $.11 (10)$ & $.12 (11)$ & $.16 (18)$ & \\ 
$7$ & $k$ & $.28 (69)$ & $.28 (53)$ & $.15 (66)$ & $.17 (27)$ & $.19 (16)$ & $.21 (18)$  \\
    & $l$ & $.17 (20)$ & $\mathbf{.11} (12)$ & $\mathbf{.11} (7)$ & $.13 (5)$ & $.16 (10)$ & $.19 (14)$  \\ \hline
\end{tabular}
\label{tab:approximation error}
\end{table}

Table \ref{tab:proportion of information retained} reports the proportion of information preserved in the low-dimensional representation. Overall, it is clear that PSM consistently achieves higher proportions of retained information in the 1-dimensional representation (ranging from \(0.81\) to \(0.99\)) compared to tPCA (ranging from \(0.43\) to \(0.94\)). Specifically, PSM outperforms tPCA by approximately \(5\%\) to \(50\%\) in terms of information preservation, demonstrating that PSM has a clear advantage in effectively preserving the primary information in low-dimensional representations across various manifold dimensions \(D\).

In summary, PSM produces a more accurate low-dimensional representation of the original data on the torus \(\mathbb{T}^D\), which is closer to the intrinsic low-dimensional geometric structure of the data while preserving more feature information. Therefore, PSM offers significant advantages over tPCA, making it particularly valuable for downstream tasks.

\begin{table}[H]
\centering
\caption{ \textit{Comparison of proportion of information retained between PSM and tPCA. Standard deviations computed based on 10 repetitions of simulations are shown in parentheses, and all standard deviations are scaled by $10^{-4}$. Bold font indicates the best performance under each experimental setting.}}

\begin{tabular}{cclllllll}
\hline
\multicolumn{9}{c}{{PSM}} \\ \hline
${D}$ & Exp1 & \phantom{X}${d=1}$ & \phantom{X}${d=2}$ & \phantom{X}${d=3}$ & \phantom{X}${d=4}$ & \phantom{X}${d=5}$ & \phantom{X}${d=6}$ & \phantom{X}${d=7}$ \\ \hline
$2$ & $a$ & $\mathbf{.81} (218)$ & $.19 (218)$ & & & & & \\
    & $b$ & $\mathbf{.90} (52)$ & $.10 (52)$ & & & & & \\ 
$3$ & $c$ & $\mathbf{.98} (4)$ & $.01 (4)$ & $.01 (8)$ & & & & \\
    & $d$ & $\mathbf{.96} (14)$ & $.02 (14)$ & $.02 (29)$ & & & & \\ 
$4$ & $e$ & $\mathbf{.99} (1)$ & $.01 (1)$ & $.00 (2)$ & $.00 (1)$ & & & \\
    & $f$ & $\mathbf{.96} (7)$ & $.02 (6)$ & $.01 (18)$ & $.01 (21)$ & & & \\ 
$5$ & $g$ & $\mathbf{.99} (1)$ & $.01 (1)$ & $.00 (1)$ & $.00 (1)$ & $.00 (3)$ & & \\
    & $h$ & $\mathbf{.95} (3)$ & $.02 (2)$ & $.01 (4)$ & $.01 (9)$ & $.01 (9)$ & & \\ 
$6$ & $i$ & $\mathbf{.99} (0)$ & $.01 (0)$ & $.00 (0)$ & $.00 (1)$ & $.00 (0)$ & $.00 (2)$ & \\
    & $j$ & $\mathbf{.95} (2)$ & $.01 (1)$ & $.01 (3)$ & $.01 (4)$ & $.01 (14)$ & $.01 (15)$ & \\ 
$7$ & $k$ & $\mathbf{.99} (0)$ & $.01 (0)$ & $.00 (0)$ & $.00 (0)$ & $.00 (1)$ & $.00 (5)$ & $.00 (22)$ \\
    & $l$ & $\mathbf{.95} (2)$ & $.01 (0)$ & $.01 (1)$ & $.01 (3)$ & $.01 (6)$ & $.00 (23)$ & $.01 (22)$ \\ \hline
    
\multicolumn{9}{c}{{tPCA}} \\ \hline
${D}$ & Exp1 & \phantom{X}${d=1}$ & \phantom{X}${d=2}$ & \phantom{X}${d=3}$ & \phantom{X}${d=4}$ & \phantom{X}${d=5}$ & \phantom{X}${d=6}$ & \phantom{X}${d=7}$\\ \hline
$2$ & $a$ & $.44 (253)$ & $.56 (253)$ & & & & & \\
    & $b$ & $.43 (356)$ & $.57 (356)$ & & & & & \\ 
$3$ & $c$ & $.94 (3)$ & $.05 (6)$ & $.01 (7)$ & & & & \\
    & $d$ & $.82 (9)$ & $.15 (67)$ & $.03 (63)$ & & & & \\ 
$4$ & $e$ & $.92 (3)$ & $.06 (31)$ & $.01 (32)$ & $.01 (62)$ & & & \\
    & $f$ & $.82 (32)$ & $.13 (21)$ & $.03 (135)$ & $.02 (89)$ & & & \\ 
$5$ & $g$ & $.89 (2)$ & $.06 (11)$ & $.03 (11)$ & $.01 (16)$ & $.01 (16)$ & & \\
    & $h$ & $.80 (16)$ & $.11 (15)$ & $.05 (26)$ & $.02 (104)$ & $.02 (66)$ & & \\ 
$6$ & $i$ & $.90 (0)$ & $.06 (1)$ & $.02 (2)$ & $.01 (7)$ & $.01 (21)$ & $.00 (25)$ & \\
    & $j$ & $.83 (2)$ & $.10 (2)$ & $.05 (1)$ & $.01 (7)$ & $.01 (21)$ & $.00 (32)$ & \\ 
$7$ & $k$ & $.90 (0)$ & $.06 (0)$ & $.03 (2)$ & $.01 (0)$ & $.00 (3)$ & $.00 (3)$ & $.00 (0)$ \\
    & $l$ & $.83 (2)$ & $.10 (2)$ & $.05 (1)$ & $.01 (2)$ & $.01 (4)$ & $.01 (24)$ & $.00 (32)$ \\ \hline
\end{tabular}

\label{tab:proportion of information retained}
\end{table}

\subsection{Experiment 2 (clustering)}
A series of subexperiments $(a-l)$ is conducted to evaluate the performance of PSM-DBSCAN. In each subexperiment, we generate 1000 samples $\mathcal{X}$ for each distinct submanifold $\psi$ on the torus $\mathbb{T}^D$, ranging from $D = 2, \ldots, 7$. The true label of each sample $x_i$ is determined by the submanifold to which it belongs. Specifically, subexperiment $(a-e)$, $(g)$, $(i)$ and $(k)$ contain $3$ submanifolds, subexperiment $(f)$, $(h)$, $(j)$ and $(l)$ contain $5$ submanifolds. Each subexperiment is repeated 10 times to assess the stability and robustness. Two types of submanifolds are considered: linear and non-linear. For the linear case, samples $\mathcal{X}$ are generated on the lines parameterized by the equation: $\mathbf{x}_{\psi} = \mathbf{p}_{\psi} + z (\mathbf{q}_{\psi} - \mathbf{p}_{\psi})$ for $z \in [0, 1]$. We randomly generate the starting point $\mathbf{p}_{\psi}$ and the ending point $\mathbf{q}_{\psi}$. For the non-linear case, a relationship among the different components is established via a sinusoidal function:$x_i^{(2j)} = \alpha_{\psi}\sin\!\bigl(\beta_{\psi} x_i^{(2j-1)}\bigr) + \gamma_{\psi}$, where $\alpha_{{\psi}} \in \{\frac{1}{3},\frac{1}{2},1,2,3\}, \beta_{{\psi}} \in \{\frac{1}{2},1,2\}, \gamma_{{\psi}} \in \{-\pi, -\frac{\pi}{2},0,\frac{\pi}{2}, \pi\}$. Gaussian noise \(\epsilon \overset{\text{i.i.d.}}{\sim} N(0, \Sigma)\) is added to $x_i$, with the covariance matrix \( \Sigma = \sigma^2 I \) ($\sigma \in \{0.3, 0.4, 0.5\}$).

\begin{figure}[H]
\centering
\rotatebox{90}{ 
\begin{minipage}{\textheight} 
\subfloat{}{
  \includegraphics[width=0.12\textwidth]{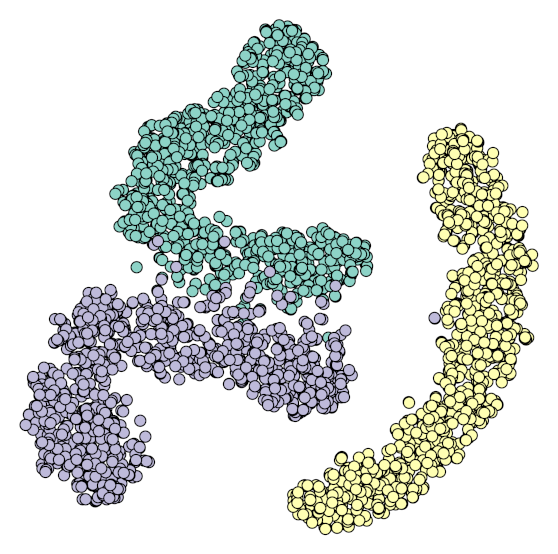}
}
\subfloat{}{
  \includegraphics[width=0.12\textwidth]{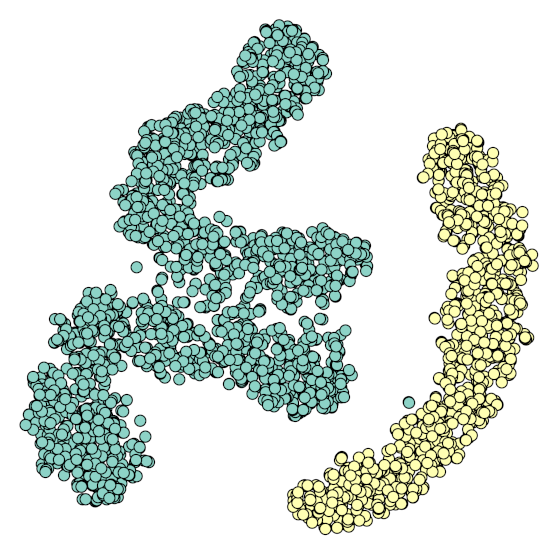}
}
\subfloat{}{
  \includegraphics[width=0.12\textwidth]{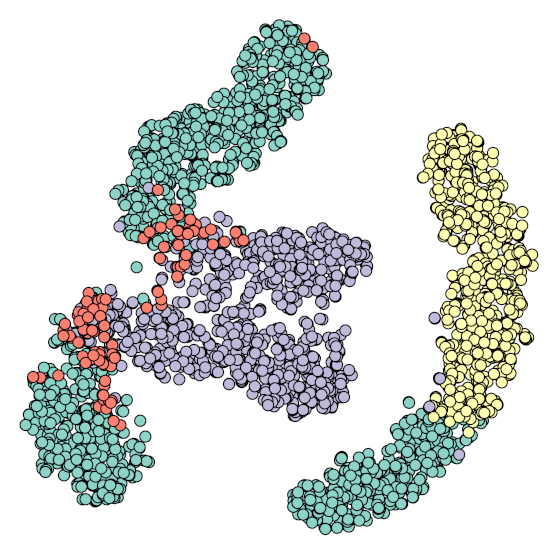}
}
\subfloat{}{
  \includegraphics[width=0.12\textwidth]{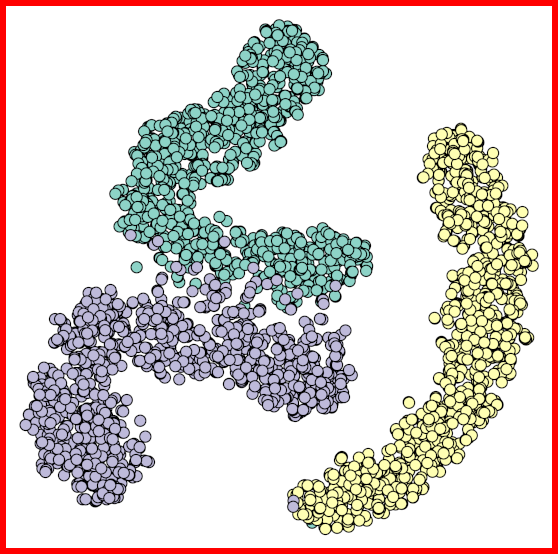}
}
\subfloat{}{
  \includegraphics[width=0.12\textwidth]{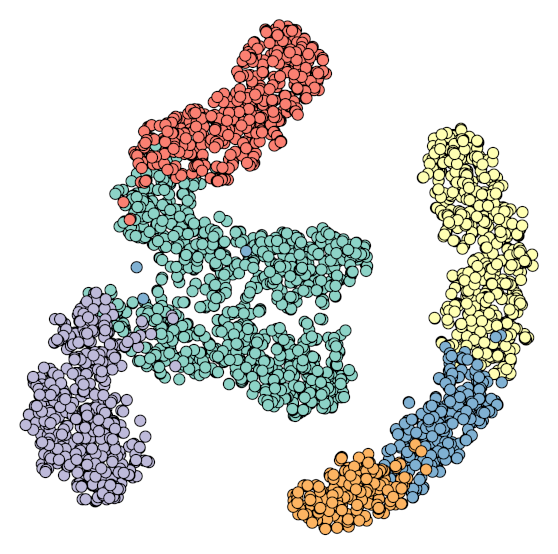}
}
\subfloat{}{
  \includegraphics[width=0.12\textwidth]{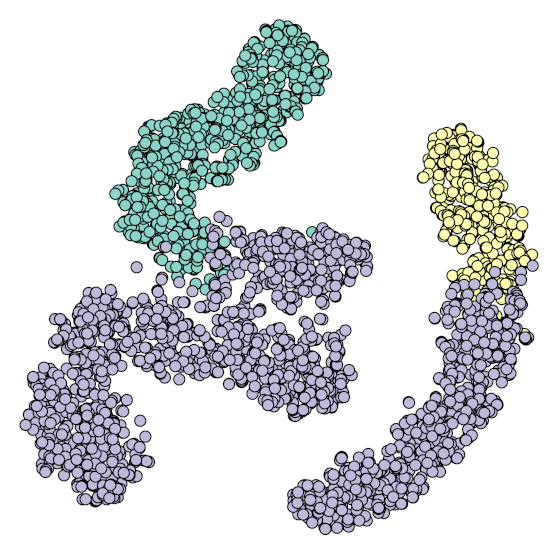}
}
\subfloat{}{
  \includegraphics[width=0.12\textwidth]{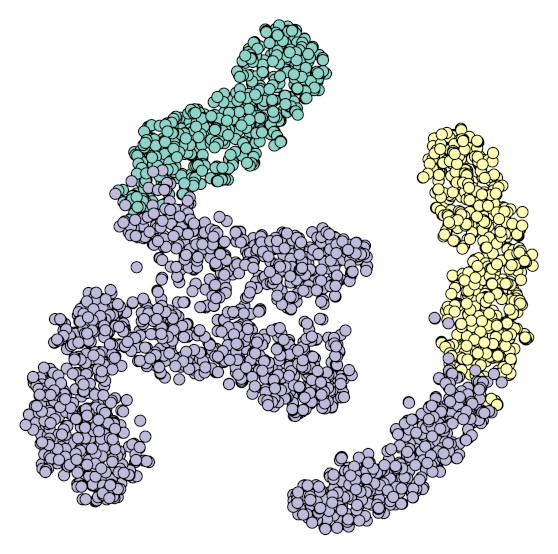}
}

\subfloat{}{
  \includegraphics[width=0.12\textwidth]{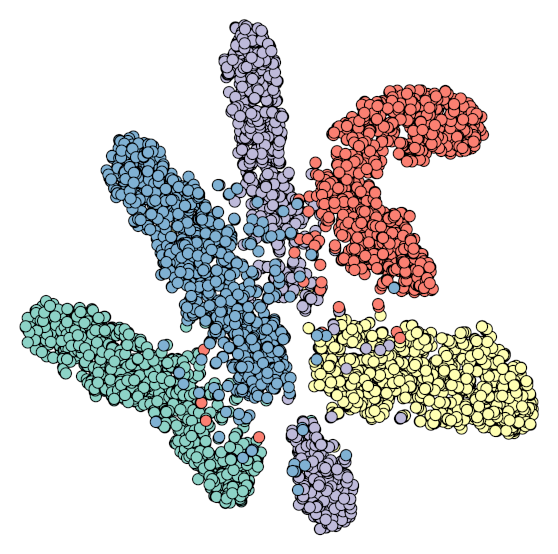}
}
\subfloat{}{
  \includegraphics[width=0.12\textwidth]{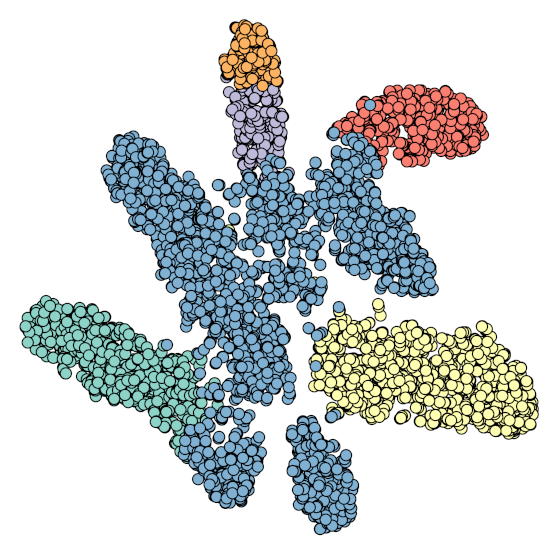}
}
\subfloat{}{
  \includegraphics[width=0.12\textwidth]{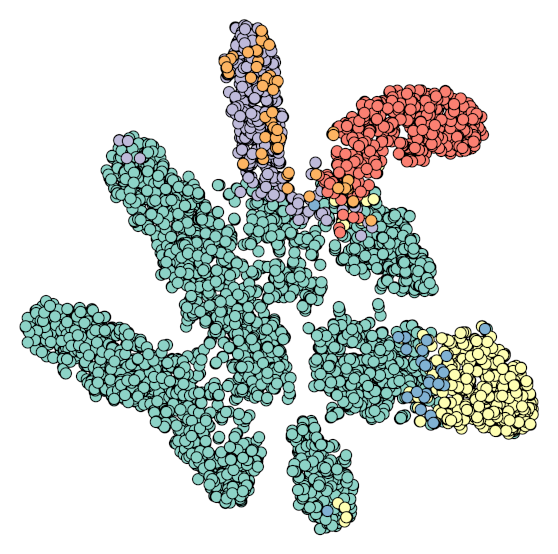}
}
\subfloat{}{
  \includegraphics[width=0.12\textwidth]{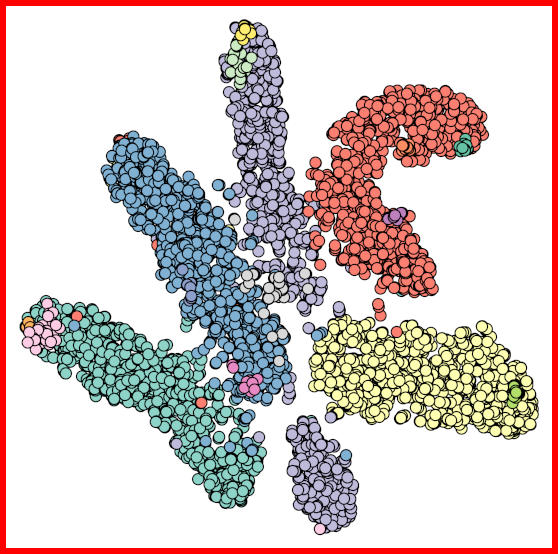}
}
\subfloat{}{
  \includegraphics[width=0.12\textwidth]{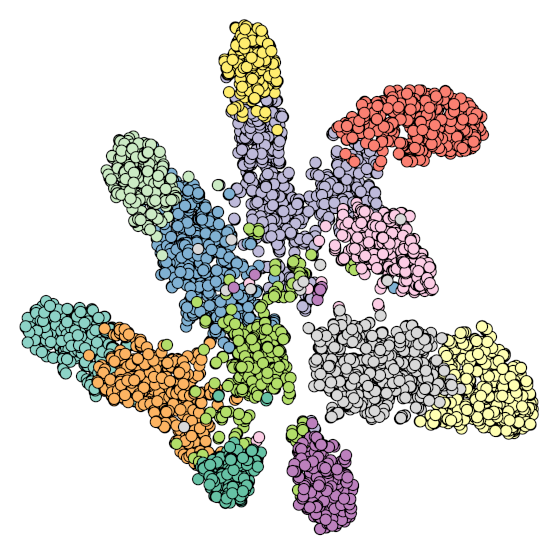}
}
\subfloat{}{
  \includegraphics[width=0.12\textwidth]{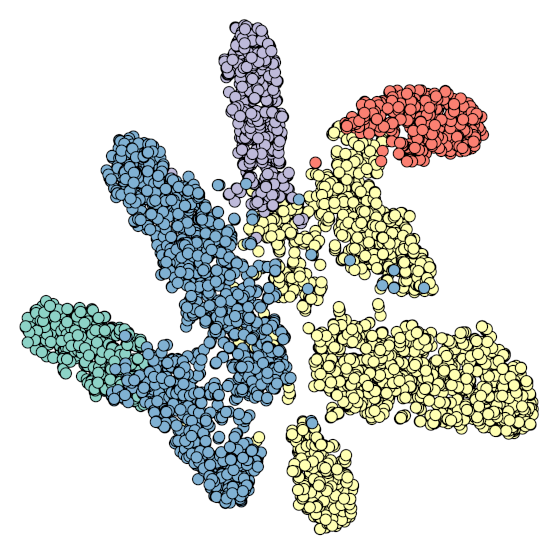}
}
\subfloat{}{
  \includegraphics[width=0.12\textwidth]{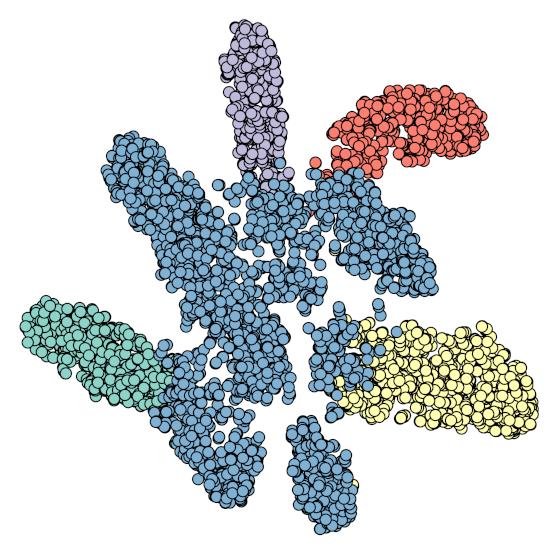}
}

\subfloat{}{
  \includegraphics[width=0.12\textwidth]{pics/clustering_7d_1.png}
}
\subfloat{}{
  \includegraphics[width=0.12\textwidth]{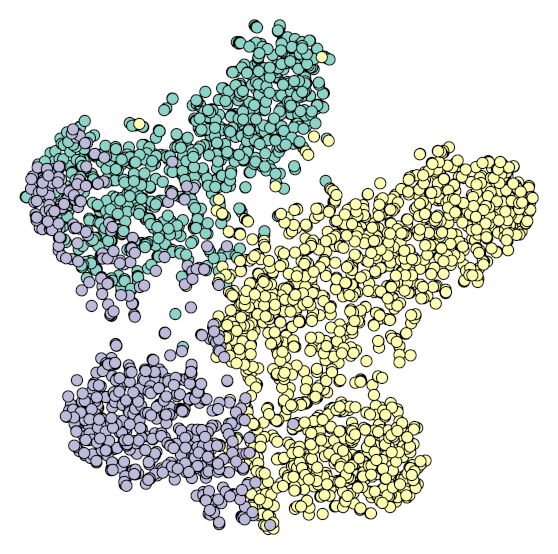}
}
\subfloat{}{
  \includegraphics[width=0.12\textwidth]{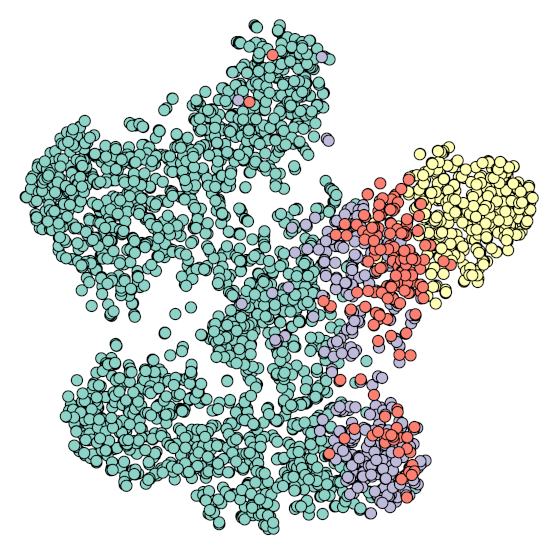}
}
\subfloat{}{
  \includegraphics[width=0.12\textwidth]{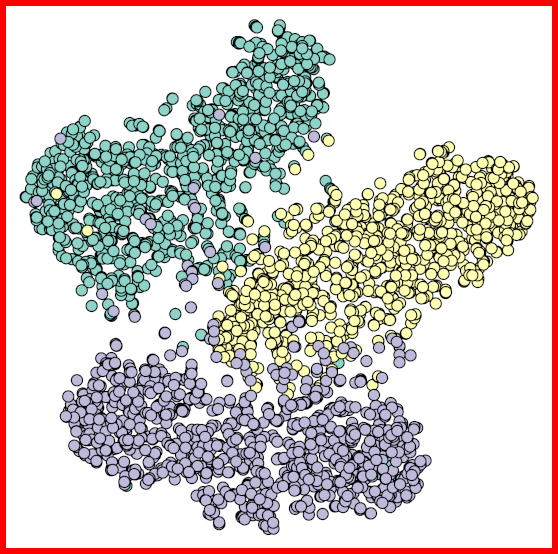}
}
\subfloat{}{
  \includegraphics[width=0.12\textwidth]{pics/clustering_tpca_7d_1.png}
}
\subfloat{}{
  \includegraphics[width=0.12\textwidth]{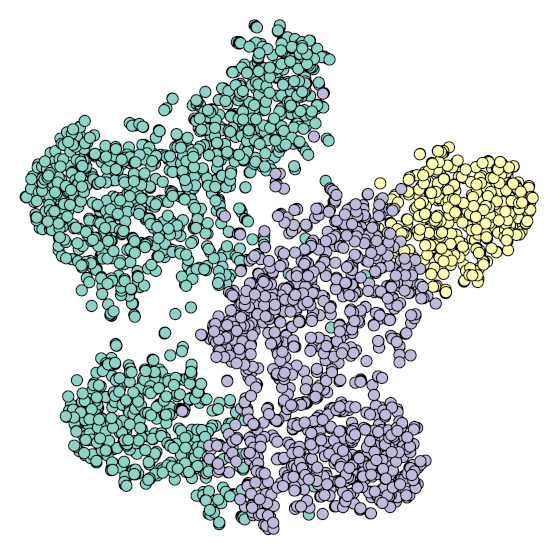}
}
\subfloat{}{
  \includegraphics[width=0.12\textwidth]{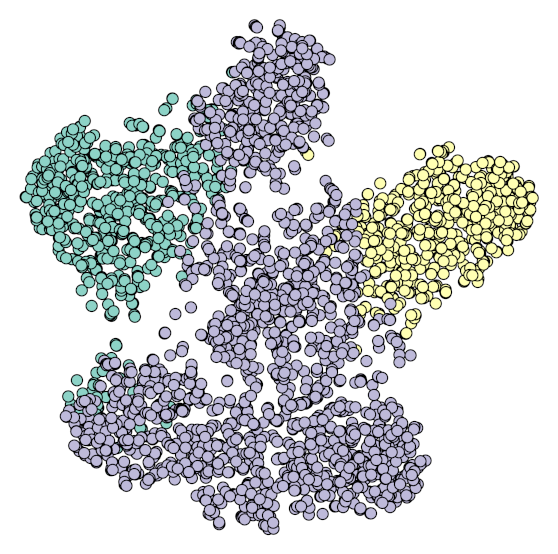}
}

\subfloat{}{
  \includegraphics[width=0.12\textwidth]{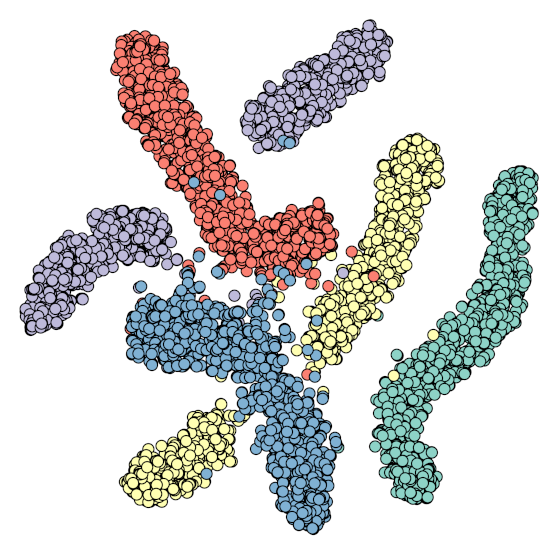}
}
\subfloat{}{
  \includegraphics[width=0.12\textwidth]{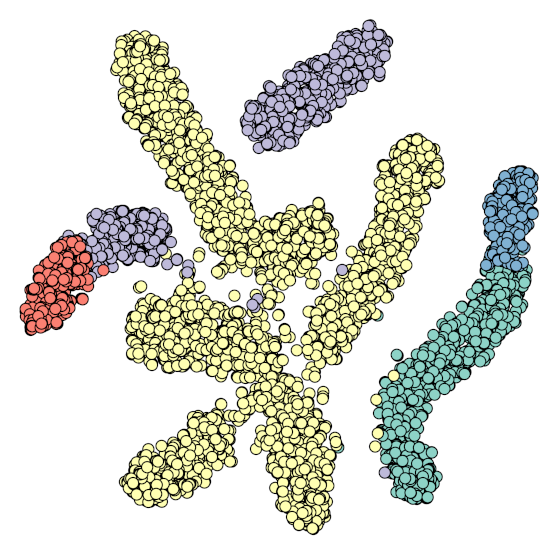}
}
\subfloat{}{
  \includegraphics[width=0.12\textwidth]{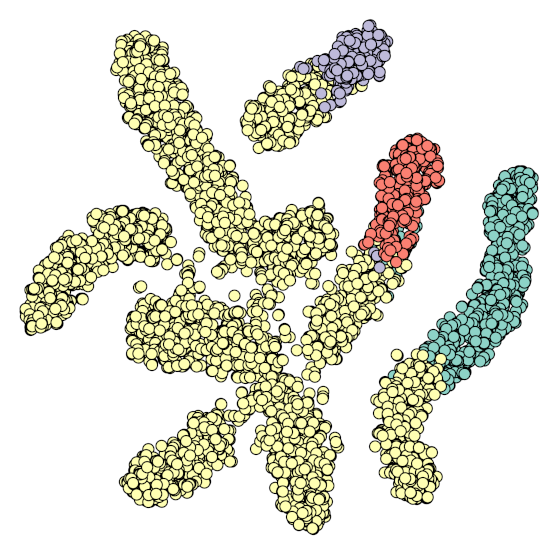}
}
\subfloat{}{
  \includegraphics[width=0.12\textwidth]{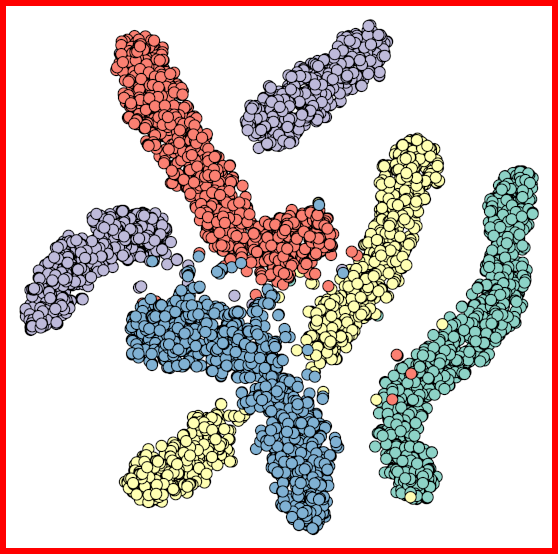}
}
\subfloat{}{
  \includegraphics[width=0.12\textwidth]{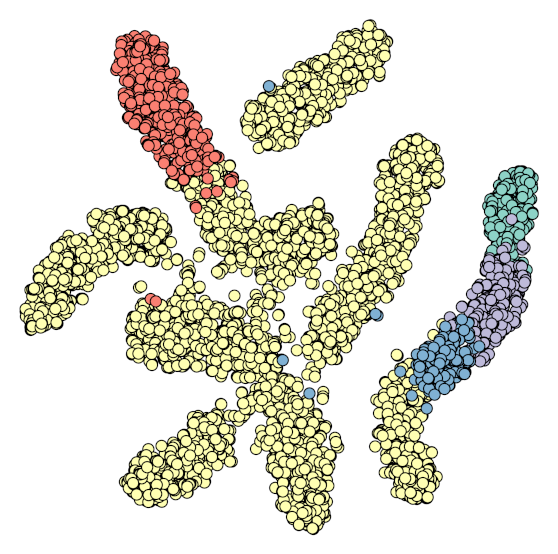}
}
\subfloat{}{
  \includegraphics[width=0.12\textwidth]{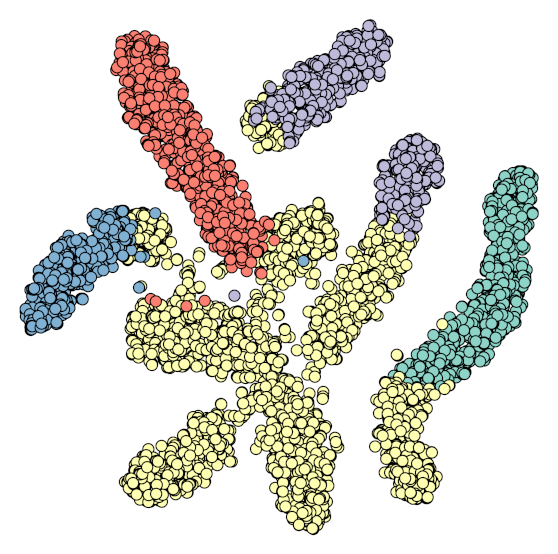}
}
\subfloat{}{
  \includegraphics[width=0.12\textwidth]{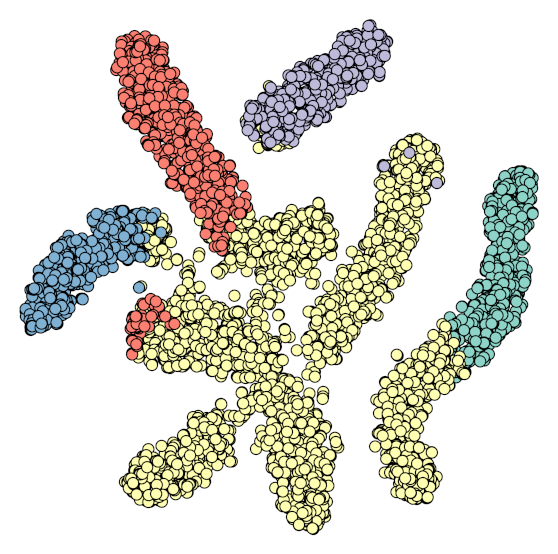}
}

\hspace{4.5em}
\raisebox{2em}{(a)} \hspace{7.5em}
\raisebox{2em}{(b)} \hspace{7.5em}
\raisebox{2em}{(c)} \hspace{7.5em}
\raisebox{2em}{(d)} \hspace{7.5em}
\raisebox{2em}{(e)} \hspace{7.5em}
\raisebox{2em}{(f)} \hspace{7.5em}
\raisebox{2em}{(g)}

\caption{\textit{An intuitive illustration of clustering performance for subexperiment $(i-l)$. Samples are shown in the angle space and visualized using t-SNE. (a) shows true labels of simulated samples generated by the different submanifold, (b) shows the clusters obtained through  DBSCAN method, (c) shows the clusters of tPCA-DBSCAN, (d) shows the clusters of PSM-DBSCAN (highlighted by the box), (e) shows the clusters of MINT-AGE, (e) shows the clusters of hierarchical clustering and (f) shows the clusters of spectral clustering. Different clusters are represented by different colors. The subexperiments are correspond to $D=6$ (rows $1$ and 2) and $D=7$ (rows $3$ and $4$).}}

\label{fig:clustering_result}
\end{minipage}
}
\end{figure}

Figure \ref{fig:clustering_result} illustrates the clustering performance of various methods. Here, we present the subexperiments $(i-l)$ with $D=6$ and $7$. The results for the remaining subexperiments are provided in Appendix C of the Supplementary document. In this simulation, we select several clustering methods to systematically evaluate the performance of PSM-DBSCAN. First, we choose DBSCAN to assess the impact of PSM for dimensionality reduction on the clustering. Next, we adopt tPCA-DBSCAN to examine the impact of using tPCA as an alternative dimensionality reduction method. We also employ MINT-AGE to contrast the clustering performance achieved via tPCA with other clustering approaches. Finally, we include two classic clustering methods: hierarchical clustering (HC) and spectral clustering (SC) to compare with these well-established methods.

Figure \ref{fig:clustering_result}(b) presents the clustering performance of DBSCAN. This indicates that DBSCAN fails to accurately identify the underlying submanifold structure in these high-dimensional settings, even though it is designed for density-based and shape-sensitive clustering. Specifically, in subexperiment $(i)$, although DBSCAN correctly clusters one of the groups (in yellow), it fails to separate the remaining two groups, as these samples are too close to each other for DBSCAN to distinguish effectively on the high-dimensional torus. Figure \ref{fig:clustering_result}(c) shows that tPCA-DBSCAN also cannot cluster accurately based on the underlying geometric structure and identifies samples from different groups into the same cluster. In contrast, as highlighted in the red box, Figure \ref{fig:clustering_result}(d) demonstrates that PSM-DBSCAN can accurately cluster samples according to the submanifold on which they reside in each subexperiment.  MINT-AGE, as shown in Figure \ref{fig:clustering_result}(e), produces fine clustering but yields an excessively large number of clusters, significantly exceeding the actual number of underlying submanifolds. Although MINT-AGE attempts to refine clustering, it still does not guarantee accurate identification of the underlying submanifold. In the subexperiment $(i)$, $(k)$ and $(l)$, MINT-AGE incorrectly groups data from distinct submanifolds into a single cluster. In Figure \ref{fig:clustering_result}(f) and (g), the two classical clustering methods have similar performances. The clustering mainly uses Euclidean distances, grouping nearby samples into the same cluster and producing several locally concentrated regions. However, it ignores the data’s underlying geometry, so it cannot reliably separate samples from different groups.

To further quantitatively compare the performance of different clustering methods, we use two classical performance metrics for clustering: ARI and NMI, which are provided in Appendix \ref{appendix: Definitions of evaluation metrics}.

Table \ref{tab:clustering_simulation} provides a quantitative comparison of clustering performance between PSM-DBSCAN and other clustering methods using ARI and NMI. Table \ref{tab:clustering_simulation} highlights that PSM-DBSCAN consistently achieves the highest ARI and NMI in all subexperiments across different dimensional scenarios, indicating that the clustering produced by PSM-DBSCAN is more closely aligned with the true labels and captures the underlying data structure more effectively. Overall, PSM-DBSCAN surpasses the second-best clustering methods by approximately from \(20\%\) to \(50\%\) in ARI and from \(10\%\) to \(40\%\) in NMI, demonstrating a significant and consistent advantage in clustering performance compared with other methods. In contrast, although standalone DBSCAN shows high ARI and NMI in the subexperiments with \(D=2\), as the dimension \(D\) increases, it achieves low ARI and NMI while PSM-DBSCAN still attains high ARI and NMI. Additionally, DBSCAN exhibits substantial variability in performance, as indicated by the large standard deviations (shown in parentheses), which suggests that this method lacks consistency and robustness. This underscores the limited ability of DBSCAN to handle high-dimensional scenarios and the necessity of dimensionality reduction via PSM. Compared to DBSCAN, tPCA-DBSCAN consistently shows lower performance in terms of ARI and NMI, suggesting that dimensionality reduction by tPCA hinders DBSCAN’s ability to capture the intrinsic structure. MINT-AGE also falls short in ARI and NMI, reflecting its limited ability to cluster accurately in challenging scenarios, which indicates that the low-dimensional representation obtained through PSM can more effectively guide clustering than that obtained via tPCA. Classical clustering methods, such as hierarchical clustering and spectral clustering, also fail to perform effectively. Hierarchical clustering consistently yields poor ARI and NMI due to its inability to handle the non-linear and high-dimensional features of the data. Similarly, spectral clustering struggles to identify the underlying submanifold structures and to cluster accurately. Compared with these widely used clustering methods, PSM-DBSCAN demonstrates the advantages of clustering on manifolds.

In summary, the simulations suggest that PSM-DBSCAN consistently exhibits improved performance relative to competing methods, demonstrating superior accuracy and robustness in clustering tasks on the torus. By effectively addressing the challenges of high dimensionality and noise through its integration with PSM, PSM-DBSCAN emerges as a powerful and reliable tool for clustering data on manifolds. The observed differences between PSM-DBSCAN and other methods underscore the critical importance of leveraging PSM to enhance clustering accuracy, stability, and adaptability.

\begin{table}[H]
\centering
\caption{ Comparison of clustering performance between DBSCAN, tPCA-DBSCAN, PSM-DBSCAN, MINT-AGE, hierarchical clustering and spectral clustering. Standard deviations computed based on 10 repetitions of simulations are shown in parentheses and scaled by $10^{-2}$. Bold font indicates the best performance under each experimental setting.}

\begin{tabular}{cclllllll}
\hline
\textbf{} & & \multicolumn{2}{c}{DBSCAN} & \multicolumn{2}{c}{tPCA-DBSCAN} & \multicolumn{2}{c}{PSM-DBSCAN} \\ \cline{1-8} 
$D$&Exp2 & \phantom{X}ARI($\uparrow$)    & \phantom{X}NMI($\uparrow$)    & \phantom{X}ARI($\uparrow$)  & \phantom{X}NMI($\uparrow$)  & \phantom{X}ARI($\uparrow$)  & \phantom{X}NMI($\uparrow$) \\ \hline

{2}                 & $a$ & .88 (10.98) & .87 (5.87) & .56 (0.20) & .73 (0.42) & \textbf{.91} (4.56) & \textbf{.89} (3.53) \\
                    & $b$ & .79 (2.50) & .79 (2.65) & .56 (0.37) & .70 (1.56) & \textbf{.95} (4.95) & \textbf{.95} (3.86) \\ 
{3}                 & $c$ & .54 (0.02) & .68 (0.03) & .80 (0.02) & .79 (0.07) & \textbf{.93} (0.01) & \textbf{.90} (0.01) \\
                    & $d$ & .68 (17.09) & .77 (10.21) & .54 (2.85) & .66 (7.00) & \textbf{.97} (0.54) & \textbf{.95} (0.74) \\ 
{4}                 & $e$ & .56 (0.25) & .70 (0.41) & .52 (3.93) & .60 (3.96) & \textbf{.95} (0.63) & \textbf{.92} (0.77) \\
                    & $f$ & .71 (4.49) & .77 (3.19) & .34 (10.50) & .49 (10.36) & \textbf{.84} (10.08) & \textbf{.90} (4.79) \\ 
{5}                 & $g$ & .63 (16.34) & .74 (9.51) & .45 (4.36) & .55 (2.33) & \textbf{.96} (0.69) & \textbf{.93} (0.94) \\
                    & $h$ & .50 (6.66) & .62 (3.90) & .22 (9.71) & .33 (8.81) & \textbf{.82} (7.51) & \textbf{.82} (3.62) \\ 
{6}                 & $i$ & .39 (10.87) & .61 (7.83) & .05 (2.95) & .21 (6.21) & \textbf{.85} (4.86) & \textbf{.82} (3.43) \\
                     & $j$ & .50 (7.74) & .65 (12.98) & .37 (8.51) & .44 (4.42) & \textbf{.74} (1.22) & \textbf{.84} (0.64) \\ 
{7}                 & $k$ & .43 (14.61) & .52 (11.30) & .24 (8.39) & .34 (11.60) & \textbf{.81} (2.05) & \textbf{.78} (0.56) \\
                    & $l$ & .22 (9.23) & .31 (12.30) & .09 (0.58) & .26 (2.39) & \textbf{.94} (3.01) & \textbf{.93} (1.75) \\ \hline
\end{tabular}

\begin{tabular}{cclllllll}
\textbf{} & & \multicolumn{2}{c}{MINT-AGE} & \multicolumn{2}{c}{HC} & \multicolumn{2}{c}{SC} \\ \cline{1-8} 
$D$&Exp2 & \phantom{X}ARI($\uparrow$)    & \phantom{X}NMI($\uparrow$)    & \phantom{X}ARI($\uparrow$)  & \phantom{X}NMI($\uparrow$)  & \phantom{X}ARI($\uparrow$)  & \phantom{X}NMI($\uparrow$) \\ \hline

{2}                 & $a$ & .39 (2.74) & .57 (1.61) & .38 (7.59) & .39 (8.46) & .24 (0.50) & .31 (0.39) \\
                    & $b$ & .46 (8.04)  & .64 (6.95)  & .08 (5.67) & .12 (8.52) & .15 (3.54) & .19 (2.74) \\ 
{3}                 & $c$ & .37 (4.17)  & .55 (3.12)  & .67 (4.56) & .71 (5.33) & .77 (9.16) & .77 (7.37) \\
                    & $d$ & .33 (6.14) & .56 (2.39) & .19 (8.41) & .25 (6.42) & .38 (10.19) & .40 (8.23) \\ 
{4}                 & $e$ & .48 (3.50) & .61 (1.54) & .56 (6.77) & .67 (7.56) & .60 (11.82) & .67 (10.55) \\
                    & $f$ & .43 (6.51) & .54 (3.90) & .43 (12.90) & .45 (9.86) & .18 (2.23) & .26 (2.35) \\ 
{5}                 & $g$ & .52 (2.68) & .62 (3.16) & .42 (7.24) & .54 (9.16) & .61 (2.36) & .66 (2.74) \\
                    & $h$ & .37 (3.66) & .58 (2.32) & .24 (3.18) & .44 (1.47) & .29 (9.34) & .51 (7.45) \\ 
{6}                 & $i$ & .39 (4.38) & .57 (4.57) & .30 (5.85) & .46 (7.74) & .28 (0.91) & .46 (1.38) \\
                     & $j$ & .35 (5.21) & .58 (4.40) & .39 (9.39) & .58 (7.76) & .32 (0.74) & .57 (0.94) \\ 
{7}                 & $k$ & .29 (1.37) & .48 (1.08) & .35 (5.85) & .44 (7.74) & .41 (0.91) & .27 (1.38) \\
                    & $l$ & .52 (4.92) & .70 (2.50) & .32 (9.39) & .53 (7.76) & .27 (0.74) & .52 (0.94) \\ \hline
\end{tabular}

\label{tab:clustering_simulation}
\end{table}

\section{Application for RNA data} \label{Application for RNA data}
In this section, we apply PSM-DBSCAN for RNA correction and compare its performance with that of the MINT-AGE-CLEAN method. We analyze classical RNA datasets from the Protein Data Bank \citep{bank1971protein}, a widely used bioinformatics database that primarily stores 3-dimensional structural data of biomolecules. These RNA datasets consist of 8,665 suites obtained from high-precision experimental X-ray measurements. Our analysis aims to demonstrate the efficacy of PSM-DBSCAN-MC in enhancing the accuracy and reliability of RNA clustering and clash correction, potentially offering substantial improvements over the MINT-AGE-CLEAN method.

\subsection{Clustering for RNA suites}

\begin{figure}[htbp]
    \centering
    \subfloat[]{\includegraphics[width=0.45\textwidth]{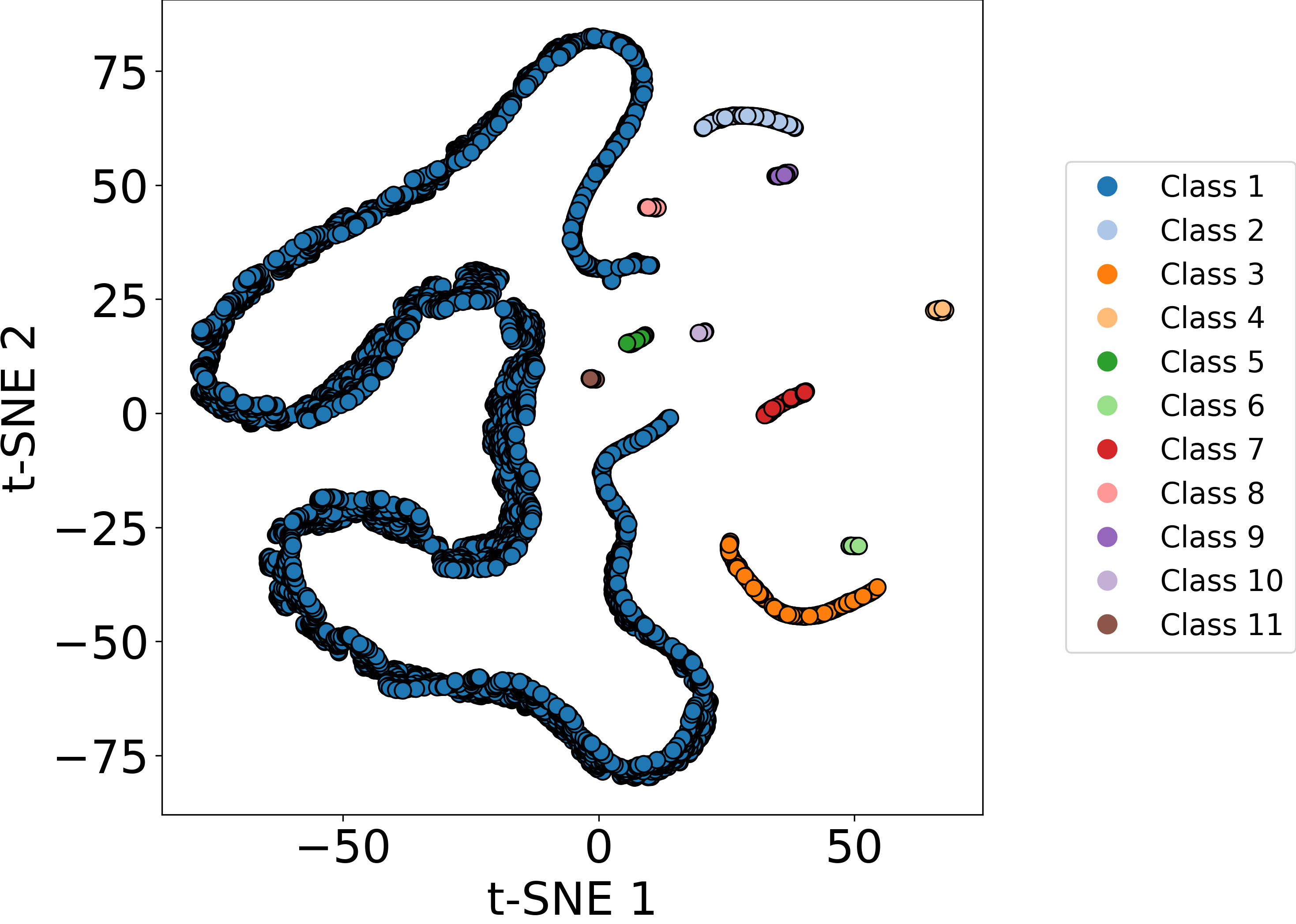}}
    \hfill
    \subfloat[]{\includegraphics[width=0.45\textwidth]{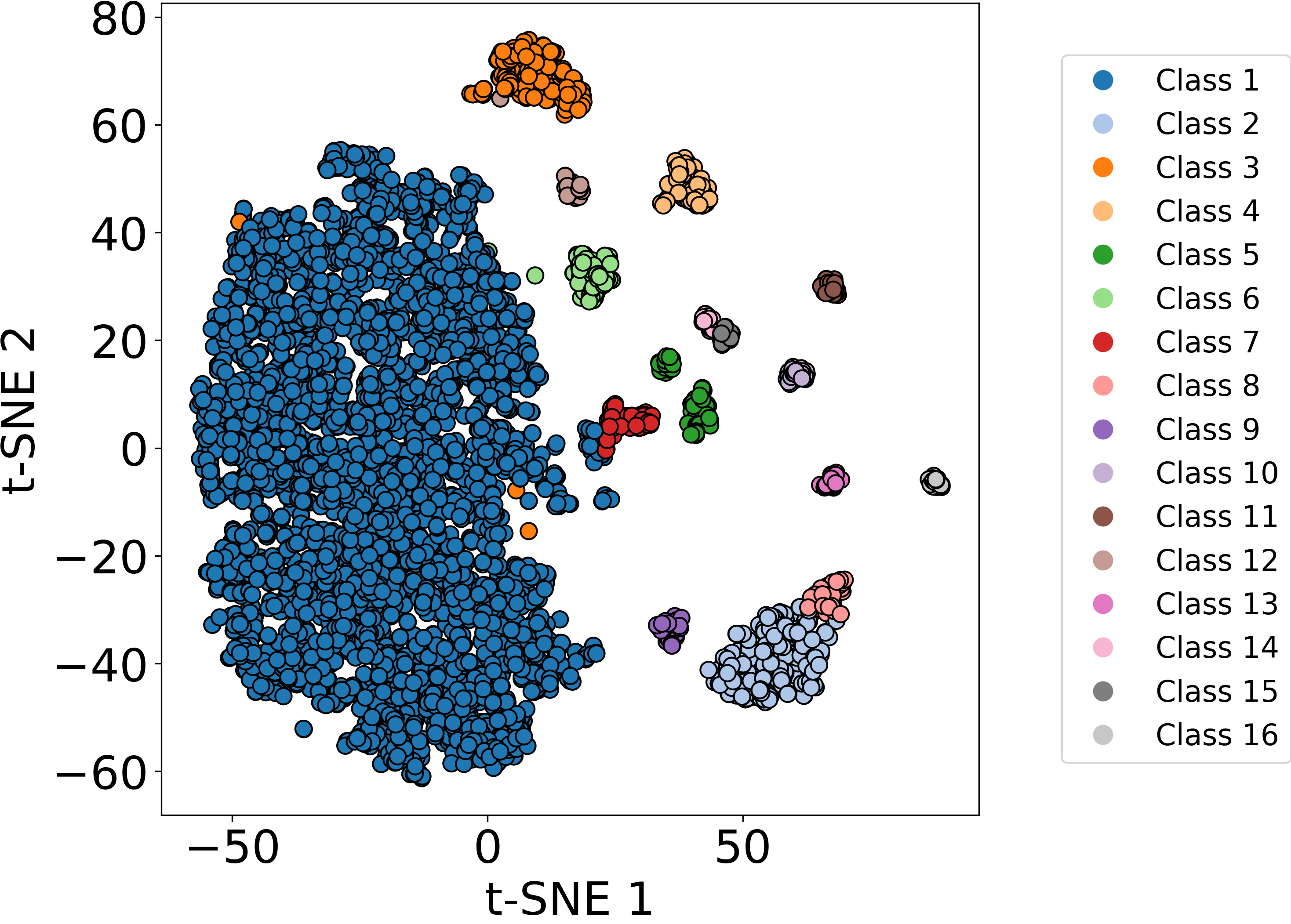}}
    \caption{\textit{The visualization of the clusters by t-SNE. (a) shows the clusters of PSM-DBSCAN, and (b) shows the clusters of MINT-AGE.}}
    \label{fig:RNA clustering}
\end{figure}

To evaluate the effectiveness of PSM-DBSCAN in clustering RNA suites, we conduct a comparative analysis with MINT-AGE. Figure \ref{fig:RNA clustering} illustrates the clustering performance achieved by these methods. Both PSM-DBSCAN and MINT-AGE successfully identify the main RNA suite clusters, such as Class 1. However, the results of PSM-DBSCAN exhibit clearer and more coherent cluster boundaries, whereas MINT-AGE tends to produce clusters with less distinct boundaries and non-contiguous structures. In addition, PSM-DBSCAN automatically detects outliers in a data-driven manner during the clustering process, without requiring a predefined outlier proportion as in MINT-AGE, which leads to more effective and robust outlier identification. Importantly, PSM-DBSCAN also reveals finer substructures within RNA suite clusters. For example, while MINT-AGE broadly groups RNA suites into Class 3, PSM-DBSCAN subdivides this set into two coherent subclusters, Class 2 and Class 9. This subdivision reflects the method’s ability to capture subtle geometric distinctions among RNA suites, which refers to the clearer boundaries, reduced noise, and ability to uncover meaningful substructures, rather than the absolute number of clusters. These properties highlight the enhanced capability of PSM-DBSCAN to exploit latent geometric characteristics, thereby achieving superior clustering performance and offering new perspectives on RNA suite organization.

Since the true labels of the RNA suites are unknown in practice, we introduce two metrics for clustering to better compare the performance of the two clustering methods on RNA suites: the silhouette score (SI) and the Davies-Bouldin index (DB), both of which are widely used to assess clustering performance. The definition of SI and DB are provided in Appendix \ref{appendix: Definitions of evaluation metrics}. The SI index measures cluster separation and cohesion, with higher values indicating better and more compact clusters. In contrast, the DB index assesses cluster similarity, with lower values indicating more distinct and well-separated clusters. Higher SI and lower DB values indicate that a clustering method performs well in terms of internal consistency and inter-cluster discrimination, thereby more accurately reflecting the intrinsic structure of the data.

Table \ref{tab:cluster} provides a quantitative comparison of the performance of PSM-DBSCAN and MINT-AGE using two key clustering evaluation metrics: SI and DB. Table \ref{tab:cluster} highlights the superior performance of PSM-DBSCAN in both metrics. PSM-DBSCAN achieves a higher SI, outperforming MINT-AGE by approximately 10\%, indicating that it forms clusters that are both well-separated and cohesive. In addition, PSM-DBSCAN attains a lower DB, reflecting more compact and distinct clusters than MINT-AGE. These results demonstrate that PSM-DBSCAN effectively groups RNA suites into intrinsic and meaningful clusters, which is critical for RNA correction. 

In summary, PSM-DBSCAN provides a more effective tool for clustering RNA data compared to MINT-AGE. It consistently outperforms MINT-AGE in both SI and DB metrics, demonstrating a superior ability to delineate and compactly represent RNA structures. These differences underscore the advantages of PSM-DBSCAN in uncovering representative and geometrically accurate clustering, thereby laying a strong foundation for subsequent RNA structure correction and analysis. By preserving the geometric properties of the data in the low-dimensional representation, effectively dealing with outliers, and forming well-defined clusters, PSM-DBSCAN establishes itself as a robust and reliable tool for RNA analysis tasks. This capability not only facilitates a deeper understanding of the relationships among RNA suites but also paves the way for accurate, data-driven corrections of structural clashes in RNA suites.

\begin{table}[H]
\centering
\caption{\textit{Comparison of clustering results between PSM-DBSCAN and MINT-AGE for RNA data, using SI and DB as metrics to measure clustering performance. Bold font indicates the best performance under each experimental setting.}}
\label{tab:cluster}
\begin{tabular}{lcccccc}
\hline
\textbf{} & {PSM-DBSCAN} & {MINT-AGE} \\ \hline
        
$\SI$($\uparrow$)     & \textbf{0.65}         & 0.59            \\
$\DB$($\downarrow$)     & \textbf{0.51}          & 0.82       \\ \hline
\end{tabular}
\end{table}

\subsection{Multiscale correction for RNA clashes}
This subsection introduces a multiscale method for RNA correction that bridges the microscopic and mesoscopic scales. The PSM-DBSCAN-MC method leverages the geometric similarities between these scales to resolve clashes in RNA suites. By incorporating the PSM-DBSCAN clustering framework, PSM-DBSCAN-MC provides a robust and comprehensive approach to RNA correction.

\begin{figure}[H]
\centering
\subfloat[]{
  \includegraphics[width=0.3\textwidth]{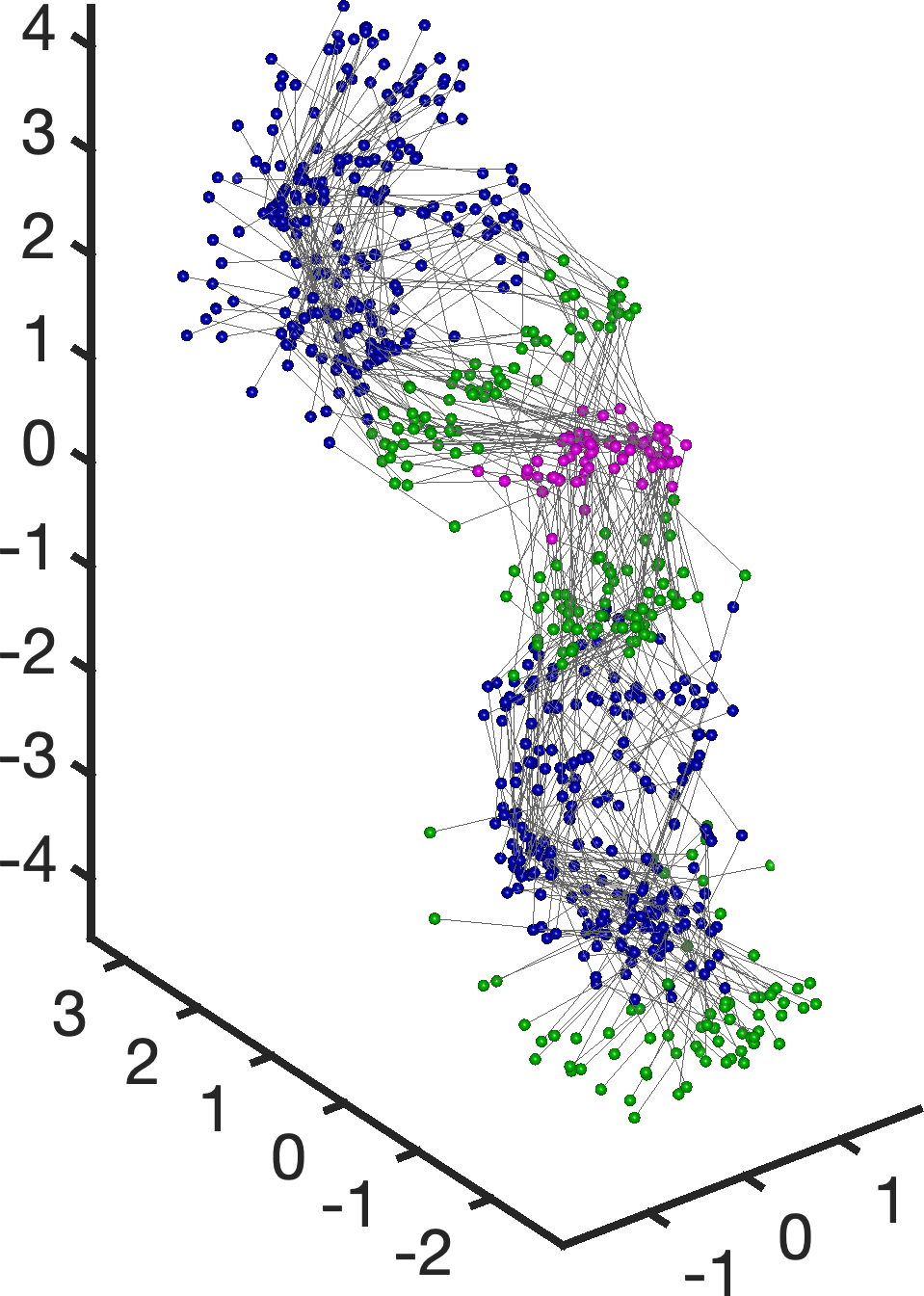}
}
\hfill
\subfloat[]{
  \includegraphics[width=0.3\textwidth]{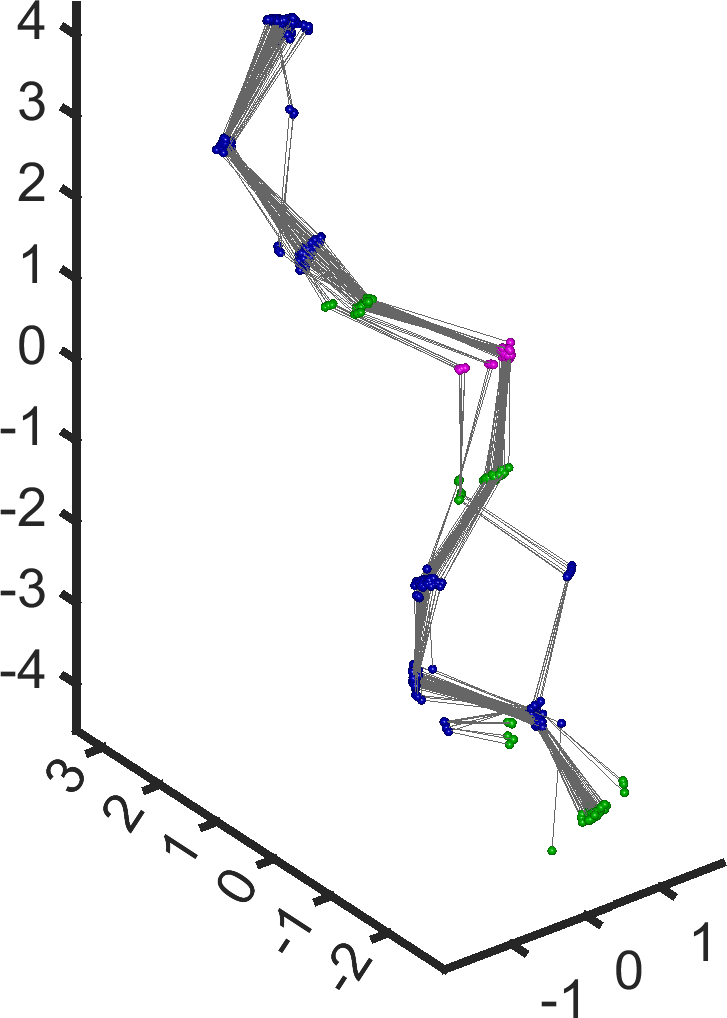}
}
\hfill
\subfloat[]{
  \includegraphics[width=0.3\textwidth]{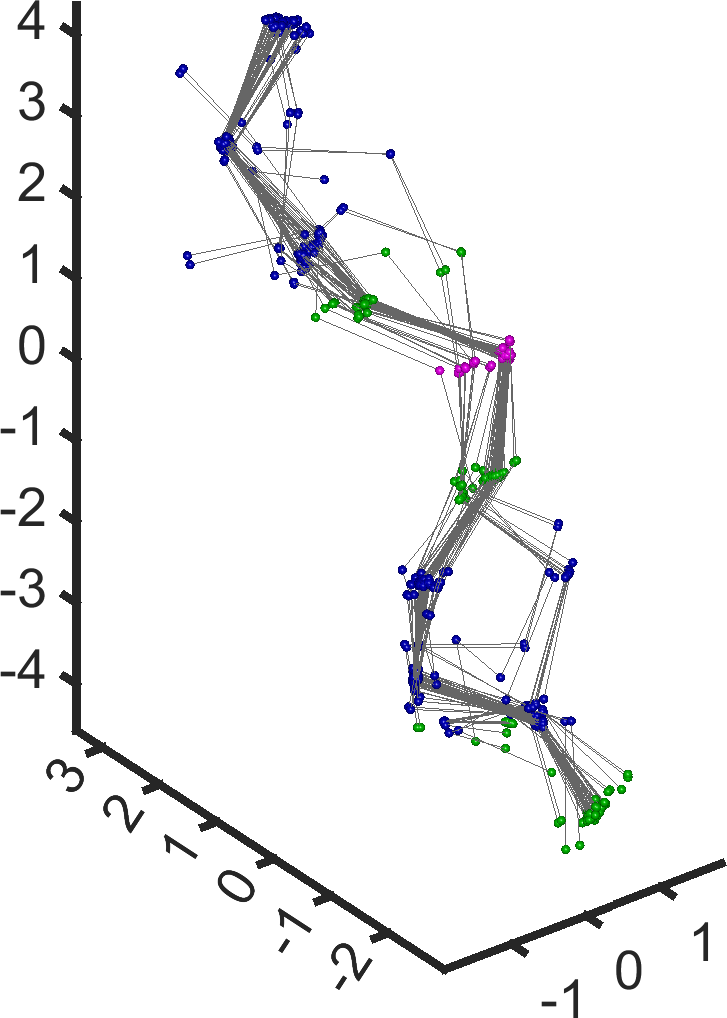}
}

\caption{\textit{Illustration of RNA clash suites and corresponding correction. (a) is the 73 clash suites, (b) is the clash correction by PSM-DBSCAN-MC and (c) is the clash correction by CLEAN-MINT-AGE.}}

\label{fig:Method comparison 1}
\end{figure}

\begin{figure}[H]
    \centering
    \includegraphics[width=0.8\textwidth]{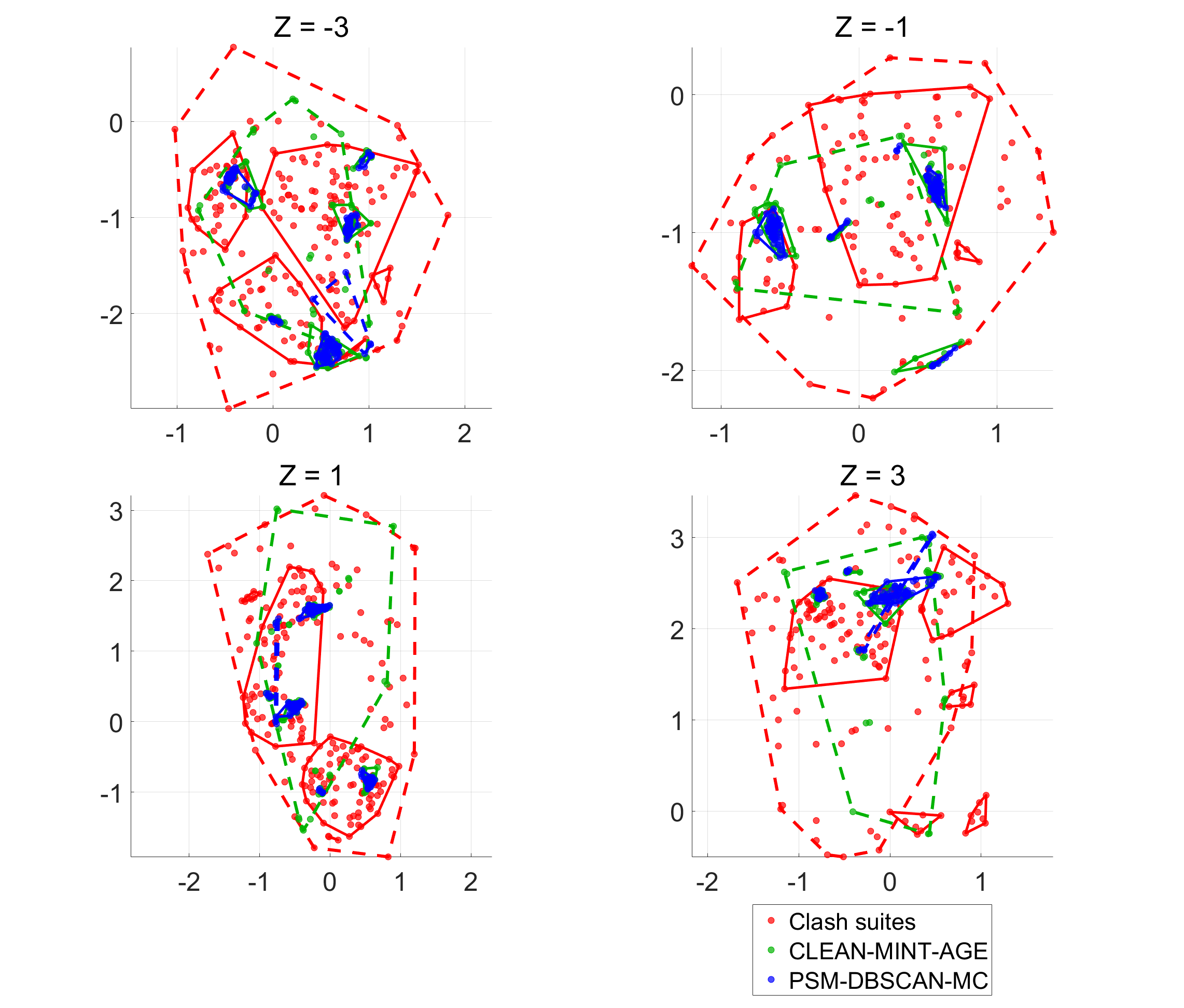}
    \caption{\textit{Illustration of the $2$-dimensional projection slides of RNA’s $3$-dimensional atomic structure. The structure is equally segmented into four intervals along the $Z$-axis, and atoms within a $Z\pm1$ range are projected onto the plane. Solid lines delineate the convex hulls of atoms in densely populated regions, while dashed lines outline the convex hulls of dispersed atoms.}}
    
    \label{fig:slides}
\end{figure}

\begin{figure}[H]
    \centering
    \includegraphics[width=1\textwidth]{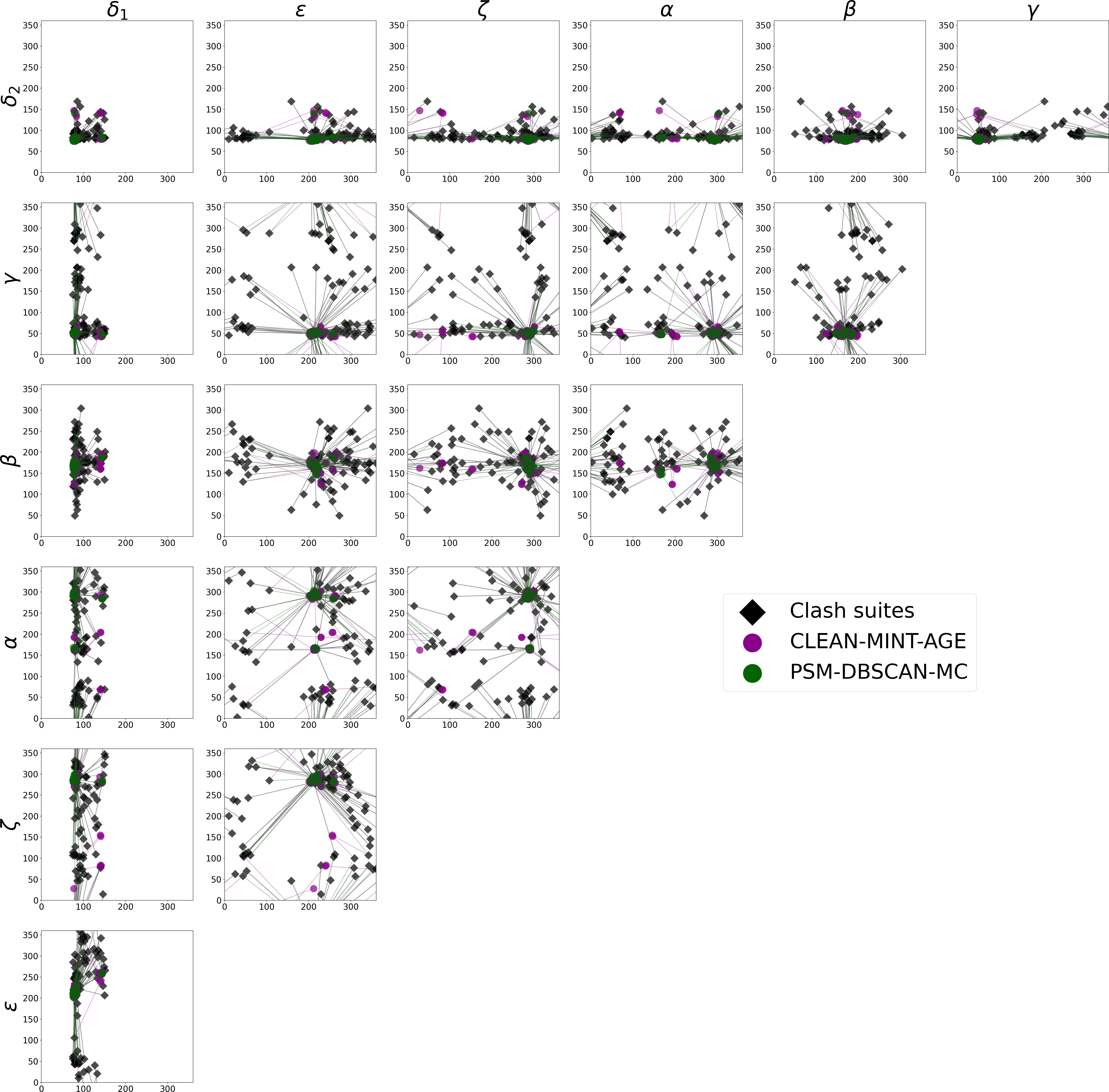}
    \caption{\textit{A comparative analysis through scatterplots showcasing all $2$-dimensional dihedral angle pairs of RNA structures. Original RNA clashes indicates the starting point prior to RNA correction.}}
    \label{fig:Method comparison 2}
\end{figure}

Figure \ref{fig:Method comparison 1} illustrates the structures of 73 original RNA clash suites along with their corresponding corrected structures, generated using PSM-DBSCAN-MC and CLEAN-MINT-AGE, respectively. This comparison highlights the different ways in which the two methods resolve clashes at the atomic level. Compared to the RNA clash suites corrected by CLEAN-MINT-AGE, PSM-DBSCAN-MC produces three-dimensional structures with reduced noise and more coherent geometric patterns, offering a new computational perspective on RNA correction. Figure \ref{fig:slides} presents four two-dimensional projection slides, obtained via convex hull analysis, that illustrate the positional distribution of atoms in the three-dimensional structures (see Figure \ref{fig:Method comparison 1}) for different methods. It is evident that the distribution of atoms in RNA clash suites is highly chaotic and spans a wide range. After applying the two correction methods, the atom distributions become more concentrated into distinct regions. The correction result of CLEAN-MINT-AGE shows that the atoms remain dispersed over a relatively wide region, whereas the correction by PSM-DBSCAN-MC yields several more compact and well-defined regions of atoms. A full movie of projection slides at different $Z$ can be found at \href{https://github.com/zhigang-yao/RNA-Clash-Correction/blob/main/RNA\_correction\_slides.gif}{https://github.com/zhigang-yao/RNA-Clash-Correction/blob/main/RNA\_correction\_slides.gif}. Meanwhile, Figure \ref{fig:Method comparison 2} provides a comparative visualization of changes in dihedral angle distributions resulting from different RNA correction methods. In the corrected two-dimensional dihedral angle distributions, PSM-DBSCAN-MC produces a more concentrated pattern with fewer deviations. In contrast, although CLEAN-MINT-AGE reduces clashes, the resulting RNA suites still exhibit broader and more sporadic dihedral angle distributions, suggesting that certain structural noise remains unresolved. In summary, while biological validation remains an open question, PSM-DBSCAN-MC demonstrates consistent improvements in simulations and provides a new and quantitatively supported perspective on RNA clash correction, offering clearer structural patterns in real data.

\section{Discussion} \label{Disccusion}

In this paper, we explore the problem of fitting effective and accurate low-dimensional representations of data on the high-dimensional torus and, based on these representations, address the challenges encountered in high-dimensional clustering on the torus. We propose PSM-DBSCAN, a robust framework that integrates PSM-based dimensionality reduction with DBSCAN. PSM accurately fits low-dimensional data on the torus while preserving its geometric features, thereby overcoming the shortcomings of existing methods such as tPCA. By producing accurate low-dimensional representations, PSM-DBSCAN significantly improves clustering performance on the high-dimensional torus.

In simulation experiments, we demonstrate the effectiveness of the proposed PSM framework, highlighting its superior capability in capturing intrinsic structures and achieving accurate clustering results on toroidal manifold data. By applying this novel clustering method to RNA suites data, we develop PSM-DBSCAN-MC for multiscale RNA correction. This work bridges the gaps in current RNA structural analysis methodologies by addressing the challenges of dimensionality reduction and clustering in high-dimensional RNA data on the torus. The proposed PSM-DBSCAN-MC not only enhances RNA correction but also provides a versatile framework for broader applications in RNA structure analysis. As our understanding of RNA continues to evolve, the principal submanifold-based analysis framework will be instrumental in advancing both theoretical insights and practical applications in RNA structure.

The current study demonstrates the effectiveness of the proposed PSM framework for toroidal data. Future research could focus on extending the applicability of the PSM framework to data on other types of manifold. In addition, improving the computational efficiency of the PSM framework by implementing parallel computing and leveraging GPU acceleration represents another promising direction for future work.

\begin{acks}
The authors acknowledge support from the Singapore Ministry of Education Tier 2 grant A-8001562-00-00 and the Tier 1 grants A-8002931-00-00 and A-8004146-00-00 at the National University of Singapore (corresponding author: Zhigang Yao, \href{mailto:zhigang.yao@nus.edu.sg}{zhigang.yao@nus.edu.sg}). The authors thank Jiaji Su for helpful discussions on algorithmic implementation.
\end{acks}

\bibliographystyle{imsart-nameyear}
\bibliography{ref} 

\appendix
\appendixpage 
\addappheadtotoc 

\section{RNA dihedral angles}
\label{appendix: RNA dihedral angles}
\begin{figure}[htbp]
    \centering
    \includegraphics[width=0.6\textwidth]{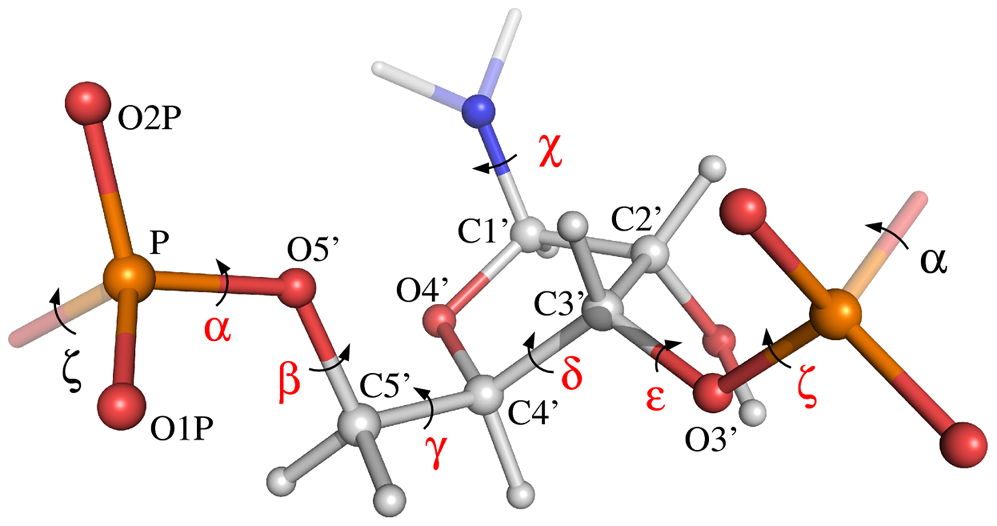}
    \caption{\textit{Structural components at a microscopic scale: $(P, O, C)$ (atoms) and bonds in a segment of the RNA backbone, with $(\alpha, \beta, \gamma, \delta, \epsilon, \zeta, \chi)$ (dihedral angles) marked to indicate the rotational freedom around each bond. Reproduced from \cite{frellsen2009probabilistic}.}}
    \label{fig: RNA backbone}
\end{figure}

In the RNA backbone, repeating sugar-phosphate units form the structural framework, with the nitrogenous bases extending from the sugar. The backbone consists of alternating phosphate groups and pentose sugars, which are connected by phosphodiester bonds. As shown in Figure \ref{fig: RNA backbone}, the angle \(\alpha\) is formed by the phosphorus atom (\(P\)), the oxygen atom in the phosphate group (\(O5^\prime\)), and the carbon atom in the sugar ring (\(C5^\prime\)). This angle influences the orientation of the phosphate group relative to the sugar, thereby affecting the flexibility and overall geometry of the RNA chain. Angles \(\beta\) and \(\gamma\) involve the carbon atoms of the sugar ring (\(C5^\prime, C4^\prime, C3^\prime\)) and are essential for determining the sugar pucker configuration, which dictates whether the sugar adopts an ‘endo’ or ‘exo’ conformation and directly influences RNA secondary structure formation. Angles \(\delta\), \(\epsilon\), and \(\zeta\) define the connections involving the oxygen atoms attached to the phosphate groups (\(O3^\prime, O5^\prime\)); these angles play a critical role in the backbone conformation between nucleotides and are vital for the overall topology and folding of the RNA molecule. Finally, angle \(\chi\) involves the nitrogenous base attached to the sugar (\(C1^\prime\)) and is essential for base stacking interactions, which are important for the stability of RNA secondary and tertiary structures.

\section{{\lowercase{t}}PCA}
\label{appendix: tPCA}

Torus PCA (tPCA) \citep{eltzner2018torus, mardia2022principal} extends principal component analysis to data on the torus, where Euclidean and geodesic PCA fail due to periodicity. The method proceeds in four steps.

\begin{figure}[H]
\centering
\subfloat[]{
  \includegraphics[width=0.33\textwidth]{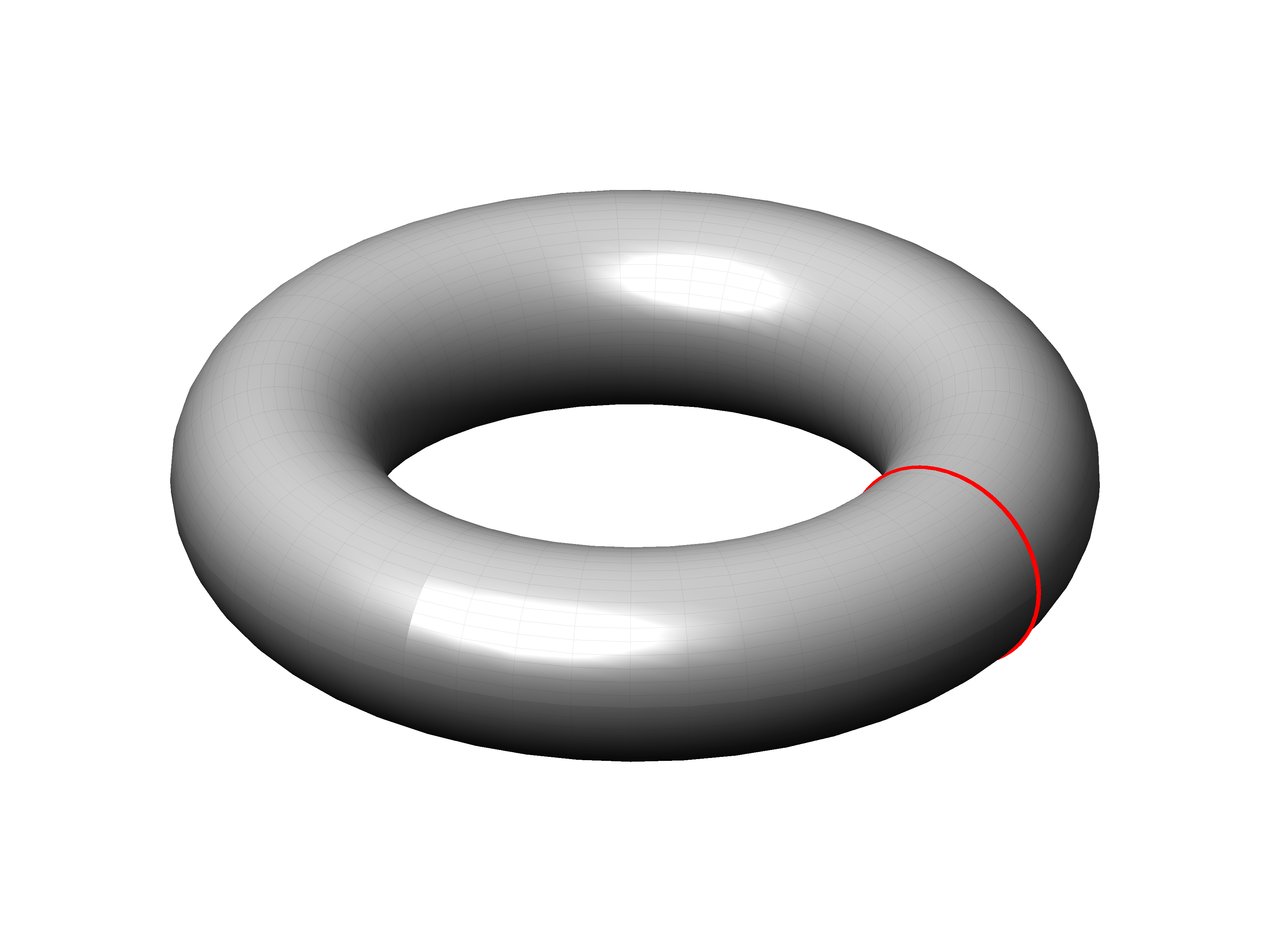}
}
\hfill
\subfloat[]{
  \includegraphics[width=0.3\textwidth]{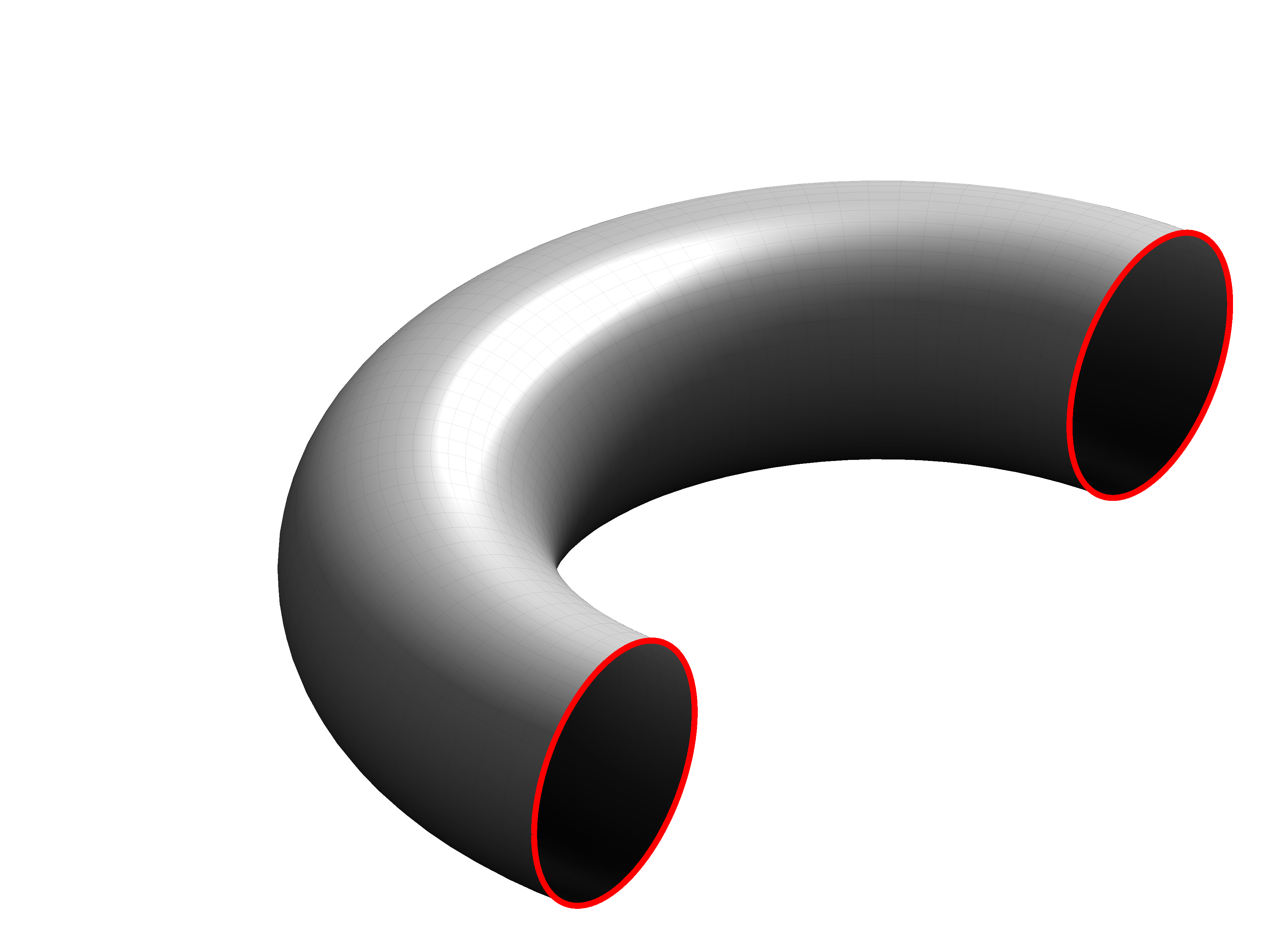}
}
\hfill
\subfloat[]{
  \includegraphics[width=0.3\textwidth]{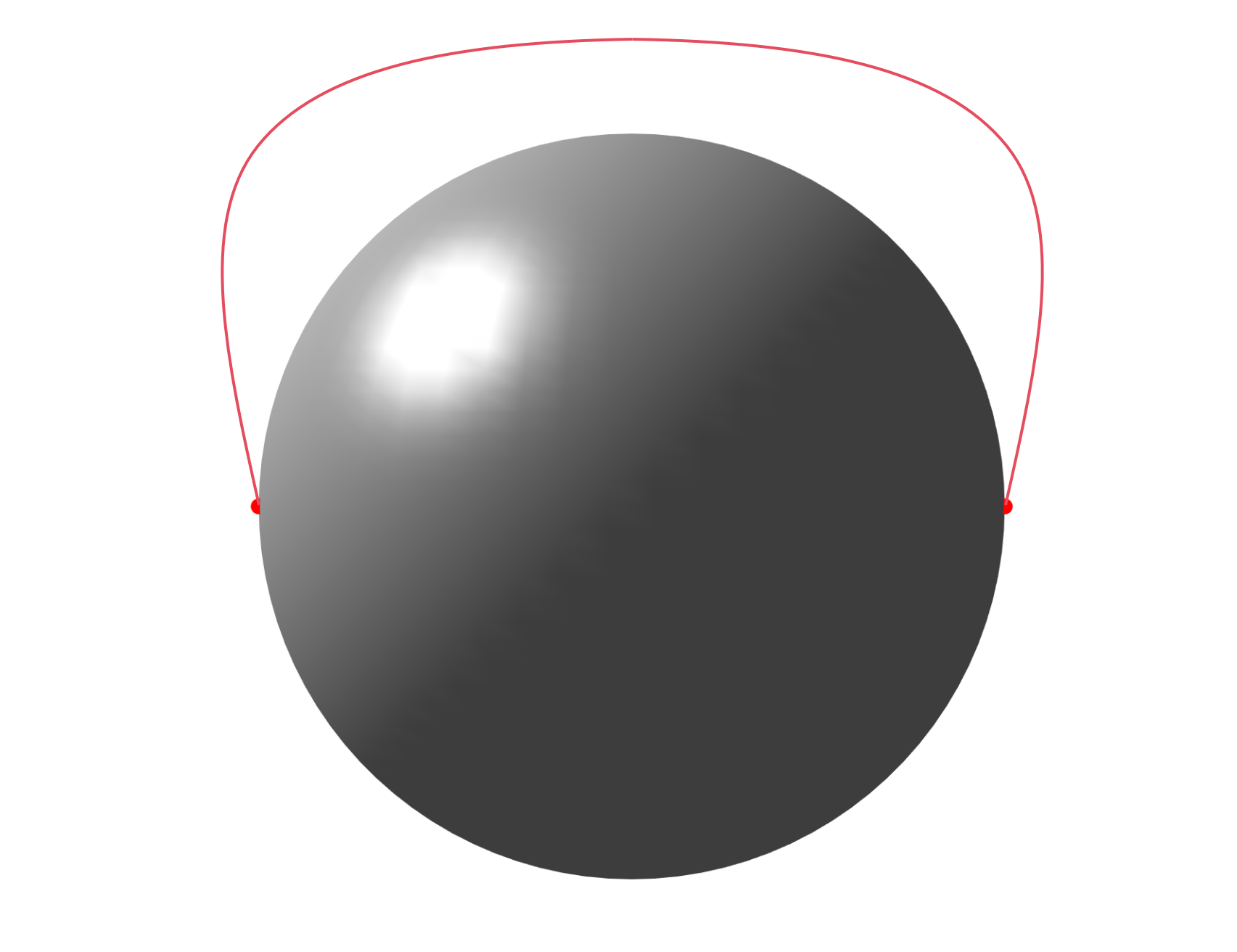}
}
\caption{\textit{Illustration of the TOSS mapping on a 2-dimensional torus. (a) The torus in $\mathbb{R}^3$. (b) Cutting along the red circles produces a cylinder. (c) Collapsing the cylinder ends identifies opposite poles, yielding a stratified sphere. This forms the first step of tPCA, followed by data-driven angle choice, PNS, and statistical testing.}}
\label{fig: toruspca}
\end{figure}

\begin{enumerate}
    \item \textbf{TOSS mapping.}  
    Given a torus $\mathbb{T}^D$, tPCA uses the torus-to-stratified-sphere (TOSS) mapping $\Phi$, as shown in Figure \ref{fig: toruspca}, which maps samples $x=(x^{(1)},\dots,x^{(D)})^T \in \mathbb{T}^D$ onto the unit sphere $\mathbb{S}^D:=\{z \in \mathbb{R}^{D+1}:\|z\|_2=1\}$. The transformed coordinates $y=(y^{(1)},\dots,y^{(D+1)})^T$ are defined as:
    \begin{equation*}
    y^{(j)} =
    \begin{cases}
    \ \cos\left(\frac{x^{(1)}}{2}\right), & j = 1, \\
    \left(\displaystyle\prod_{k=1}^{j-1}\sin\frac{x^{(k)}}{2}\right)\cos \frac{x^{(j)}}{2}, & j = 2, \ldots, D-1, \\
    \left(\displaystyle\prod_{k=1}^{D-1}\sin\frac{x^{(k)}}{2}\right)\cos x^{(D)}, & j = D,
    \\
    \left(\displaystyle\prod_{k=1}^{D-1}\sin\frac{x^{(k)}}{2}\right)\sin x^{(D)}, & j = D+1.
    \end{cases}
    \end{equation*}

    This mapping preserves torus periodicity but produces a stratified sphere, where certain poles are identified as singular strata.
    
    \item \textbf{Data-driven torus angles.}  
    To keep singularities away from observed data, torus coordinates are re-centered and reordered. Options include \emph{mean centering} (MC) or \emph{gap centering} (GC), combined with ordering by variance (\emph{spread inside}, SI, or \emph{spread outside}, SO), leading to four possible parametrizations: (MC, SI), (MC, SO), (GC, SI), (GC, SO).

    \item \textbf{Principal nested spheres (PNS).}  
    After TOSS mapping, dimension reduction is carried out via PNS. An $\ell$-dimensional subsphere is
    \[
      \mathbb{S}_{V,\phi} = \{ y \in \mathbb{S}^k : V^T y = \phi \},
    \]
    where $V \in \mathbb{R}^{(k+1)\times(k-\ell)}$ has orthonormal columns and $\phi \in \mathbb{R}^{k-\ell}$ gives offsets. The projection of $y\in \mathbb{S}^k$ onto this subsphere is denoted $\pi_{\mathbb{S}_{V,\phi}}(y)$. By iteratively fitting subspheres of decreasing dimension, one obtains a nested sequence
    \[
      \mathbb{S}^k \supset \mathbb{S}^{k-1} \supset \cdots \supset \mathbb{S}^1.
    \]
    
    For $x \in \mathbb{T}^D$, write $y=\Phi(x)$. The residual distance to a subsphere is defined by
    \[
      \rho_\ell(\mathbb{S}_{V,\phi}, x) := d_{\mathbb{S}^k}\!\bigl(y, \pi_{\mathbb{S}_{V,\phi}}(y)\bigr),
    \]
    where $d_{\mathbb{S}^k}$ is geodesic distance. When $y$ lies near a singular stratum, suitable projected versions $\bar{y}^d$ (projections of $y$ to lower-dimensional strata $H_d/\!\sim$) are considered.

    \item \textbf{Statistical safeguard.}  
    To prevent overfitting, each fitted small subsphere is statistically compared against the corresponding great subsphere. A likelihood ratio test determines whether the smaller subsphere is genuinely supported by the data.
\end{enumerate}

In summary, tPCA maps torus data to a stratified sphere via TOSS, reparametrizes angles to avoid singularities, and applies principal nested spheres with statistical safeguards. This yields a PCA-like dimension reduction method that respects torus periodicity and geometry.

\section{Additional results of experiment 2} 
\label{appendix:simulation_results}

\begin{figure}[H]
\centering
\rotatebox{90}{ 
\begin{minipage}{\textheight} 
\subfloat{}{
  \includegraphics[width=0.12\textwidth]{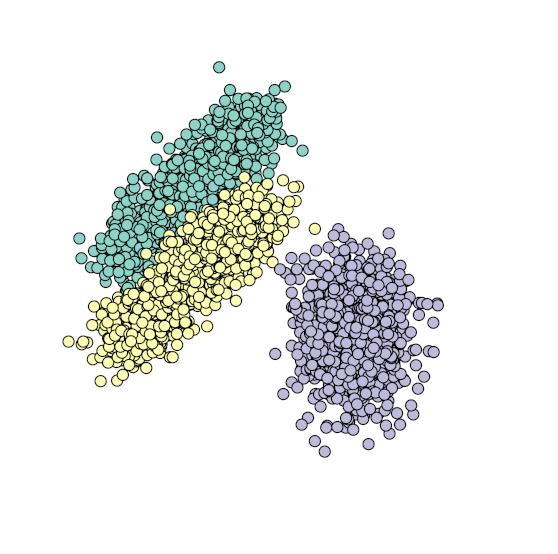}
}
\subfloat{}{
  \includegraphics[width=0.12\textwidth]{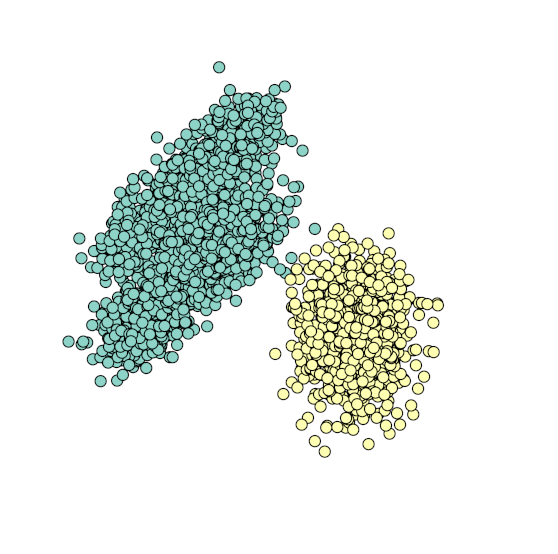}
}
\subfloat{}{
  \includegraphics[width=0.12\textwidth]{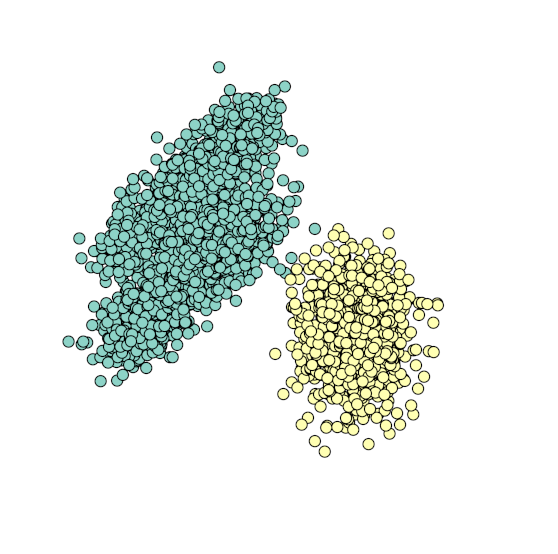}
}
\subfloat{}{
  \includegraphics[width=0.12\textwidth]{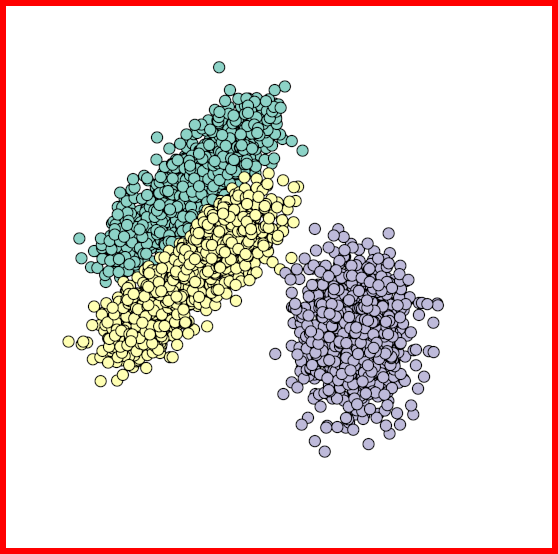}
}
\subfloat{}{
  \includegraphics[width=0.12\textwidth]{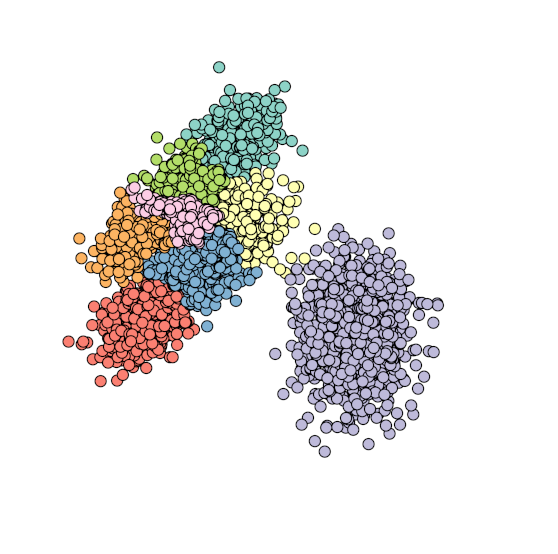}
}
\subfloat{}{
  \includegraphics[width=0.12\textwidth]{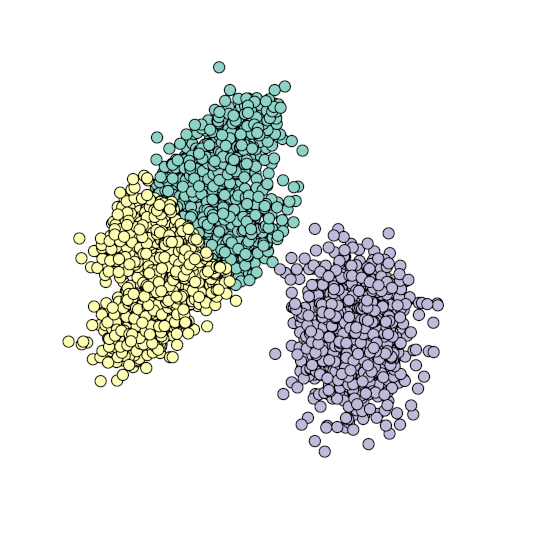}
}
\subfloat{}{
  \includegraphics[width=0.12\textwidth]{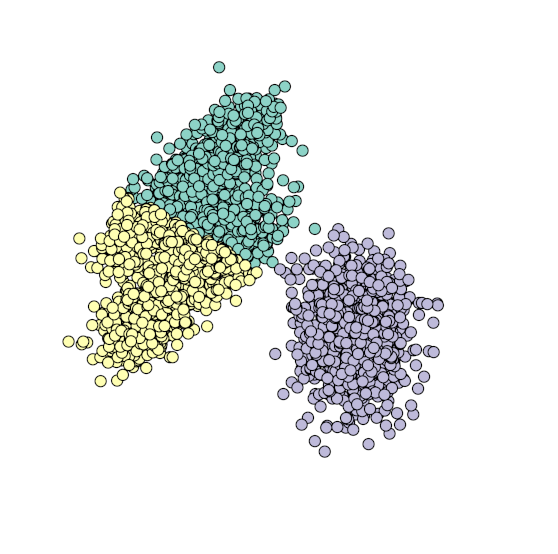}
}

\subfloat{}{
  \includegraphics[width=0.12\textwidth]{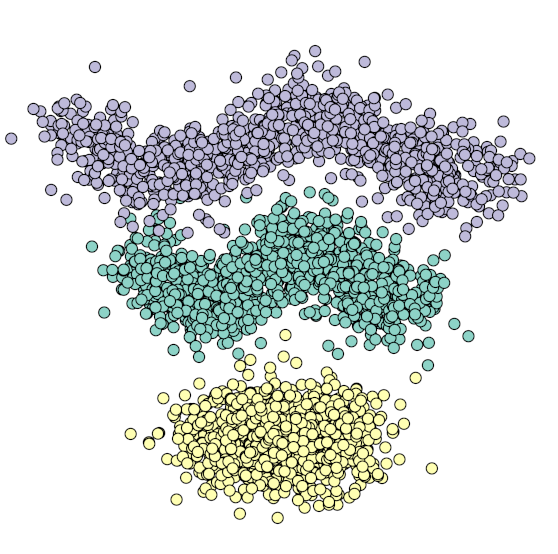}
}
\subfloat{}{
  \includegraphics[width=0.12\textwidth]{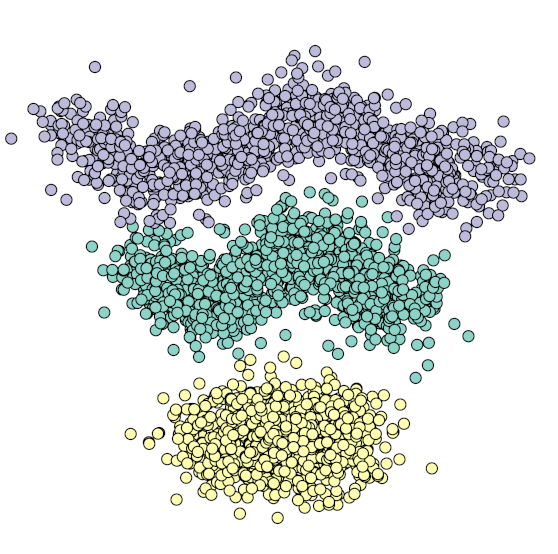}
}
\subfloat{}{
  \includegraphics[width=0.12\textwidth]{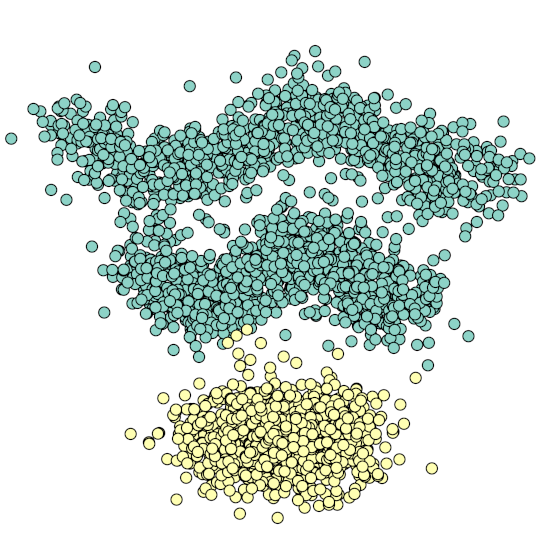}
}
\subfloat{}{
  \includegraphics[width=0.12\textwidth]{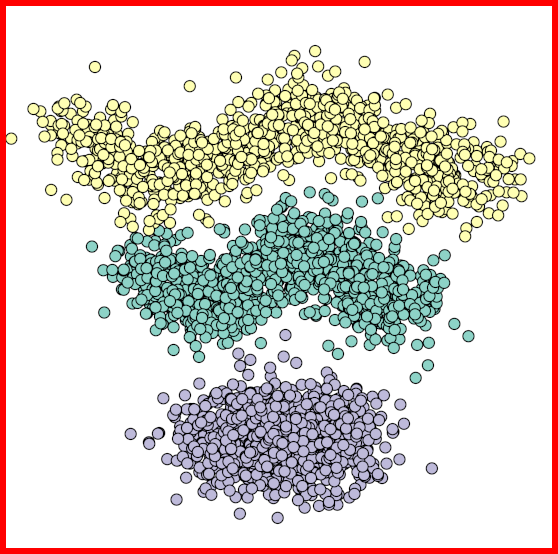}
}
\subfloat{}{
  \includegraphics[width=0.12\textwidth]{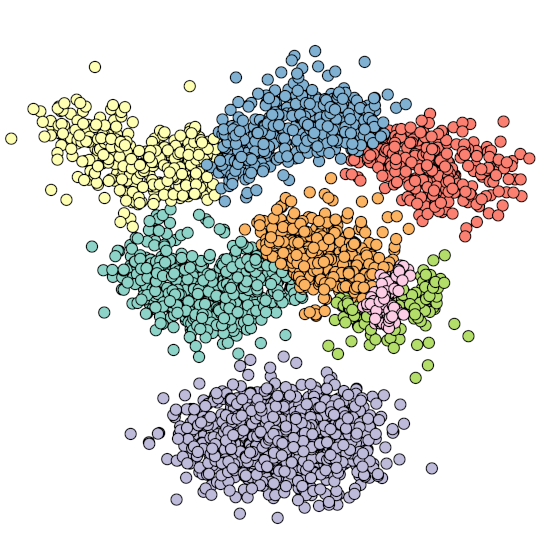}
}
\subfloat{}{
  \includegraphics[width=0.12\textwidth]{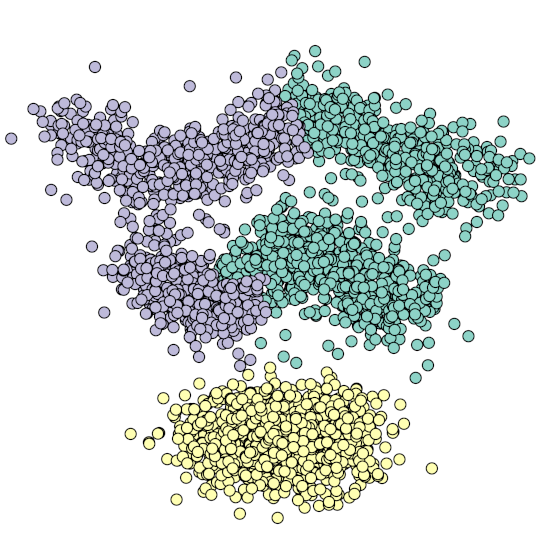}
}
\subfloat{}{
  \includegraphics[width=0.12\textwidth]{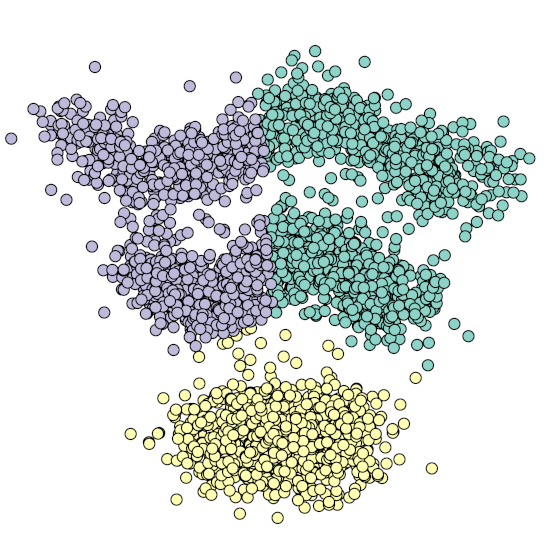}
}

\subfloat{}{
  \includegraphics[width=0.12\textwidth]{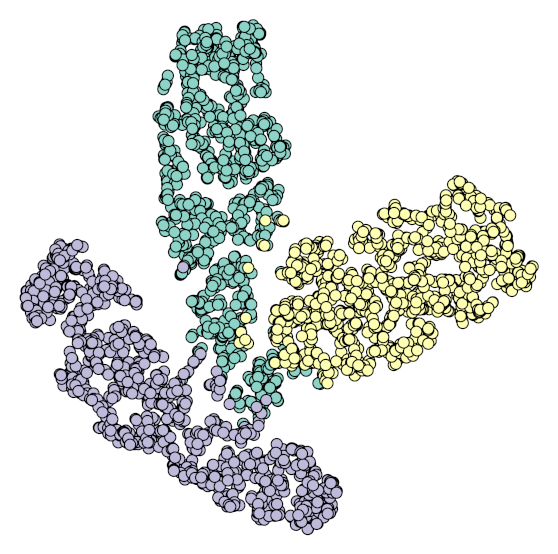}
}
\subfloat{}{
  \includegraphics[width=0.12\textwidth]{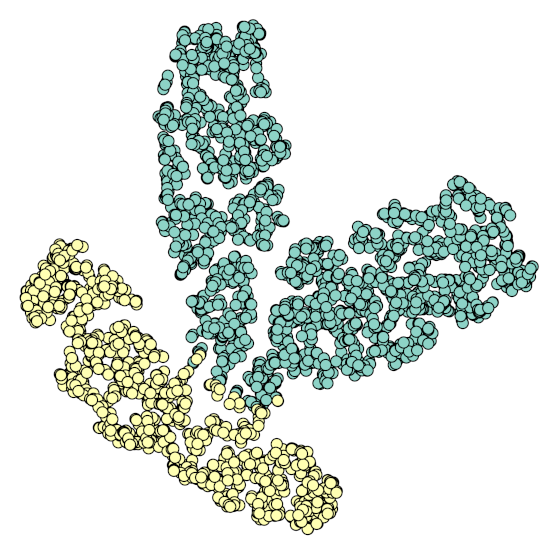}
}
\subfloat{}{
  \includegraphics[width=0.12\textwidth]{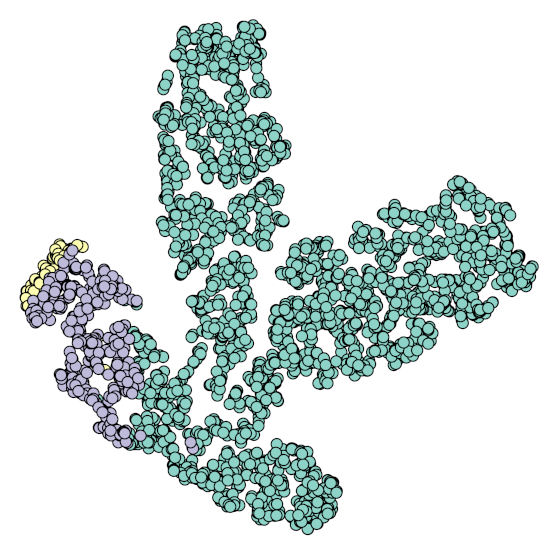}
}
\subfloat{}{
  \includegraphics[width=0.12\textwidth]{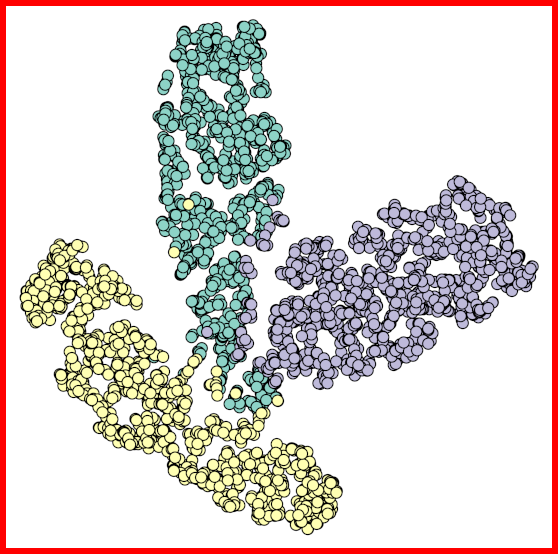}
}
\subfloat{}{
  \includegraphics[width=0.12\textwidth]{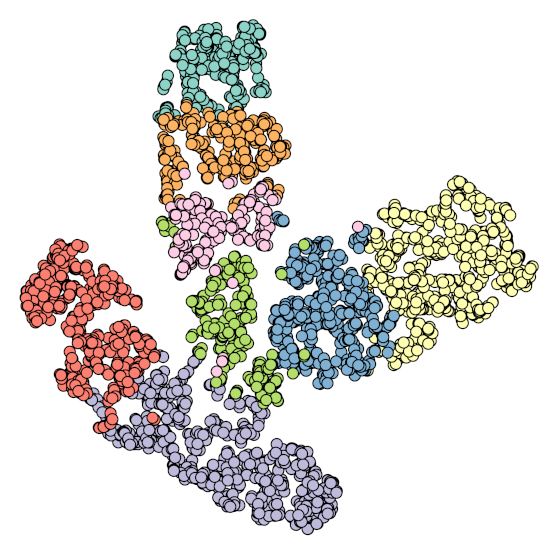}
}
\subfloat{}{
  \includegraphics[width=0.12\textwidth]{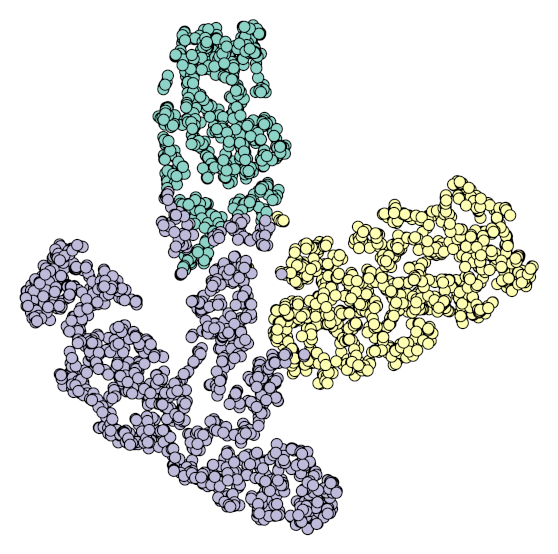}
}
\subfloat{}{
  \includegraphics[width=0.12\textwidth]{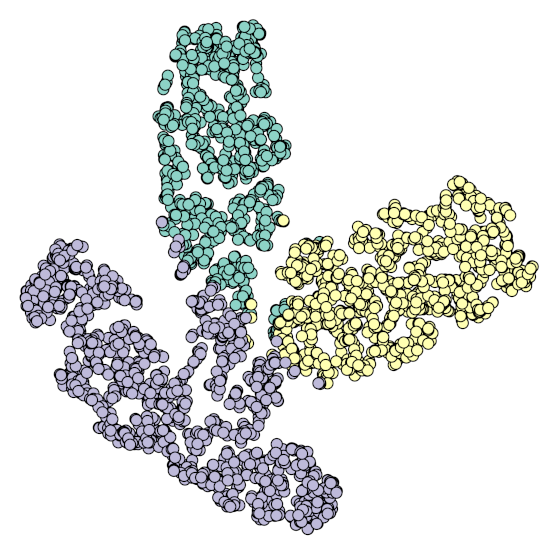}
}

\subfloat{}{
  \includegraphics[width=0.12\textwidth]{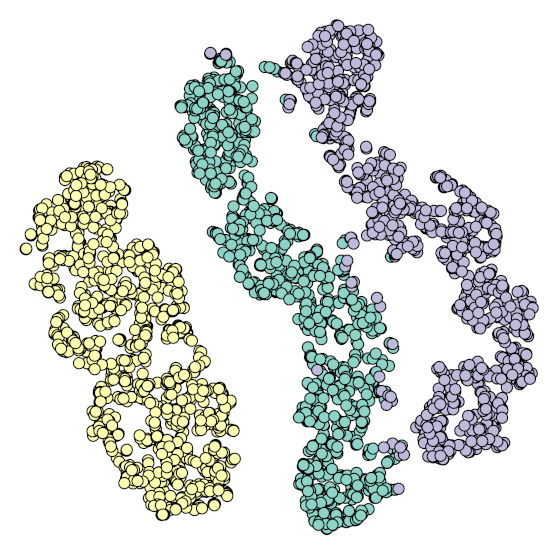}
}
\subfloat{}{
  \includegraphics[width=0.12\textwidth]{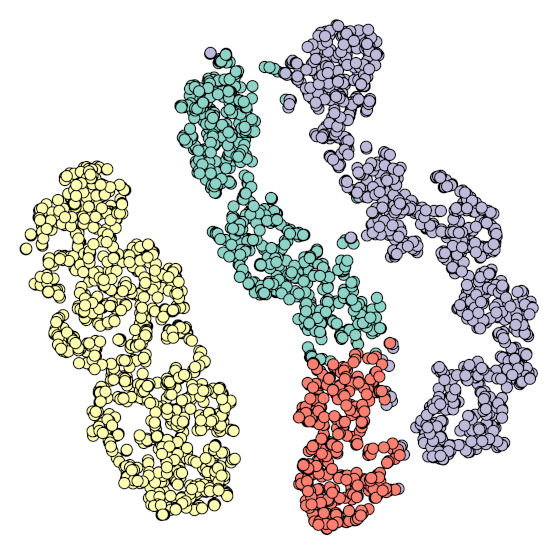}
}
\subfloat{}{
  \includegraphics[width=0.12\textwidth]{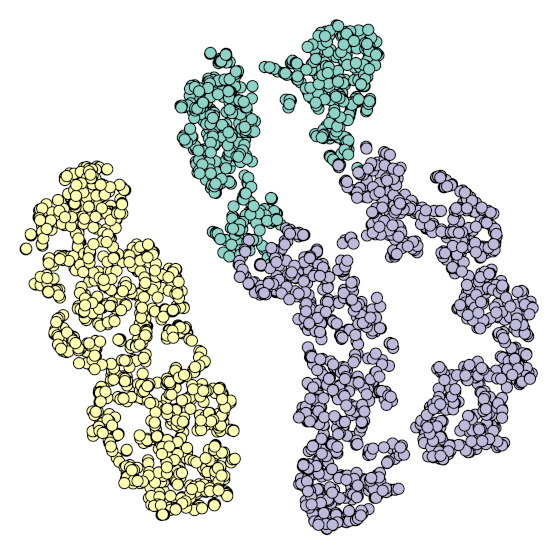}
}
\subfloat{}{
  \includegraphics[width=0.12\textwidth]{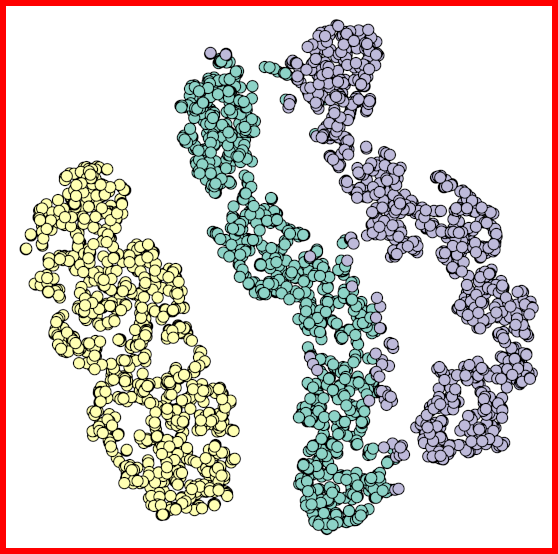}
}
\subfloat{}{
  \includegraphics[width=0.12\textwidth]{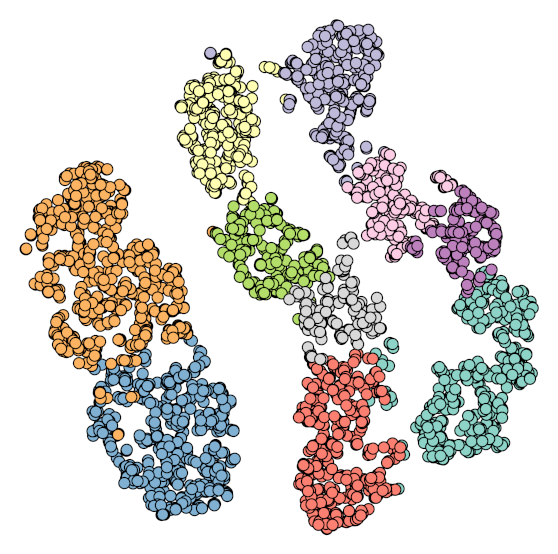}
}
\subfloat{}{
  \includegraphics[width=0.12\textwidth]{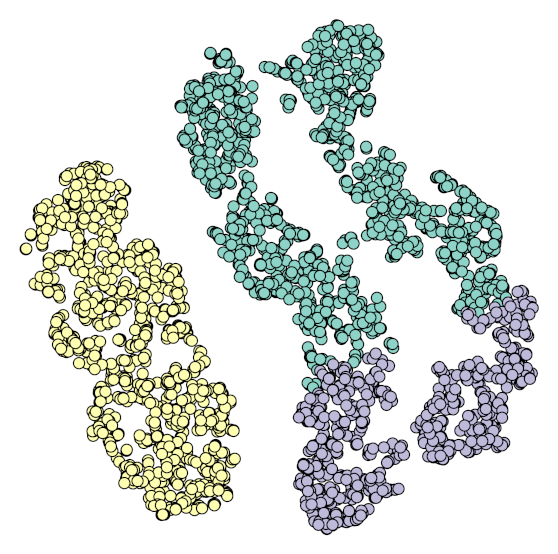}
}
\subfloat{}{
  \includegraphics[width=0.12\textwidth]{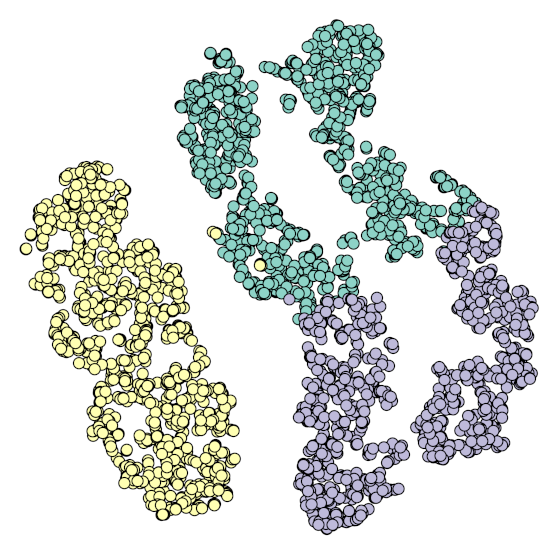}
}

\hspace{4.5em}
\raisebox{2em}{(a)} \hspace{7.5em}
\raisebox{2em}{(b)} \hspace{7.5em}
\raisebox{2em}{(c)} \hspace{7.5em}
\raisebox{2em}{(d)} \hspace{7.5em}
\raisebox{2em}{(e)} \hspace{7.5em}
\raisebox{2em}{(f)} \hspace{7.5em}
\raisebox{2em}{(g)}

\caption{\textit{An intuitive illustration of clustering performance for subexperiment $(a-d)$. Rows $1$ and $2$ correspond to subexperiments for $D=2$, rows $3$ and $4$ correspond to $D=3$. }}
\label{fig: simulation_appendix_1}
\end{minipage}
}
\end{figure}

\begin{figure}[H]
\centering
\rotatebox{90}{ 
\begin{minipage}{\textheight} 
\subfloat{}{
  \includegraphics[width=0.12\textwidth]{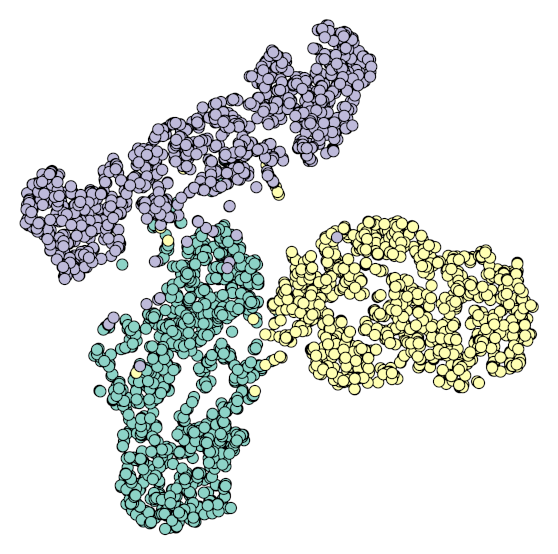}
}
\subfloat{}{
  \includegraphics[width=0.12\textwidth]{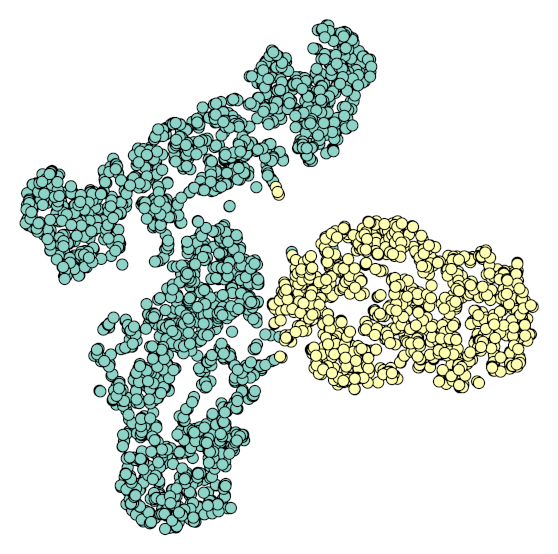}
}
\subfloat{}{
  \includegraphics[width=0.12\textwidth]{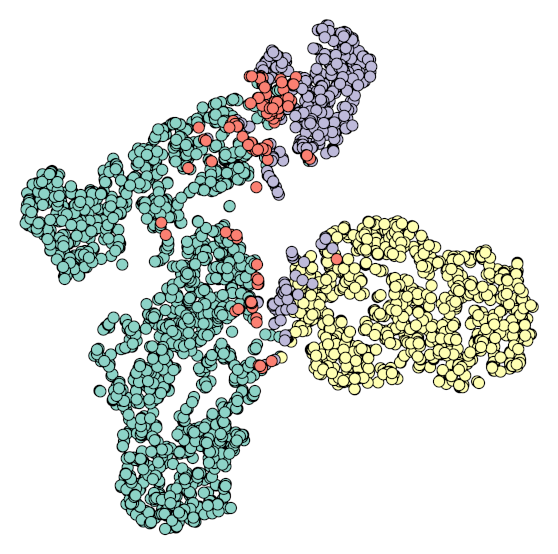}
}
\subfloat{}{
  \includegraphics[width=0.12\textwidth]{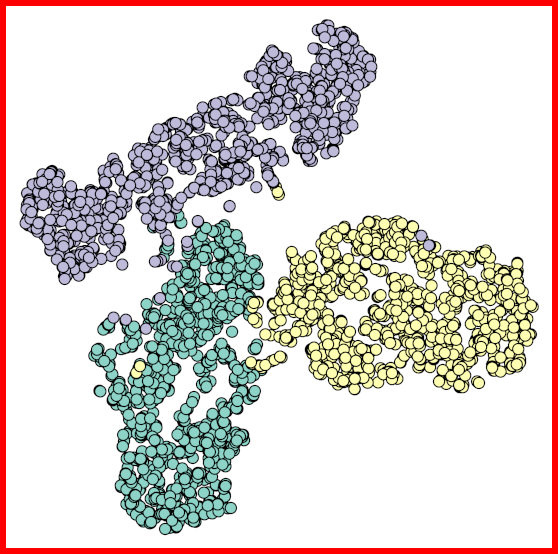}
}
\subfloat{}{
  \includegraphics[width=0.12\textwidth]{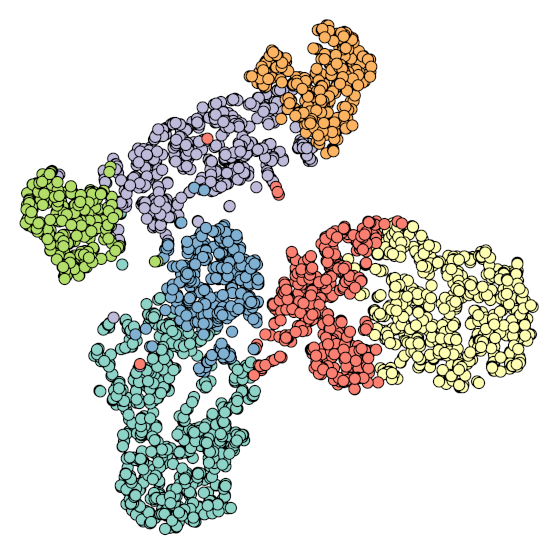}
}
\subfloat{}{
  \includegraphics[width=0.12\textwidth]{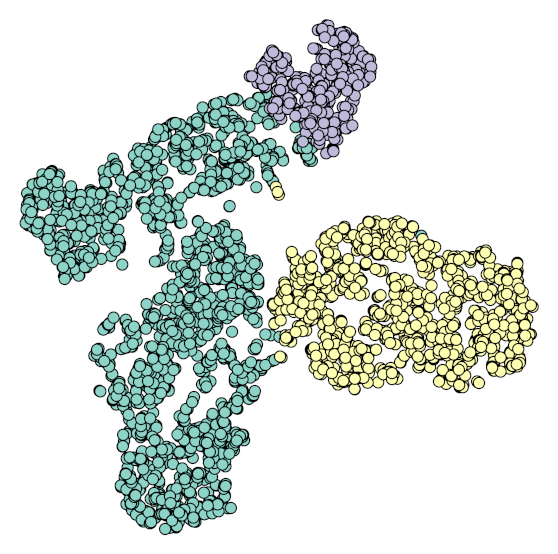}
}
\subfloat{}{
  \includegraphics[width=0.12\textwidth]{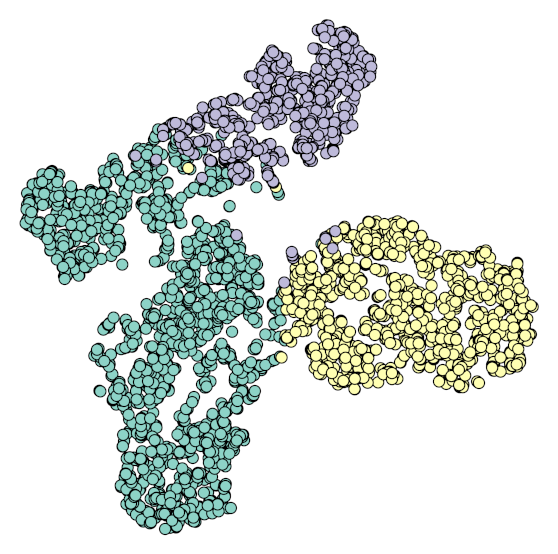}
}

\subfloat{}{
  \includegraphics[width=0.12\textwidth]{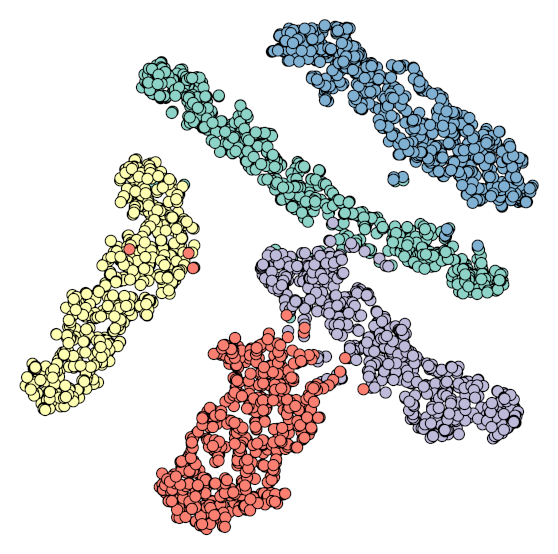}
}
\subfloat{}{
  \includegraphics[width=0.12\textwidth]{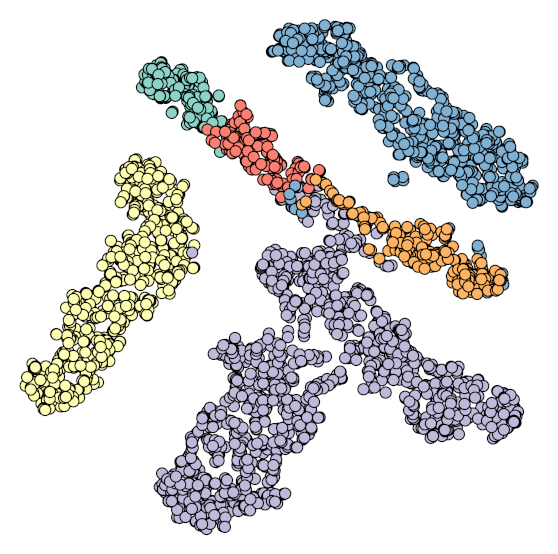}
}
\subfloat{}{
  \includegraphics[width=0.12\textwidth]{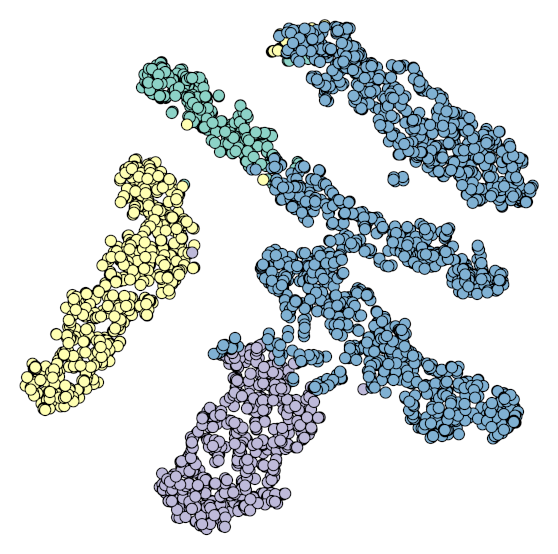}
}
\subfloat{}{
  \includegraphics[width=0.12\textwidth]{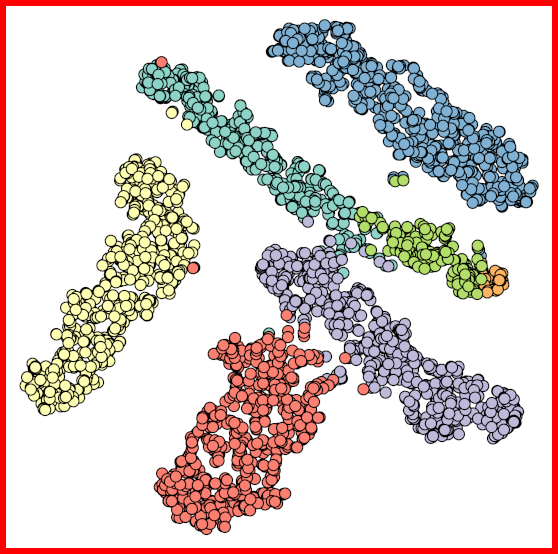}
}
\subfloat{}{
  \includegraphics[width=0.12\textwidth]{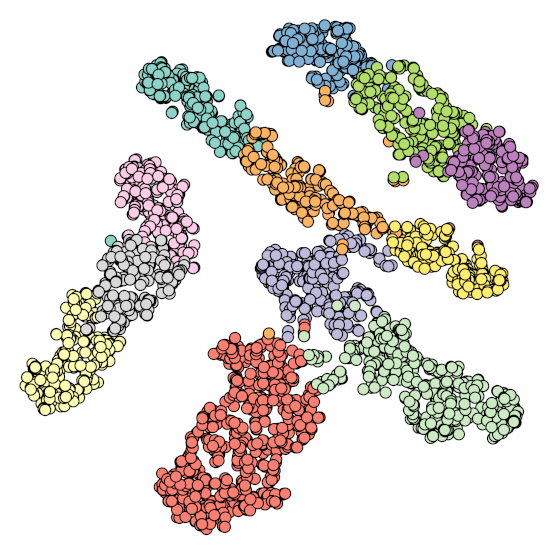}
}
\subfloat{}{
  \includegraphics[width=0.12\textwidth]{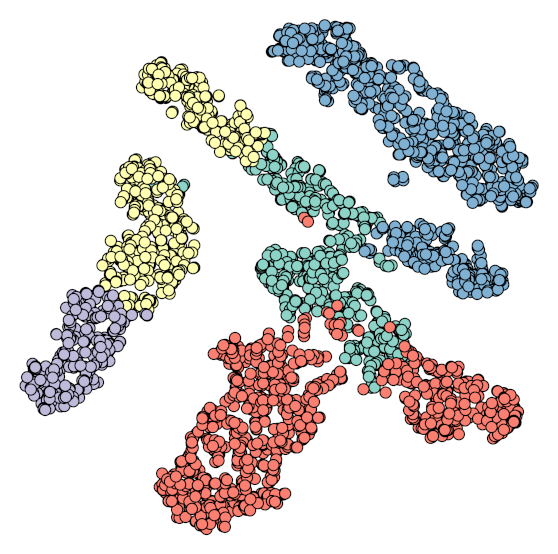}
}
\subfloat{}{
  \includegraphics[width=0.12\textwidth]{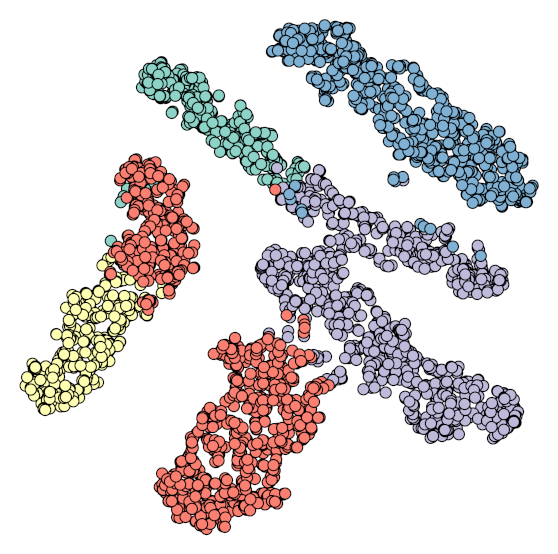}
}

\subfloat{}{
  \includegraphics[width=0.12\textwidth]{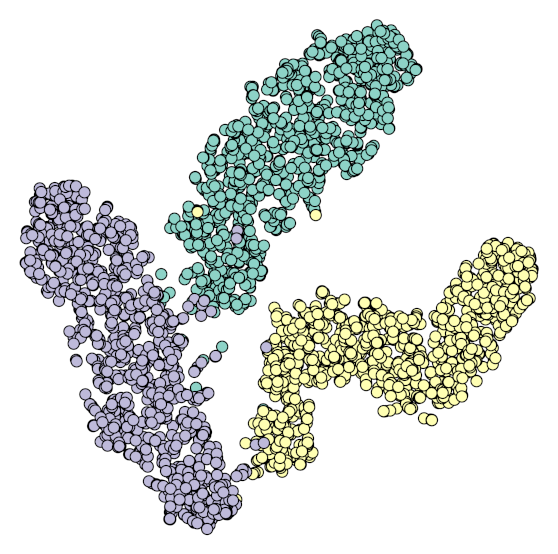}
}
\subfloat{}{
  \includegraphics[width=0.12\textwidth]{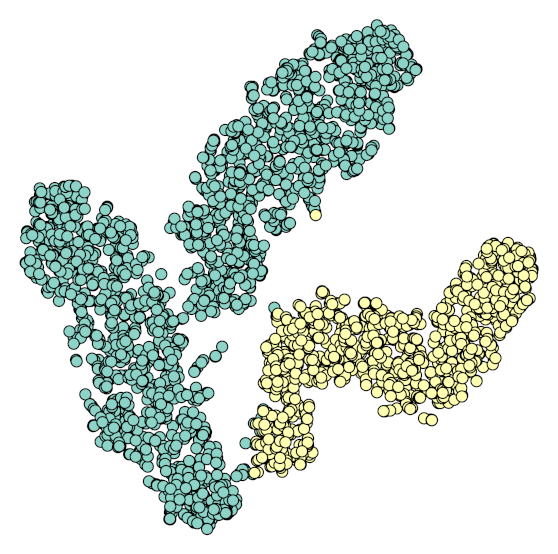}
}
\subfloat{}{
  \includegraphics[width=0.12\textwidth]{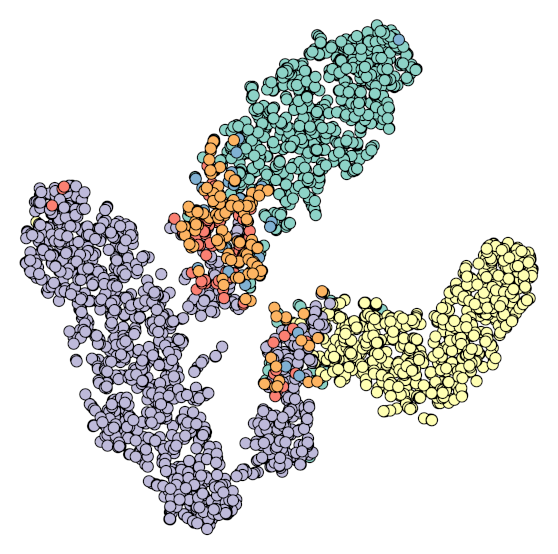}
}
\subfloat{}{
  \includegraphics[width=0.12\textwidth]{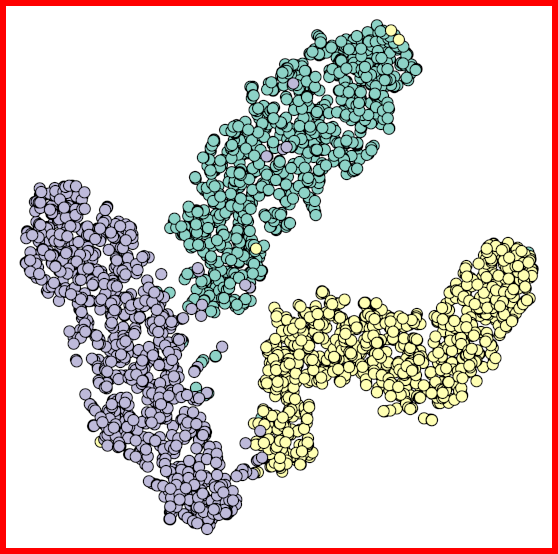}
}
\subfloat{}{
  \includegraphics[width=0.12\textwidth]{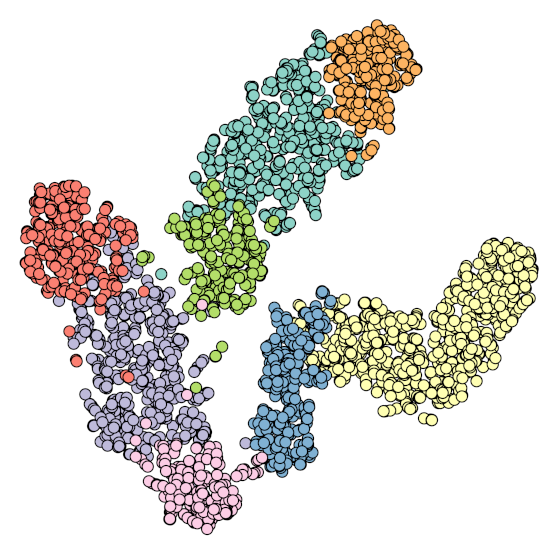}
}
\subfloat{}{
  \includegraphics[width=0.12\textwidth]{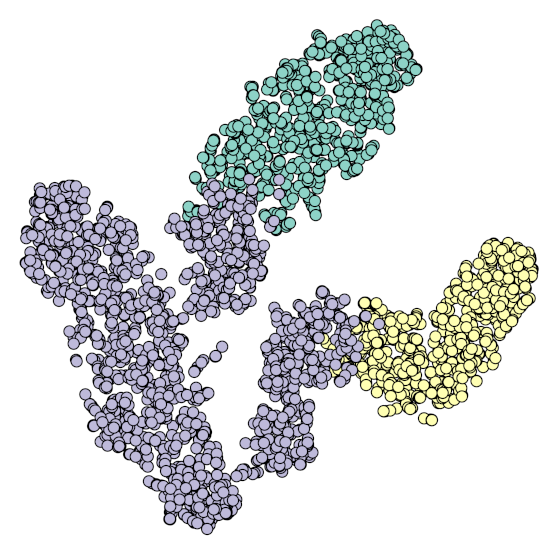}
}
\subfloat{}{
  \includegraphics[width=0.12\textwidth]{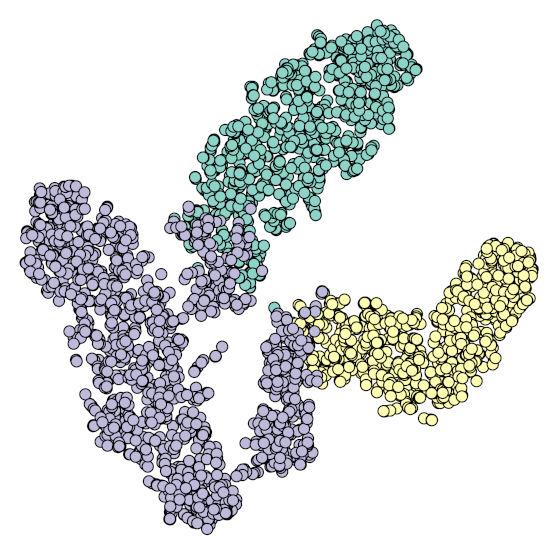}
}

\subfloat{}{
  \includegraphics[width=0.12\textwidth]{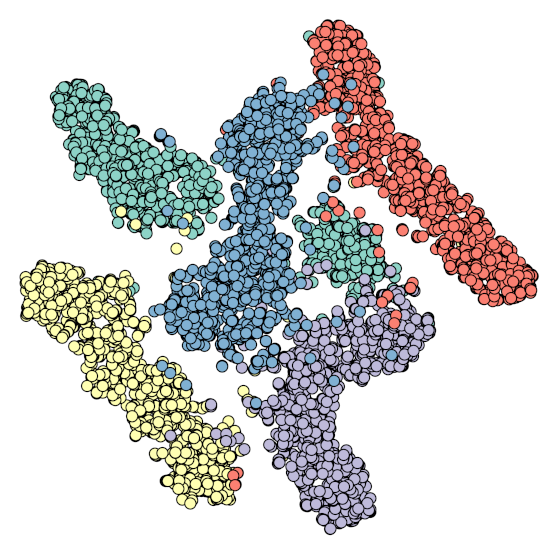}
}
\subfloat{}{
  \includegraphics[width=0.12\textwidth]{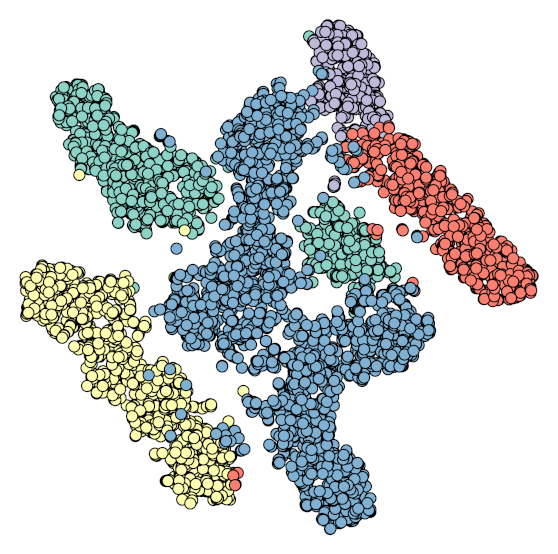}
}
\subfloat{}{
  \includegraphics[width=0.12\textwidth]{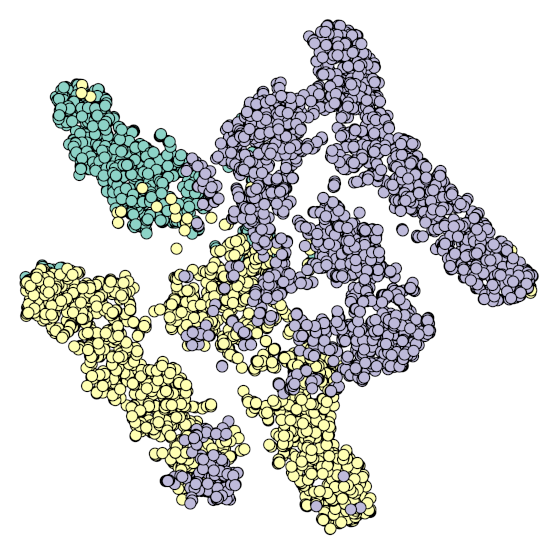}
}
\subfloat{}{
  \includegraphics[width=0.12\textwidth]{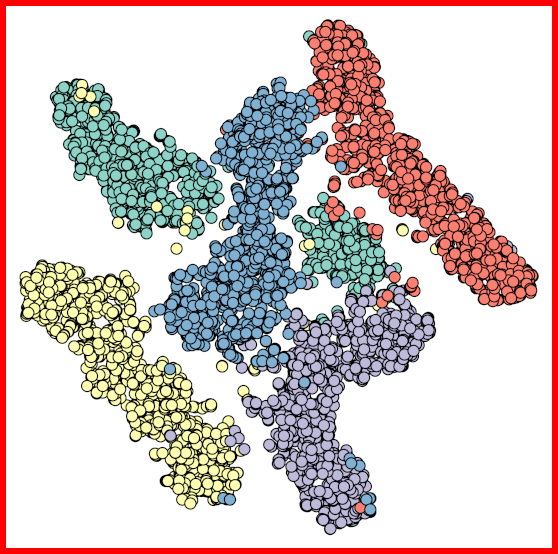}
}
\subfloat{}{
  \includegraphics[width=0.12\textwidth]{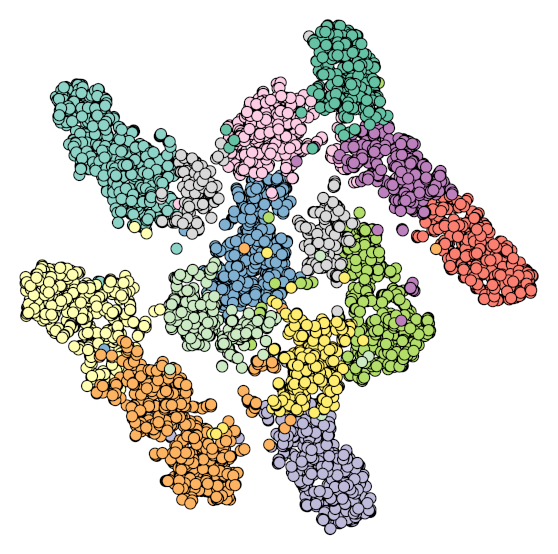}
}
\subfloat{}{
  \includegraphics[width=0.12\textwidth]{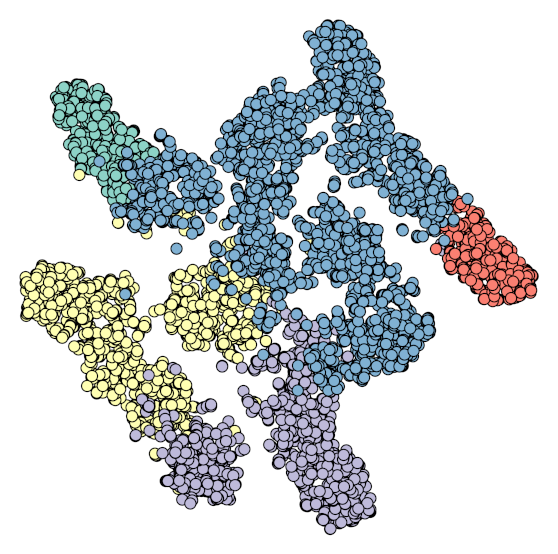}
}
\subfloat{}{
  \includegraphics[width=0.12\textwidth]{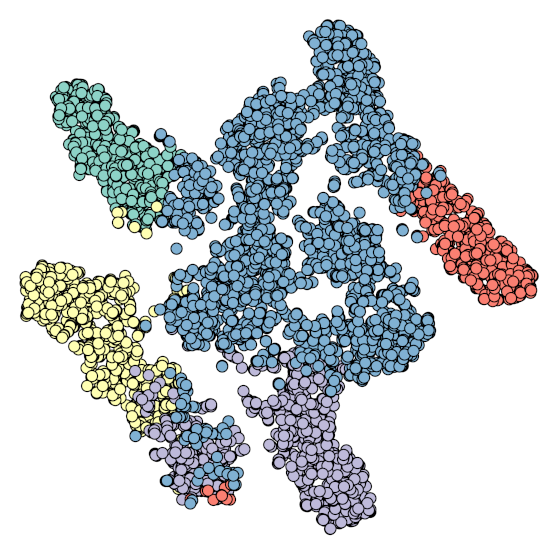}
}

\hspace{4em}
\raisebox{2em}{(a)} \hspace{7.5em}
\raisebox{2em}{(b)} \hspace{7.5em}
\raisebox{2em}{(c)} \hspace{7.5em}
\raisebox{2em}{(d)} \hspace{7.5em}
\raisebox{2em}{(e)} \hspace{7.5em}
\raisebox{2em}{(f)} \hspace{7.5em}
\raisebox{2em}{(g)}

\caption{\textit{An intuitive illustration of clustering performance for subexperiment $(e-h)$. Rows $1$ and $2$ correspond to subexperiments for $D=4$, rows $3$ and $4$ correspond to $D=5$. }}
\label{fig: simulation_appendix_2}
\end{minipage}
}
\end{figure}

In this section, we show the additional subexperiments of experiment 2 for dimensions $D=2$, $D=3$, $D=4$, and $D=5$, which are illustrated in Figure \ref{fig: simulation_appendix_1} and Figure \ref{fig: simulation_appendix_2}. These figures provide detailed clustering performance, enabling a comprehensive comparison between PSM-DBSCAN and other clustering methods in various dimensional settings. 

\section{Definitions of evaluation metrics}
\label{appendix: Definitions of evaluation metrics}
\subsection{Definitions of $\Ae$ and $P_d$}
\begin{definition}\label{de1}
\textit{The approximation error ($\Ae$) for the fitted samples $\widetilde{\mathcal{X}}$ is defined as:}

\begin{equation*}
\Ae = \sum_{i=1}^N \inf_{\psi \in \Psi} \operatorname{d}_{\mathbb{T}^D}(\widetilde{x}_{i}, \psi).
\end{equation*}
\end{definition}

\begin{definition}\label{de2}
\textit{
Given the $d$-dimensional fitted samples $\widetilde{\mathcal{X}}^d = \{\widetilde{x}^d_i\}_{i=1}^N$ and their corresponding $(d-1)$-dimensional fitted samples $\{\widetilde{x}^{d-1}_i\}_{i=1}^N$, the residuals $R_d$ is defined as:}

\begin{equation*}
\mathrm{R}_d = \frac{1}{N}\sum_{i=1}^N \operatorname{d}_{\psi^d}(\widetilde{x}^d_i, \widetilde{x}^{d-1}_i)^2,
\end{equation*}
\textit{where $\operatorname{d}_{\psi^d}$ represents the geodesic distance on the $d$-dimensional submanifold $\psi^d$.}
\end{definition}

\begin{definition}\label{de3}
\textit{
The proportion of information retained $\mathrm{P}_d$ in dimension $d$ is defined as: }

\begin{equation*}
\mathrm{P}_d  = \frac{\mathrm{R}_d}{\sum_{j=2}^D\mathrm{R}_j}.
\end{equation*}
\end{definition}

\subsection{Definitions of ARI and NMI}
\begin{definition}\label{de4}
\textit{
Given the samples $\mathcal{X}$ divided into two different sets of clusters $\mathcal{U}$ and $\mathcal{V}$, where $\mathcal{U}$ is the set of true labels and $\mathcal{V}$ is the set of labels produced by a clustering algorithm. Define the contingency table $\mathbf{L}$ where $n_{ij}$ is the number of samples that are in both cluster $\mathcal{U}_i$ in $\mathcal{U}$ and cluster $\mathcal{V}_j$ in $\mathcal{V}$. Let $a_i$ and $b_j$ be the sums of the rows and columns of $\mathbf{L}$, respectively:
}
\begin{equation*}
a_i = \sum_{j} n_{ij}, \quad b_j = \sum_{i} n_{ij}.
\end{equation*}
\textit{
The Rand Index (RI) is then calculated as:
}
\begin{equation*}
\text{RI} = \frac{\sum_{ij} \binom{n_{ij}}{2} + \sum_{ij} \binom{a_i - n_{ij}}{2} + \sum_{ij} \binom{b_j - n_{ij}}{2}}{\binom{n}{2}}.
\end{equation*}
\textit{
The Expected Rand Index (ERI) under random labeling is given by:
}
\begin{equation*}
\text{ERI} = \frac{\sum_{i} \binom{a_i}{2} \sum_{j} \binom{b_j}{2}}{\binom{n}{2}^2}.
\end{equation*}
\textit{
ARI is then defined as the normalized difference between RI and ERI:
}
\begin{equation*}
\text{ARI} = \frac{\text{RI} - \text{ERI}}{\max(\text{RI}) - \text{ERI}}.
\end{equation*}
\end{definition}

\begin{definition}\label{de5}
\textit{
 Define $H(\mathcal{U})$ and $H(\mathcal{V})$ as the entropies of $\mathcal{U}$ and $\mathcal{V}$, respectively, and $I(\mathcal{U}; \mathcal{V})$ as the mutual information between $\mathcal{U}$ and $\mathcal{V}$:
}
\begin{equation*}
H(\mathcal{U}) = -\sum_{i} \frac{\#(\mathcal{U}_i)}{N} \log \frac{\#(\mathcal{U}_i)}{N}, \quad H(\mathcal{V}) = -\sum_{j} \frac{\#(\mathcal{V}_j)}{N} \log \frac{\#(\mathcal{V}_j)}{N},
\end{equation*}
\begin{equation*}
I(\mathcal{U}; \mathcal{V}) = \sum_{i,j} \frac{\#(\mathcal{U}_i \cap \mathcal{V}_j)}{N} \log \frac{N \#(\mathcal{U}_i \cap \mathcal{V}_j)}{\#(\mathcal{U}_i) \#(\mathcal{V}_j)}.
\end{equation*}
\textit{
NMI is then defined as the ratio of mutual information and the mean of the entropies of $\ \mathcal{U}$ and $\mathcal{V}$:
}
\begin{equation*}
\text{NMI} = \frac{2 I(\mathcal{U}; \mathcal{V})}{H(\mathcal{U}) + H(\mathcal{V})}.
\end{equation*}
\end{definition}

\subsection{Definitions of DB and SI}
\begin{definition}\label{de6}
\textit{
Given a set of samples $\mathcal{X}$. For each sample \(x_i\) assigned to cluster \(C_m\):}

\begin{itemize}
    \item \textit{Let \(\mu^{a}(x_i)\) be the mean distance between \(x_i\) and all other samples in the same cluster \(C_j\):}
    \begin{equation*}
    \mu^{a}(x_i) = \frac{1}{\#(\mathcal{C}_m) - 1} \sum_{\substack{x_j \in \mathcal{C}_j \\ j \neq i}} \operatorname{d}(x_i, x_j),
    \end{equation*}
    \item \textit{Let \(\mu^{b}(x_i)\) be the minimum mean distance between \(x_i\) and all samples in any other cluster \(C_l\), for \(l \neq m\):}
    \begin{equation*}
    \mu^{b}(x_i) = \min_{l \neq m} \left\{ \operatorname{d}(x_i, \mathcal{C}_l) \right\} = \min_{l \neq m} \left\{ \frac{1}{\#(\mathcal{C}_l)} \sum_{x_j \in \mathcal{C}_l} \operatorname{d}(x_i, x_j) \right\}.
    \end{equation*}
\end{itemize}
\textit{
For each sample \(x_i\), \(\si(x_i)\) is defined as:
}
\begin{equation*}
\si(x_i) = \frac{\mu^{b}(x_i) - \mu^{a}(x_i)}{\max\{\mu^{a}(x_i), \mu^{b}(x_i)\}}.
\end{equation*}
\textit{
\(\SI\) for a clustering method is then defined as the mean of \(\si(x_i)\) of all samples:
}
\begin{equation*}
\SI = \frac{1}{N} \sum_{i=1}^N \si(x_i).
\end{equation*}
\end{definition}

\begin{definition}\label{de7}
\textit{
Given a set of samples \(\mathcal{X}\), let \(r\) denote the number of clusters. Let \(\mathcal{C}
_j\) be the \(j\)-th cluster set with centroid \(u_j = \frac{1}{|\mathcal{C}_j|} \sum_{\mathbf{x}_i \in \mathcal{C}_j} \mathbf{x}_i,
\ j=1,\dots,r\). For each cluster set \(\mathcal{C}
_j\), define:}
\textit{The average intra-cluster distance:}
    \begin{equation*}
    \Theta_j = \frac{1}{\#(\mathcal{C}
_j)} \sum_{x_i \in \mathcal{C}
_j} \|x_i - u_j\|_2.
    \end{equation*}

\textit{
Then the $\DB$ index is defined as:}
\begin{equation*}
\DB = \frac{1}{r} \sum_{j=1}^r \max_{m \neq j} \left\{ \frac{\Theta_j + \Theta_m}{\|u_j - u_m\|_2} \right\}.
\end{equation*}

\end{definition}

\end{document}